\documentclass{jpp}

\usepackage{graphicx}
\usepackage{natbib}

\usepackage{graphics}
\usepackage{epsfig}
\usepackage{bm}
\usepackage{amsmath,amssymb}
\usepackage{stmaryrd}

\def\ov#1{\overline{#1}}

\def\vb#1{\mbox{\boldmath$#1$}}
\def\pd#1#2{\frac{\partial #1}{\partial #2}}
\def\fd#1#2{\frac{\delta #1}{\delta #2}}
\def\wh#1{\widehat{#1}}
\def\bdot{\,\vb{\cdot}\,}
\def\btimes{\,\vb{\times}\,}

\def\bhat{\wh{{\sf b}}}
\def\cal#1{\mathcal{#1}}

\def\exd{{\sf d}}
\def\bhat{\wh{{\sf b}}}

\newcommand{\bc}{\begin{center}}
\newcommand{\ec}{\end{center}}
\newcommand{\bt}{\begin{tabbing}}
\newcommand{\et}{\end{tabbing}}
\newcommand{\be}{\begin{equation}}
\newcommand{\ee}{\end{equation}}
\newcommand{\ba}{\begin{eqnarray}}
\newcommand{\ea}{\end{eqnarray}}

\begin{document}

\title{Lifting of the Vlasov-Maxwell Bracket by Lie-transform Method}

\author[Brizard, Morrison, Burby, de Guillebon, and Vittot]{A.~J.~Brizard$^1$, P.~J.~Morrison$^2$, J.~W.~Burby$^3$, L.~de Guillebon$^4$, and M.~Vittot$^4$}
\affiliation{$^1$Department of Physics, Saint Michael's College, Colchester, VT 05439, USA\\ 
$^2$Department of Physics and Institute for Fusion Studies, University of Texas at Austin, Austin, TX 78712, USA \\ 
$^3$Courant Institute of Mathematical Sciences, New York, New York 10012, USA \\
$^4$Aix Marseille Univ., Univ.~Toulon, CNRS, CPT, Marseille, France}

\maketitle

\begin{abstract}
The Vlasov-Maxwell equations possess a Hamiltonian structure expressed in terms of a Hamiltonian functional and a functional bracket. In the present paper, the transformation (``lift'') of the Vlasov-Maxwell bracket induced by the dynamical reduction of single-particle dynamics is investigated when the reduction is carried out by Lie-transform perturbation methods.  The ultimate goal of this work is to derive explicit Hamiltonian formulations for the guiding-center and gyrokinetic Vlasov-Maxwell equations that have important applications in our understanding of turbulent magnetized plasmas.
\end{abstract}

\section{Introduction}

Reduced plasma models play an important role in the analytical and numerical investigations of the complex nonlinear dynamics of magnetized plasmas. The process of dynamical reduction is generally based on the elimination of fast time scales from either kinetic plasma equations or fluid plasma equations. In the kinetic case, the dynamical reduction is usually carried out by considering  a sequence of phase-space transformations designed to eliminate a fast orbital time scale (e.g., the time scale associated with the fast gyromotion of a charged particle about a magnetic-field line) from the plasma kinetic equations. In the fluid case, on the other hand, the fast time scale is often of dynamical origin (e.g., the fast compressional Alfv\'{e}n wave time scale in a  strongly-magnetized plasma), and the dynamical reduction involves the identification of a small number of fluid moments and electromagnetic-field components that capture the desired reduced fluid dynamics (e.g., reduced magnetohydrodynamics (MHD)).

Reduced plasma models can be either  dissipationless or dissipative, depending on the dynamical time scales of interest. When considering the complex nonlinear dynamics of high-temperature magnetized plasmas, e.g., magnetized fusion plasmas,  dissipationless reduced plasma models can offer great mathematical simplicity since the time scales of interest may be much shorter than collisional (dissipative) time scales. Furthermore, when a dissipationless reduced plasma model is shown to possess a Hamiltonian structure, the powerful methods of Hamiltonian field theory can be brought to bear in understanding its analytical and numerical solutions.

 A fundamental question thus arises when reducing a system of Hamiltonian field equations: Will the dissipationless form of the reduced system possess a Hamiltonian structure? In the case of several reduced dissipationless fluid plasma models derived from Hamiltonian plasma models, the underlying reduced Hamiltonian structure can be constructed from the reduced equations themselves, as was done for MHD \citep{pjmG80} and many other plasma models \citep{pjm82}.  The purpose of the present work is to investigate whether the Hamiltonian structure of reductions of the Vlasov-Maxwell equations can be constructed directly from the original Hamiltonian structure of the these equations.  This will be accomplished by the process of {\it lifting} introduced by \cite{pjm13} and further developed by \cite{pjmVG13}.  The present work is a continuation of these earlier works that allows for greater generality, while casting the formalism in the language of Lie transforms that  is commonplace in gyrokinetic theory \citep{Brizard_Hahm_2007}. 


The remainder of this paper is organized as follows. First we consider some preliminary basics  in Sec.~\ref{sec:prelim}; viz.,   the definition of what constitutes a Hamiltonian field theory is given and followed by some general comments on coordinate  changes.  Then, in Sec.~\ref{sec:Ham_VM}, we review the Hamiltonian structure of the Vlasov-Maxwell equations, where the Hamiltonian functional and the Vlasov-Maxwell bracket are presented. In Sec.~\ref{sec:general_lift}, we derive the general transformation (lift) procedure of the Hamiltonian structure of the Vlasov-Maxwell equations based on a general class of phase-space transformations that depends on the electromagnetic fields $({\bf E}, {\bf B})$.  These transformations are complicated because they depend on both independent and dependent variables; therefore, we  introduce  operators on functions and meta-operators on functionals to facilitate the  transformation of  the Vlasov-Maxwell equations as well as the Vlasov-Maxwell Poisson bracket. Following this general phase-space transformation, we therefore show how functional derivatives appearing in the Vlasov-Maxwell bracket are lifted to a new function space. In Sec.~\ref{sec:local}, we demonstrate this general lifting procedure by considering a preliminary transformation from particle phase space to local phase-space coordinates that depends on the local magnetic field ${\bf B}({\bf x})$ only. As a result of the preliminary local phase-space transformation, however, the evolution of the local Vlasov function now depends explicitly on the fast gyromotion time scale, which must be removed by the near-identity guiding-center phase-space transformation (see \cite{Cary_Brizard_2009} for a recent review). 

In Sec.~\ref{sec:Dyn_Red}, we review the process of dynamical reduction of the phase-space particle dynamics by Lie-transform perturbation methods. Here, the dynamical reduction requires that a near-identity transformation be applied to the local phase-space coordinates. In Sec.~\ref{sec:Red_VM}, we construct the reduced Vlasov-Maxwell equations by Lie-transform and meta-operator methods and, in Sec.~\ref{sec:Lift_VM}, we derive the reduced Vlasov-Maxwell bracket by the application of meta-operators on the Vlasov-Maxwell bracket. We also verify that the reduced Vlasov-Maxwell equations can be expressed as Hamiltonian field equations in terms of the reduced Hamiltonian functional and the reduced Vlasov-Maxwell bracket. 
Lastly,  Apps.~\ref{sec:Jacobi_PB}-\ref{sec:Jacobi_MV} present explicit proofs of the Jacobi identity for the single-particle Poisson bracket and the Vlasov-Maxwell bracket, while Apps.~\ref{sec:comm_op}-\ref{sec:explicit_bracket} present near-identity operator identities used in the text.

\section{\label{sec:prelim}Preliminaries}

Although  Hamiltonian descriptions of plasma dynamical systems are discussed in several sources  \citep{pjm82, pjm98, pjm05},  for self-containedness we  briefly review here some basics of  Hamiltonian field theory, before continuing on to the Vlasov-Maxwell theory.

\subsection{General Hamiltonian field theory}

The Hamiltonian formulation of a general field theory involving an $N$-component field $\vb{\Psi} = (\psi^{1},\cdots,\psi^{N})$ is expressed in terms of a Hamiltonian functional ${\cal H}[\vb{\Psi}]$, identified from the energy conservation law of the field equations, and a bracket structure 
\begin{equation}
\left[{\cal F},\frac{}{} {\cal G}\right]_{\Psi} \;\equiv\; \int_{\bf r}\;\fd{\cal F}{\psi^{a}({\bf r})}\;\mathbb{J}^{ab}(\vb{\Psi}; {\bf r})\;
\fd{\cal G}{\psi^{b}({\bf r})},
\label{eq:bracket_def}
\end{equation}
where $\int_{\bf r}$ denotes an integration over the base space for the fields $\vb{\Psi}$. Here, the $N\times N$ matrix-operator $\mathbb{J}^{ab} = -\,\mathbb{J}^{ba}$ is antisymmetric, while ${\cal F}[\vb{\Psi}]$ and ${\cal G}[\vb{\Psi}]$ are arbitrary functionals; summation over repeated indices is implied throughout the manuscript and explicit time dependence is not displayed unless necessary. In addition, functional derivatives 
$\delta{\cal F}/\delta\psi^{a}$ are defined in terms of the Fr\'{e}chet derivative:
\[ \delta{\cal F} \;\equiv\; \left.\frac{d}{d\varepsilon}\right|_{\varepsilon = 0}\;{\cal F}[\vb{\Psi} + \varepsilon\,\delta\vb{\Psi}] \;=\;
\int_{\bf r}\;\fd{\cal F}{\psi^{a}({\bf r})}\;\delta\psi^{a}({\bf r}), \]
which may involve integration by parts if the functional ${\cal F}$ depends on $\nabla\psi^{a}$.

Since the operator $\mathbb{J}^{ab} = -\,\mathbb{J}^{ba}$ is antisymmetric, the bracket \eqref{eq:bracket_def} is also antisymmetric:
\begin{equation}
\left[{\cal F},\frac{}{} {\cal G}\right]_{\Psi} \;=\; -\;\left[{\cal G},\frac{}{} {\cal F}\right]_{\Psi}.
\label{eq:antisym}
\end{equation} 
The bracket also possesses the Leibniz property
\begin{equation}
\left[{\cal F},\frac{}{} {\cal G}\,{\cal K}\right]_{\Psi} \;=\; \left[{\cal F},\frac{}{} {\cal G}\right]_{\Psi}\;{\cal K} \;+\; {\cal G}\;\left[{\cal F},\frac{}{} {\cal K}\right]_{\Psi},
\label{eq:Leibniz}
\end{equation}
and it satisfies the Jacobi identity
\begin{equation}
\left[{\cal F},\frac{}{} [{\cal G},\;{\cal K}]_{\Psi}\right]_{\Psi} \;+\; \left[{\cal G},\frac{}{} [{\cal K},\;{\cal F}]_{\Psi}\right]_{\Psi} \;+\;
\left[{\cal K},\frac{}{} [{\cal F},\;{\cal G}]_{\Psi}\right]_{\Psi} \;\equiv\; 0,
\label{eq:Jacobi_def}
\end{equation}
where ${\cal F}$, ${\cal G}$, and ${\cal K}$ are arbitrary functionals. Using the Hamiltonian functional ${\cal H}[\vb{\Psi}]$ and the bracket 
\eqref{eq:bracket_def}, the field equations for $\vb{\Psi}$ are expressed in Hamiltonian form as
\begin{equation}
\pd{\cal F}{t} \;\equiv\; \left[{\cal F},\frac{}{} {\cal H}\right]_{\Psi} \;=\; \int_{\bf r}\;\fd{\cal F}{\psi^{a}({\bf r})}\;\left(
\mathbb{J}^{ab}(\vb{\Psi};{\bf r})\;\fd{\cal H}{\psi^{b}({\bf r})} \right) \;\equiv\; \int_{\bf r}\;\fd{\cal F}{\psi^{a}({\bf r})}\;
\pd{\psi^{a}({\bf r})}{t}.
\label{eq:psi_dot}
\end{equation}
While an antisymmetric matrix-operator $\mathbb{J}^{ab}$ that satisfies the Leibniz property  \eqref{eq:Leibniz} is relatively easy to construct, the Jacobi identity \eqref{eq:Jacobi_def} is generally difficult to satisfy.  (See \cite{pjm82} where the Jacobi identity is discussed in generality.)

The purpose of the present paper is to investigate how the bracket structure \eqref{eq:bracket_def} is affected by a  field transformation $\vb{\Psi} \rightarrow  \ov{\vb{\Psi}}$, where  the transformation depends on both independent and dependent variables, as noted above,  with the new fields $\ov{\vb{\Psi}}$ having  desirable  properties making them amenable to  theoretical and/or numerical reduction. 

\subsection{Functional transformation of a Hamiltonian bracket}

As noted above, the question of how Hamiltonian functional brackets transform has recently been studied in the ``lift'' context by \cite{pjm13} and \cite{pjmVG13} for the Vlasov-Maxwell equations.  In particular, \cite{pjmVG13} studied the process of lifting associated with a change of momentum-space coordinates ${\bf p} \rightarrow \ov{\bf p}$ (at a fixed particle position ${\bf x}$) that is dependent on the local magnetic field; this case is revisited here  in Sec.~\ref{sec:local}.  Alternatively,  one may consider transformations in the context of Dirac constraint theory \citep{pjmLB09, pjmCT12, pjmCGBT13}.  In particular,  \cite{squire} considered the construction of a reduced bracket for the gyrokinetic Vlasov-Poisson equations by using  this method.  Another context is that of ``beatification'' \citep{pjmV16, pjmVC16} where one uses transformations to perturbatively remove nonlinearity from the Poisson bracket and place it in the Hamiltonian functional.  

In the present paper, we study the lifting of the Vlasov-Maxwell bracket associated with a phase-space coordinate transformation 
\begin{equation}
{\cal T}_{\epsilon} \;\equiv\; {\cal T}^{\epsilon}\,\circ\,{\cal T}_{0}:\;\; {\bf z} \;\rightarrow\; {\bf z}_{0} \;\equiv\; {\cal T}_{0}{\bf z} \;\rightarrow\; \ov{\bf Z} \;\equiv\; {\cal T}^{\epsilon}{\bf z}_{0} \;\equiv\; {\cal T}_{\epsilon}{\bf z},
\label{eq:total_pst} 
\end{equation}
which represents the composition of a preliminary phase-space transformation ${\cal T}_{0}$ to local phase-space coordinates ${\bf z}_{0}$ followed by a near-identity phase-space transformation ${\cal T}^{\epsilon} \equiv \cdots{\cal T}_{3}\,{\cal T}_{2}\,{\cal T}_{1}$ to reduced phase-space coordinates $\ov{\bf Z}$ that is generated by Lie-transform perturbation methods.  The phase-space coordinate transformation \eqref{eq:total_pst} will be assumed to be invertible, with ${\cal T}_{\epsilon}^{-1}\equiv {\cal T}_{0}^{-1}\,{\cal T}^{-\epsilon}$. (See, e.g., \cite{Brizard_2008} and \cite{Brizard_Hahm_2007} for further discussion.) 

As by \cite{pjm13}, our construction will explicitly guarantee that the reduced Vlasov-Maxwell bracket satisfies the Jacobi identity \eqref{eq:Jacobi_def}, while the reduced functionals 
will depend on the reduced Vlasov distribution function $\ov{F} \equiv {\sf T}_{\epsilon}^{-1}f$, defined as the push-forward ${\sf T}_{\epsilon}^{-1} \equiv {\sf T}^{-\epsilon}\,
{\sf T}_{0}^{-1}$ of the particle Vlasov distribution function $f$, and the electromagnetic fields $({\bf E},{\bf B})$. The present reduced Vlasov-Maxwell Hamiltonian formulation uses exclusively  the variable ${\bf E}$, while the earlier work of \cite{pjm13} used both ${\bf E}$ and the electric displacement field ${\bf D}_{\epsilon} \equiv {\bf E} + 4\pi\,{\bf P}_{\epsilon}$ as a field variable, where ${\bf P}_{\epsilon}$ denotes the reduced polarization \citep{Brizard_2008, Brizard_2013, Tronko_Brizard_2015}.

\section{\label{sec:Ham_VM}Hamiltonian Structure of the Vlasov-Maxwell Equations}

We now proceed with a brief review of the Hamiltonian structure of the Vlasov-Maxwell equations introduced by  \cite{pjm80} with corrections given in \cite{pjm82} and \cite{MW}. The Hamiltonian functional is simply represented as the energy invariant of the Vlasov-Maxwell equations:
\begin{equation}
{\cal H}[f,{\bf E},{\bf B}] \;=\; \frac{1}{8\pi}\;\int_{\bf r}\;\left( |{\bf E}({\bf r})|^{2} \;+\frac{}{} |{\bf B}({\bf r})|^{2} \right) \;+\; 
\int_{\bf z}\,{\cal J}({\bf z})\;f({\bf z})\;K,
\label{eq:Ham_def}
\end{equation}
where the Jacobian ${\cal J}$ is shown explicitly, and $K = |{\bf p}|^{2}/(2m)$ denotes the kinetic energy of a particle of 
mass $m$ and charge $e$ (summation over particle species is implied whenever applicable). The Vlasov-Maxwell bracket is a bilinear operator on arbitrary functionals ${\cal F}[f,{\bf E},{\bf B}]$ and 
${\cal G}[f,{\bf E},{\bf B}]$:
\begin{eqnarray}
\left[{\cal F},\frac{}{}{\cal G}\right] & = & \int_{\bf z}\,{\cal J}\;f \left\{ \frac{1}{{\cal J}}\,\fd{\cal F}{f},\; \frac{1}{{\cal J}}\,
\fd{\cal G}{f} \right\} \;+\; 4\pi\,c\;\int_{\bf r}\;\left( \fd{\cal F}{{\bf E}}\bdot\nabla\btimes\fd{\cal G}{{\bf B}} \;-\; \fd{\cal G}{{\bf E}}\bdot\nabla\btimes
\fd{\cal F}{{\bf B}} \right) \nonumber \\
 &  &-\;4\pi\,e\;\int_{\bf z}\,{\cal J}\;f \left( \fd{\cal F}{{\bf E}}\bdot\left\{ {\bf x},\; \frac{1}{{\cal J}}\,\fd{\cal G}{f} \right\} \;-\; 
\fd{\cal G}{{\bf E}}\bdot\left\{ {\bf x},\; \frac{1}{{\cal J}}\,\fd{\cal F}{f} \right\} \right),
\label{eq:MV_bracket}
\end{eqnarray}
where the variation of an arbitrary functional ${\cal F}[f,{\bf E},{\bf B}]$ is defined in terms of the Fr\'{e}chet derivative
\begin{equation}
\delta{\cal F} \equiv \frac{d}{d\varepsilon}\left({\cal F}\left[f + \varepsilon\,\delta f, {\bf E} + \varepsilon\,
\delta{\bf E},\frac{}{} {\bf B} + \varepsilon\,\delta{\bf B}\right]\right)_{\varepsilon = 0} = \int_{\bf z}\delta f\;\fd{\cal F}{f} + \int_{\bf r}\left(\delta{\bf E}\bdot
\fd{\cal F}{\bf E} + \delta{\bf B}\bdot\fd{\cal F}{\bf B} \right).
\label{eq:Frechet}
\end{equation}
While the Jacobian ${\cal J}$ associated with the particle noncanonical phase-space coordinates ${\bf z} = ({\bf x}, {\bf p})$ is constant in Eqs.~\eqref{eq:Ham_def}-\eqref{eq:MV_bracket} (i.e., ${\cal J} = 1$), we 
continue displaying it as a place holder in Eqs.~\eqref{eq:Ham_def}-\eqref{eq:MV_bracket} so that the case of the reduced Vlasov-Maxwell equations is interpreted correctly. For the sake of our presentation, the three terms appearing on the right of Eq.~\eqref{eq:MV_bracket} are called, respectively, the Vlasov, Maxwell, and interaction sub-brackets. 

The single-particle noncanonical Poisson bracket $\{\;,\;\}$ appearing in the Vlasov and interaction sub-brackets is defined in terms of functions  $f$ and $g$ on (noncanonical) particle phase space as
\begin{equation}
\{ f,\; g\} \;=\; \left( \nabla f\bdot\pd{g}{\bf p} \;-\; \pd{f}{\bf p}\bdot\nabla g \right) \;+\; \frac{e}{c}\,{\bf B}\bdot\pd{f}{\bf p}
\btimes\pd{g}{\bf p}.
\label{eq:PB_def}
\end{equation}
The Poisson bracket \eqref{eq:PB_def} can also be written in divergence form as
\begin{equation}
\{ f,\; g\} \;\equiv\; \pd{f}{z^{\alpha}}\;J^{\alpha\beta}\;\pd{g}{z^{\beta}} \;=\; \frac{1}{\cal J}\;\pd{}{z^{\alpha}}\left({\cal J}\,f\;
J^{\alpha\beta}\,\pd{g}{z^{\beta}} \right) \;=\; \frac{1}{\cal J}\;\pd{}{z^{\alpha}}\left({\cal J}\,f\frac{}{} \{z^{\alpha},\; g\} \right),
\label{eq:PB_def_div}
\end{equation}
where the antisymmetric Poisson matrix components $J^{\alpha\beta}({\bf z}) \equiv \{ z^{\alpha},\; z^{\beta}\}$ satisfy the Liouville identities
\begin{equation}
\frac{1}{\cal J}\;\pd{}{z^{\alpha}}\left({\cal J}\frac{}{}J^{\alpha\beta} \right) \;\equiv\; 0.
\label{eq:Liouville_id}
\end{equation}
The divergence form \eqref{eq:PB_def_div} implies the phase-space integral identity
\begin{equation}
\int_{\bf z}\;{\cal J}\,f\;\{g,\; h\} \;\equiv\; -\;\int_{\bf z}\;{\cal J}\,g\;\{f,\; h\},
\label{eq:fgh_id}
\end{equation}
where $(f,g,h)$ are three arbitrary functions on particle phase space. In what follows, explicit and implicit time dependences are assumed for the Vlasov distribution $f$ and the electromagnetic fields $({\bf E},{\bf B})$ and time is unaffected by the phase-space transformations considered here. The proof of the Jacobi identity for the Poisson bracket \eqref{eq:PB_def}:
\begin{equation}
\left\{ f,\frac{}{} \{g,\; h\}\right\} \;+\; \left\{ g,\frac{}{} \{h,\; f\}\right\} \;+\; \left\{ h,\frac{}{} \{f,\; g\}\right\} \;=\; 0 
\label{eq:Jacobi_PB}
\end{equation}
is given in App.~\ref{sec:Jacobi_PB} [see Eq.~\eqref{eq:d_omega}], where the magnetic field ${\bf B}$ is required to be divergenceless, i.e.,  $\nabla\bdot{\bf B} \equiv 0$.

The Vlasov-Maxwell bracket \eqref{eq:MV_bracket} is easily seen to satisfy  the properties  of antisymmetry $[{\cal G},\;{\cal F}] = -\,[{\cal F},\;{\cal G}]$  and the Leibniz property $[{\cal F},\;{\cal G}\,{\cal K}] = [{\cal F},\;{\cal G}]\,{\cal K} + {\cal G}\,[{\cal F},\;{\cal K}]$. In addition, it was given in  \cite{pjm82} by direct calculation that the Jacobi identity condition yields
\begin{equation}
\big[ {\cal F},  [{\cal G}, {\cal K}]\big] +  \big[ {\cal G}, [{\cal K}, {\cal F}]\big] + \big[ {\cal K},  [{\cal F}, {\cal G}]\big] 
 \equiv 
  \int_{\bf z}\!f({\bf z})\, \nabla\cdot {\bf B}\; \left(\frac{\partial }{\bf \partial p}\frac{\delta {\cal F}}{\delta f}\cdot 
\frac{\partial }{\bf \partial p}\frac{\delta {\cal G}}{\delta f}
\btimes\frac{\partial }{\bf \partial p}\frac{\delta {\cal H}}{\delta f}\right)\,.
\label{eq:MV_Jacobi}
\end{equation}
Thus, for the Jacobi identity to hold,  it was seen that the domain of functionals must be restricted to divergence free $\mathbf{B}$-fields.  Details of the original (onerous and lengthy) calculation were recorded in an appendix of  \cite{pjm13}, while a  simplified version of this calculation is presented here in App.~\ref{sec:Jacobi_MV}. 

We note in passing that it was shown by \cite{pjmGe82} and \cite{pjm82} that the $\nabla\cdot\mathbf{B}=0$ condition can be removed for MHD, while attempts to remove this condition  for Vlasov-Maxwell have not been as successful.  However,  \cite{pjmCGBT13} used Dirac-constraint theory to replace this condition for the Vlasov-Maxwell equations by a boundary condition at infinity. 

\subsection{Vlasov-Maxwell equations}

The Hamiltonian evolution of a generic functional ${\cal F}[f,{\bf E},{\bf B}]$ is expressed in terms of the Hamiltonian functional \eqref{eq:Ham_def} and the Vlasov-Maxwell bracket \eqref{eq:MV_bracket} as
\begin{eqnarray}
\pd{\cal F}{t} \equiv \left[{\cal F},\frac{}{} {\cal H}\right] & = & -\int_{\bf z}\;\fd{\cal F}{f} \left( \left\{ f,\; K\;\right\} \;+\frac{}{} e{\bf E}\bdot\left\{{\bf x},\; f\right\}\right) \;-\; \int_{\bf r}\;\fd{\cal F}{{\bf B}}\bdot\left(c\frac{}{}\nabla\btimes
{\bf E} \right) \nonumber \\
 & &+\; \int_{\bf r}\;\fd{\cal F}{{\bf E}}\bdot\left( c\,\nabla\btimes{\bf B} \;-\; 4\pi\,e\;\int_{\bf z}\,{\cal J}\;f\,\delta^{3}({\bf x} - {\bf r})\;\left\{ {\bf x},\; K\right\} \right),
\label{eq:F_dot}
\end{eqnarray}
which becomes
\begin{equation}
\pd{\cal F}{t} \;\equiv\;  \int_{\bf z}\;\fd{\cal F}{f({\bf z})} \;\pd{f}{t} \;+\; \int_{\bf r}\left(\fd{\cal F}{{\bf E}({\bf r})}\bdot\pd{\bf E}{t} 
\;+\; \fd{\cal F}{{\bf B}({\bf r})}\bdot\pd{\bf B}{t}\right),
\label{eq:F_dot_def}
\end{equation}
where the Vlasov equation for $\partial f/\partial t$ is
\begin{equation}
\pd{f}{t} \;=\; -\,{\bf v}\bdot\nabla f({\bf z}) \;-\; e\left( {\bf E}({\bf x}) \;+\; \frac{\bf v}{c}\btimes{\bf B}({\bf x})\right)\bdot\pd{f({\bf z})}{\bf p},
\label{eq:Vlasov_eq}
\end{equation}
with the particle's velocity ${\bf v} \equiv {\bf p}/m$, and the Maxwell equations are
\begin{eqnarray}
\pd{\bf E}{t} & = & c\,\nabla\btimes{\bf B} \;-\; 4\pi\,e\;\int_{\bf z}\,{\cal J}\;f\,\delta^{3}({\bf x} - {\bf r})\;{\bf v} \;\equiv\; 
c\,\nabla\btimes{\bf B} \;-\; 4\pi\;{\bf J}({\bf r}), \label{eq:E_dot} \\
\pd{\bf B}{t} & = & -\;c\,\nabla\btimes{\bf E}. \label{eq:B_dot}
\end{eqnarray}
The remaining Maxwell equations
\begin{eqnarray}
\nabla\bdot{\bf E} & = & 4\pi\,e\;\int_{\bf z}\,{\cal J}\;f\;\delta^{3}({\bf x} - {\bf r}) \;\equiv\; 4\pi\;\varrho({\bf r}), \label{eq:div_E} \\
\nabla\bdot{\bf B} & = & 0 \label{eq:div_B}
\end{eqnarray}
can be seen as initial conditions for the electromagnetic fields $({\bf E}, {\bf B})$ for $\nabla\bdot(\partial{\bf B}/\partial t) = 0$ and 
$\nabla\bdot(\partial{\bf E}/\partial t) = -\,4\pi\;\nabla\bdot{\bf J} = 4\pi\;\partial\varrho/\partial t$, which represents the charge conservation law and follows from Eq.~\eqref{eq:E_dot}. We note that the charge density $\varrho({\bf r})$ and the current density ${\bf J}({\bf r})$ are functionals of the Vlasov distribution $f({\bf z})$, both labeled by the field position ${\bf r}$, and the presence of the delta function $\delta^{3}({\bf x} - {\bf r})$ in Eqs.~\eqref{eq:E_dot} and \eqref{eq:div_E} implies that only particles located at the field position ${\bf x} = {\bf r}$ contribute to the electromagnetic fields $({\bf E}, {\bf B})$. 

\subsection{Hamiltonian properties}

The characteristics of the Vlasov equation \eqref{eq:Vlasov_eq} are the equations of motion
\begin{equation}
\frac{d{\bf x}}{dt} \;=\; \frac{\bf p}{m} \;\;\;{\rm and}\;\; \frac{d{\bf p}}{dt} \;=\; e\;{\bf E}({\bf x}) \;+\; \frac{e\,{\bf v}}{c}\btimes{\bf B}({\bf x}),
\label{eq:motion}
\end{equation}
which can be expressed in (noncanonical) Hamiltonian form as \citep{pjm13}
\begin{equation}
\frac{dz^{\alpha}}{dt} \;\equiv\; \left\{ z^{\alpha},\; K\right\} \;+\; e\,{\bf E}({\bf x})\bdot\left\{ {\bf x},\; z^{\alpha}\right\}.
\label{eq:dotz_alpha}
\end{equation}
Hence, the Vlasov equation \eqref{eq:Vlasov_eq} may be expressed as
\begin{equation}
\pd{f}{t} \;+\; \frac{dz^{\alpha}}{dt}\;\pd{f}{z^{\alpha}} \;=\; 0,
\label{eq:Vlasov_Ham}
\end{equation}
which may also be expressed in divergence form as follows.

We note that the noncanonical Hamilton equations \eqref{eq:dotz_alpha} and the Jacobian ${\cal J}$ satisfy the Liouville theorem 
\begin{equation}
\pd{\cal J}{t} \;+\; \pd{}{z^{\alpha}}\left({\cal J}\;\frac{dz^{\alpha}}{dt}\right) \;=\; 0,
\label{eq:Liouville_th}
\end{equation}
so that the Vlasov equation \eqref{eq:Vlasov_Ham} may also be written in divergence form as
\begin{equation}
\pd{({\cal J}\,f)}{t} \;=\; -\;\pd{}{z^{\alpha}}\left({\cal J}\,f\;\frac{dz^{\alpha}}{dt}\right).
\label{eq:Vlasov_div}
\end{equation}
This form proves useful when evaluating the time derivative of a general Vlasov functional ${\cal G} \equiv \int {\cal J}\,f\;g\;d^{6}z$ in particle phase space:
\begin{equation}
\pd{\cal G}{t} \;=\; \int \left( \pd{({\cal J}\,f)}{t}\;g \;+\; {\cal J}\,f\;\pd{g}{t} \right) d^{6}z \;\equiv\; \int {\cal J}\,f\;\frac{dg}{dt}\;
d^{6}z,
\label{eq:dgdt}
\end{equation}
where $g({\bf z},t)$ is an arbitrary function on particle phase space and integration by parts was performed in order to obtain the last expression.

Lastly, using Eq.~\eqref{eq:F_dot_def} with Eqs.~\eqref{eq:E_dot}-\eqref{eq:B_dot} and \eqref{eq:Vlasov_div}, we easily show that the Hamiltonian functional \eqref{eq:Ham_def} is itself an invariant of the Vlasov-Maxwell dynamics:
\begin{eqnarray}
\pd{\cal H}{t} & = & \int_{\bf r}\left[ \frac{\bf E}{4\pi}\bdot\left(c\,\nabla\btimes{\bf B} \;-\frac{}{} 4\pi\,{\bf J} \right) \;-\; \frac{\bf B}{4\pi}\bdot\left(c\frac{}{}\nabla\btimes{\bf E}\right) \right] \;-\; 
\int_{\bf z}\,\pd{}{z^{\alpha}}\left({\cal J}\,f\;\frac{dz^{\alpha}}{dt}\right)\;K \nonumber \\
 & = & \int_{\bf z}\,{\cal J}\,f\; \left( \frac{dK}{dt} \;-\; e\,{\bf E}\bdot{\bf v} \right) \;=\; 0,
\label{eq:H_dot}
\end{eqnarray}
which vanishes since $dK/dt = {\bf v}\bdot d{\bf p}/dt = e\,{\bf v}\bdot{\bf E}$. Equation \eqref{eq:H_dot} can also be expressed as $\partial{\cal H}/\partial t = 
[{\cal H},\;{\cal H}] \equiv 0$, which immediately follows from the antisymmetry of the Vlasov-Maxwell bracket \eqref{eq:MV_bracket}.

\section{\label{sec:general_lift}General Transformation Lift of the Vlasov-Maxwell Equations}

In this section, we consider the transformation of the Hamiltonian structure \eqref{eq:Ham_def}-\eqref{eq:MV_bracket} of the Vlasov-Maxwell equations (introduced in Sec.~\ref{sec:Ham_VM}) associated with a general phase-space transformation that depends on the electromagnetic field $({\bf E},{\bf B})$ as well of their spatial gradients. Hence, we consider a time-dependent phase-space transformation ${\cal T}: {\bf z} \rightarrow \ov{\bf Z}$ that is also invertible ${\cal T}^{-1}: \ov{\bf Z} \rightarrow {\bf z}$. 

In the next sections, we will consider the two-step transformation process \eqref{eq:total_pst} generally encountered in the dynamical reduction of charged-particle motion in a strong magnetic field (e.g., guiding-center transformation \citep{Cary_Brizard_2009}). In the first step (see Sec.~\ref{sec:local}), we present the preliminary transformation ${\cal T}_{0}$ from particle phase-space coordinates $({\bf x}, {\bf p})$ to {\it local} phase-space coordinates 
$({\bf x}, p_{\|}, \mu,\zeta)$, where $p_{\|}$ denotes the component of the particle's momentum along the local magnetic field ${\bf B} \equiv B\;\bhat$: $p_{\|} \equiv {\bf p}\bdot\bhat$, $p_{\bot}(\mu,{\bf x})$ denotes the magnitude of the components of the particle's momentum perpendicular to the local magnetic field: $p_{\bot}(\mu,{\bf x}) \equiv |{\bf p}\btimes\bhat|$, where $\mu$ denotes the local magnetic moment, and the gyroangle $\zeta$ denotes the orientation of the perpendicular vector-component of the particle's momentum about the magnetic field: ${\bf p}_{\bot} \equiv {\bf p} - p_{\|}\,\bhat$. 

The preliminary phase-space transformation introduces explicit dependence on the fast gyromotion time scale (through the local gyroangle $\zeta$). In the second step (see Sec.~\ref{sec:Dyn_Red}), we present the near-identity phase-space transformation ${\cal T}^{\epsilon} \equiv \cdots {\cal T}_{2}
{\cal T}_{1}$ from the local phase-space coordinates $z_{0}^{\alpha} \equiv ({\bf x}, p_{\|}, \mu,\zeta)$ to the reduced phase-space coordinates $\ov{Z}^{\alpha} \equiv (\ov{\bf X}, \ov{p}_{\|}, \ov{\mu}, \ov{\zeta})$ generated by Lie-transform perturbation methods.

\subsection{General operators on functions and functionals}

A general phase-space transformation ${\cal T}$ induces operators on functions and meta-operators on functionals. First, we introduce the push-forward operator ${\sf T}^{-1}: f \rightarrow \ov{f} \equiv {\sf T}^{-1}f$ that transforms a function $f$ on particle phase space into a function $\ov{f}$ on the new phase space. We also introduce the pull-back operator ${\sf T}: \ov{f} \rightarrow f \equiv {\sf T}\ov{f}$ that transforms a function $\ov{f}$ on the new phase space into a function $f$ on particle phase space. These operator definitions ensure that the scalar-covariance property is satisfied: 
\begin{equation}
\left. \begin{array}{rcl}
f({\bf z}) & = & f({\cal T}^{-1}\ov{\bf Z}) \;\equiv\; {\sf T}^{-1}f(\ov{\bf Z}) \;=\; \ov{f}(\ov{\bf Z}) \\
 & & \\
\ov{f}(\ov{\bf Z}) & = & \ov{f}({\cal T}{\bf z}) \;\equiv\; {\sf T}\ov{f}({\bf z}) \;=\; f({\bf z})
\end{array} \right\}. 
\end{equation}
The transformed Jacobian $\ov{\cal J}$, on the other hand, is defined from the push-forward relation ${\sf T}^{-1}({\cal J}\,d^{6}z) \equiv \ov{\cal J}\,
d^{6}\ov{Z}$, where $d^{6}z$ and $d^{6}\ov{Z}$ denote differential six-forms in their respective phase spaces. 

Secondly, we introduce the meta-push-forward functional operator $\mathbb{T}: {\cal F} \rightarrow \ov{\cal F} \equiv \mathbb{T}{\cal F}$ and the meta-pull-back functional operator $\mathbb{T}^{-1}: \ov{\cal F} \rightarrow {\cal F} \equiv \mathbb{T}^{-1}\ov{\cal F}$, which satisfy the functional-covariance property 
\begin{equation}
\left. \begin{array}{rcl}
\ov{\cal F}[\ov{f}] & = & \mathbb{T}{\cal F}[\ov{f}] \;\equiv\; {\cal F}[{\sf T}\ov{f}] \;=\; {\cal F}[f] \\
 &  & \\
{\cal F}[f] & = & \mathbb{T}^{-1}\ov{\cal F}[f] \;\equiv\; \ov{\cal F}[{\sf T}^{-1}f] \;=\; \ov{\cal F}[\ov{f}]
\end{array} \right\}.
\label{eq:meta_general}
\end{equation} 
For example, we consider the simple functional transformations:
\begin{eqnarray*} 
{\cal F}[f] & \equiv & \int_{\bf z} {\cal J}f({\bf z})\,g({\bf z}) \;=\; \int_{\ov{\bf Z}}\ov{\cal J}\;{\sf T}^{-1}f(\ov{\bf Z})\;
{\sf T}^{-1}g(\ov{\bf Z}) \;=\; \int_{\ov{\bf Z}}\ov{\cal J}\;\ov{f}(\ov{\bf Z})\;{\sf T}^{-1}g(\ov{\bf Z}) \\
 & = & \int_{\bf z} {\cal J}\;{\sf T}\ov{f}({\bf z})\,g({\bf z}) \;\equiv\; \mathbb{T}{\cal F}[\ov{f}], \\
\ov{\cal F}[\ov{f}] & \equiv & \int_{\ov{\bf Z}} \ov{\cal J}\,\ov{f}(\ov{\bf Z})\,\ov{g}(\ov{\bf Z}) \;=\; \int_{\bf z}{\cal J}\;{\sf T}\ov{f}({\bf z})\;
{\sf T}\ov{g}({\bf z}) \;=\; \int_{\bf z}{\cal J}\;f({\bf z})\;{\sf T}\ov{g}({\bf z}) \\
 & = &  \int_{\ov{\bf Z}} \ov{\cal J}\;{\sf T}^{-1}f(\ov{\bf Z})\;\ov{g}(\ov{\bf Z}) \;\equiv\; \mathbb{T}^{-1}\ov{\cal F}[f],
\end{eqnarray*}
where a sequence of phase-space coordinate transformations and integrations by parts lead to the relations \eqref{eq:meta_general}.

\subsection{Transformed Vlasov-Maxwell equations}

The transformation of the Vlasov equation \eqref{eq:Vlasov_Ham} proceeds through the push-forward transformation of each of its parts:
\begin{equation}
{\sf T}^{-1}\left(\pd{f}{t}\right) \;=\; -\;{\sf T}^{-1}\left( \{ f,\; K \} \;+\frac{}{} e\,{\bf E}\bdot\{ {\bf x},\; f \} \right).
\label{eq:Vlasov_push-forward}
\end{equation}
While time is unaffected by the phase-space transformations considered here, we note that the partial time derivative $\partial/\partial t$ does not commute with the push-forward operator ${\sf T}^{-1}$ and, thus, the commutation relation $[{\sf T}^{-1},\,\partial/\partial t]$ must be calculated carefully.

\subsubsection{Transformed partial time derivative}

On the left side of Eq.~\eqref{eq:Vlasov_push-forward}, we introduce the transformed partial time derivative $\ov{\partial}/\partial t$:
\begin{equation}
{\sf T}^{-1}\left(\pd{f}{t}\right) \;=\; \left[{\sf T}^{-1}\left(\pd{}{t}\,{\sf T}\right)\right]{\sf T}^{-1}f \;\equiv\; 
\frac{\ov{\partial}\ov{f}}{\partial t},
\label{eq:ov_partial_t-1}
\end{equation}
which is defined by the operator-commutation identity
\begin{equation} 
\frac{\ov{\partial}\ov{f}}{\partial t} \;\equiv\; \pd{\ov{f}}{t} \;+\; \left[ {\sf T}^{-1}\left(\pd{}{t}\;{\sf T}\right) \;-\; \pd{}{t}\right]\ov{f} \;\equiv\; \pd{\ov{f}}{t} \;+\; \frac{\ov{\partial}\,\ov{Z}^{\alpha}}{\partial t}\;\pd{\ov{f}}{\ov{Z}^{\alpha}},
\label{eq:partial_T_def}
\end{equation}
where the second term is expressed as a partial-differential operator in the new phase space, with
\begin{equation} 
\frac{\ov{\partial}\,\ov{Z}^{\alpha}}{\partial t} \;\equiv\; {\sf T}^{-1}\left( \pd{({\sf T}\ov{Z}^{\alpha})}{t}\right).
\label{eq:ov_partial_t}
\end{equation}
Here, the right side of Eq.~\eqref{eq:ov_partial_t} involves a three-step procedure: first, the pull-back ${\sf T}\ov{Z}^{\alpha}$ of each new phase-space coordinate 
$\ov{Z}^{\alpha}$ is calculated as a function on particle phase space ${\sf T}\ov{Z}^{\alpha}({\bf z})$; second, the partial time derivative 
$\partial({\sf T}\ov{Z}^{\alpha})/\partial t$ of the pull-back is calculated in particle phase space in terms of the electromagnetic-field derivatives 
$\partial{\bf E}/\partial t$ and $\partial{\bf B}/\partial t$; and, third, $\partial({\sf T}\ov{Z}^{\alpha})/\partial t$ is then pushed-forward into the new phase space: ${\sf T}^{-1}[\partial({\sf T}\ov{Z}^{\alpha})/\partial t]$. Note that, for a time-independent phase-space transformation, the operators $\partial/\partial t$ and ${\sf T}^{-1}$ commute and ${\sf T}^{-1}(\partial f/\partial t) = \partial({\sf T}^{-1}f)/\partial t = \partial\ov{f}/\partial t$.

\subsubsection{Transformed Poisson bracket}

On the right side of Eq.~\eqref{eq:Vlasov_push-forward}, we define the transformed Poisson bracket $\ov{\{\;,\;\}}$ on the new phase space from the identity \citep{Brizard_2008}
\begin{equation}
\ov{\left\{ \ov{f},\frac{}{} \ov{g}\right\}} \;\equiv\; {\sf T}^{-1}\left(\left\{ f,\frac{}{} g\right\}\right) \;\equiv\; {\sf T}^{-1}\left( \left\{ 
{\sf T}\ov{f},\frac{}{} {\sf T}\ov{g}\right\} \right) \;\equiv\; \pd{\ov{f}}{\ov{Z}^{\mu}}\;\ov{J}^{\mu\nu}\;\pd{\ov{g}}{\ov{Z}^{\nu}},
\label{eq:new_PB}
\end{equation}
where, by construction, the new Poisson bracket automatically satisfies the Jacobi identity and the transformed Poisson matrix $\ov{J}^{\mu\nu}$ is defined in terms of the particle Poisson matrix $J^{\alpha\beta} \equiv \{ z^{\alpha},\; z^{\beta}\}$ as
\begin{equation}
\ov{J}^{\mu\nu} \;\equiv\; {\sf T}^{-1}\left(\pd{({\sf T}\ov{Z}^{\mu})}{z^{\alpha}}\;J^{\alpha\beta}\;\pd{({\sf T}\ov{Z}^{\nu})}{z^{\beta}}\right)
\;=\; {\sf T}^{-1}\left(\left\{ {\sf T}\ov{Z}^{\mu},\frac{}{} {\sf T}\ov{Z}^{\nu}\right\}\right) \;\equiv\; \ov{\left\{\ov{Z}^{\mu},\; \ov{Z}^{\nu}\right\}}.
\label{eq:ovJ_munu}
\end{equation}
The transformed Poisson bracket \eqref{eq:new_PB} is guaranteed to satisfy the Jacobi identity because it is derived from a transformed Lagrange two-form
$\ov{\omega} \equiv {\sf T}^{-1}\omega = {\sf T}^{-1}(\exd\gamma) = \exd({\sf T}^{-1}\gamma) \equiv \exd\ov{\gamma}$ that is closed $\exd\ov{\omega} \equiv 0$.

The transformed Poisson bracket \eqref{eq:new_PB} can also be expressed in phase-space divergence form as
\begin{equation}
\ov{\left\{ \ov{f},\frac{}{} \ov{g}\right\}} \;\equiv\; \frac{1}{\ov{\cal J}}\,\pd{}{\ov{Z}^{\alpha}} \left( \ov{\cal J}\,\ov{f}\frac{}{} \ov{\left\{
\ov{Z}^{\alpha},\; \ov{g}\right\}} \right),
\label{eq:PB_def_div_new}
\end{equation}
which follows from the Liouville identities
\begin{equation}
\frac{1}{\ov{\cal J}}\,\pd{}{\ov{Z}^{\alpha}} \left( \ov{\cal J}\,\ov{J}^{\alpha\beta} \right) \;\equiv\; 0.
\label{eq:ov_Liouville}
\end{equation}
The divergence form \eqref{eq:PB_def_div_new} yields the phase-space integral identity
\begin{equation}
\int_{\ov{\bf Z}}\;\ov{\cal J}\,\ov{f}\;\ov{\left\{\ov{g},\frac{}{} \ov{h}\right\}} \;\equiv\; -\;\int_{\ov{\bf Z}}\;\ov{\cal J}\,\ov{g}\;
\ov{\left\{\ov{f},\frac{}{} \ov{h}\right\}}.
\label{eq:fgh_id_new}
\end{equation}
Using the definition \eqref{eq:new_PB}, the right side of Eq.~\eqref{eq:Vlasov_push-forward} becomes
\begin{equation} 
-\;{\sf T}^{-1}\left( \{ f,\; K \} \;+\frac{}{} e\,{\bf E}\bdot\{ {\bf x},\; f \} \right) \;=\; -\;\ov{\left\{ \ov{f},\frac{}{} \ov{K}\right\}} \;-\; 
e\,{\sf T}^{-1}{\bf E}\bdot\ov{\left\{{\sf T}^{-1}{\bf x},\frac{}{} \ov{f}\right\}}, 
\label{eq:push_Newton}
\end{equation}
where ${\sf T}^{-1}{\bf x}$ denotes the push-forward of the particle position ${\bf x}$ and ${\sf T}^{-1}{\bf E}$ denotes the push-forward of the electric field as it appears in the reduced particle dynamics. 

\subsubsection{Transformed Vlasov equation}

By combining the left and right sides of Eq.~\eqref{eq:Vlasov_push-forward}, we obtain the transformed Vlasov equation
\begin{equation}
\pd{\ov{f}}{t} \;=\; -\;\left[ \ov{\left\{ \ov{Z}^{\alpha},\frac{}{} \ov{K}\right\}} \;+\; e\,{\sf T}^{-1}{\bf E}\bdot\ov{\left\{{\sf T}^{-1}{\bf x},
\frac{}{} \ov{Z}^{\alpha} \right\}} \;+\; \frac{\ov{\partial}\ov{Z}^{\alpha}}{\partial t}\right]\;\pd{\ov{f}}{\ov{Z}^{\alpha}} \;\equiv\;
-\;\frac{\ov{d}\ov{Z}^{\alpha}}{dt}\;\pd{\ov{f}}{\ov{Z}^{\alpha}},
\label{eq:transformed_Vlasov}
\end{equation}
where the transformed phase-space dynamics $\ov{d}\ov{Z}^{\alpha}/dt$ includes the transformed partial-time derivative \eqref{eq:ov_partial_t} as well as the transformed Hamilton equations 
\begin{equation}
\dot{\ov{Z}}^{\alpha} \;\equiv\; \ov{\left\{ \ov{Z}^{\alpha},\frac{}{} \ov{K}\right\}} \;+\; e\,{\sf T}^{-1}{\bf E}\bdot\ov{\left\{
{\sf T}^{-1}{\bf x},\frac{}{} \ov{Z}^{\alpha}\right\}}.
\label{eq:ov_Ham_def}
\end{equation}
Next, the transformed Liouville Theorem requires that the transformed Jacobian $\ov{\cal J}$ satisfies the evolution equation
\[ \pd{\ov{\cal J}}{t} \;+\; \pd{}{\ov{Z}^{\alpha}}\left( \ov{\cal J}\;\frac{\ov{d}\ov{Z}^{\alpha}}{dt} \right) \;=\; 0. \]
Since the transformed Hamilton equations \eqref{eq:ov_Ham_def} satisfy the identity
\[ \pd{}{\ov{Z}^{\alpha}}\left( \ov{\cal J}\;\frac{}{} \dot{\ov{Z}}^{\alpha} \right) \;\equiv\; 0, \]
however, the transformed Jacobian therefore satisfies the equation
\begin{equation}
\pd{\ov{\cal J}}{t} \;=\; -\; \pd{}{\ov{Z}^{\alpha}}\left( \ov{\cal J}\;\frac{\ov{\partial}\ov{Z}^{\alpha}}{\partial t} \right) \;=\; -\;
\pd{}{\ov{Z}^{\alpha}}\left[ \ov{\cal J}\;{\sf T}^{-1}\left(\pd{({\sf T}\ov{Z}^{\alpha})}{t} \right) \right].
\label{eq:time_ovJac}
\end{equation}
Moreover, the transformed Vlasov equation \eqref{eq:transformed_Vlasov} can be written in divergence form as
\begin{equation}
\pd{(\ov{\cal J}\,\ov{f})}{t} \;+\; \pd{}{\ov{Z}^{\alpha}}\left( \ov{\cal J}\,\ov{f}\;\frac{\ov{d}\ov{Z}^{\alpha}}{dt} \right) \;=\; 0,
\label{eq:transformed_div}
\end{equation}
which leads to the integral identity for $\ov{\cal G} \equiv \int_{\ov{\bf Z}}\ov{\cal J}\ov{f}\;\ov{g}$:
\[ \pd{\ov{\cal G}}{t} \;=\; \int_{\ov{\bf Z}}\left[ \pd{(\ov{\cal J}\,\ov{f})}{t}\;\ov{g} + 
\ov{\cal J}\,\ov{f}\;\left(\pd{\ov{g}}{t} \right) \right] \;=\; \int_{\ov{\bf Z}}\,\ov{\cal J}\,\ov{f}\;\frac{\ov{d}\ov{g}}{dt} \;\equiv\; 
\int_{\ov{\bf Z}}\,\ov{\cal J}\,\ov{f}\;{\sf T}^{-1}\left(\frac{dg}{dt}\right). \]
For example, this equation guarantees that the particle-number conservation law (with $g = 1$) is satisfied by the transformed Vlasov equation.

\subsubsection{Transformed Maxwell equations}

Lastly, we turn our attention to the transformation of the Maxwell equations \eqref{eq:E_dot}-\eqref{eq:div_B}. Of course, Eqs.~\eqref{eq:B_dot} and 
\eqref{eq:div_B} are unchanged by the phase-space transformation since they are source-less. The Maxwell equations Eqs.~\eqref{eq:E_dot} and 
\eqref{eq:div_E}, however, become
\begin{eqnarray}
\pd{\bf E}{t} & = & c\,\nabla\btimes{\bf B} - 4\pi\,e\int_{\ov{\bf Z}}\ov{\cal J}\,\ov{f}\,\delta^{3}({\sf T}^{-1}{\bf x} - {\bf r})\;
\frac{\ov{d}({\sf T}^{-1}{\bf x})}{dt} \equiv c\,\nabla\btimes{\bf B} - 4\pi\,\mathbb{T}{\bf J}({\bf r}), \label{eq:Edot_new} \\
\nabla\bdot{\bf E} & = & 4\pi\,e\int_{\ov{\bf Z}}\ov{\cal J}\,\ov{f}\,\delta^{3}({\sf T}^{-1}{\bf x} - {\bf r}) \equiv 4\pi\,
\mathbb{T}\varrho({\bf r}), 
\label{eq:divE_new}
\end{eqnarray}
where we have applied the meta-push-forward $\mathbb{T}$ on the particle current density ${\bf J}$ in Eq.~\eqref{eq:E_dot} and the particle charge density $\varrho$ in 
Eq.~\eqref{eq:div_E}, which are both treated as functionals labeled by the field position ${\bf r}$. In Eq.~\eqref{eq:Edot_new}, the push-forward of the particle velocity is defined as
\begin{equation}
\frac{\ov{d}({\sf T}^{-1}{\bf x})}{dt} \;=\; {\sf T}^{-1}\left(\frac{d{\bf x}}{dt}\right) \;=\; \ov{\left\{ {\sf T}^{-1}{\bf x},\frac{}{} \ov{K}\right\}},
\label{eq:push_x_dot}
\end{equation}
where we used the identities $\ov{\partial}({\sf T}^{-1}{\bf x})/\partial t \equiv {\sf T}^{-1}(\partial{\bf x}/\partial t) \equiv 0$ (i.e., the particle position ${\bf x}$ does not explicitly depend on time) and $\ov{\{ {\sf T}^{-1}{\bf x},\;{\sf T}^{-1}{\bf x}\}} \equiv {\sf T}^{-1}(\{{\bf x},\; 
{\bf x}\}) \equiv 0$.

\subsection{Transformed functional variations}

The variation of an arbitrary functional $\ov{\cal F}[\ov{f},{\bf E},{\bf B}]$ is defined in terms of the Fr\'{e}chet derivative \eqref{eq:Frechet},
which can be transformed to yield
\begin{eqnarray}
\mathbb{T}\left[\delta\left(\mathbb{T}^{-1}\ov{\cal F}\right)\right] & = & \int_{\ov{\bf Z}}\ov{\cal J}\;{\sf T}^{-1}(\delta f)\;\;
{\sf T}^{-1}\left[ \frac{1}{\cal J}\fd{(\mathbb{T}^{-1}\ov{\cal F})}{f}\right] \nonumber \\
 &  &+\; \int_{\bf r} \left( \delta{\bf E}\bdot\fd{(\mathbb{T}^{-1}\ov{\cal F})}{{\bf E}} + \delta{\bf B}\bdot\fd{(\mathbb{T}^{-1}\ov{\cal F})}{{\bf B}} \right).
\label{eq:meta_ovF}
\end{eqnarray}
Here, the push-forward of the variation $\delta f$ on particle phase space:
\begin{equation}
{\sf T}^{-1}(\delta f) \;=\; \delta\left({\sf T}^{-1}f\right) \;+\; \left( \left[{\sf T}^{-1},\; \delta\right]{\sf T}\right){\sf T}^{-1}f \;\equiv\;
\delta\ov{f} \;+\; \left( \left[{\sf T}^{-1},\; \delta\right]{\sf T}\right)\ov{f}
\label{eq:push_delta_f_def}
\end{equation}
is expressed in terms of the variation $\delta\ov{f}$ on the new phase space and the commutation operator 
\begin{equation}
([{\sf T}^{-1},\; \delta]{\sf T}) \;=\; {\sf T}^{-1}(\delta{\sf T}) - \delta \;\equiv\; {\sf T}^{-1}\left[\delta\left({\sf T}\ov{Z}^{\alpha}\right)\right]
\pd{}{\ov{Z}^{\alpha}},
\label{eq:delta_T_def}
\end{equation}
which is similar to the commutation operation \eqref{eq:partial_T_def}, with $\partial/\partial t$ replaced by $\delta$. Hence, 
Eq.~\eqref{eq:push_delta_f_def} can be expressed as
\begin{eqnarray}
{\sf T}^{-1}(\delta f) & = & \delta\ov{f} \;+\; {\sf T}^{-1}\left[\delta\frac{}{}\left({\sf T}\ov{Z}^{\alpha}\right)\right]\; \pd{\ov{f}}{\ov{Z}^{\alpha}}\nonumber \\
 & \equiv & \delta\ov{f} \;+\; \int_{\bf r}\left[ \delta{\bf E}({\bf r})\bdot\partial_{\bf E}\ov{f} \;+\frac{}{} 
\delta{\bf B}({\bf r})\bdot\partial_{\bf B}\ov{f}\right],
\label{eq:push_deltaf_EB}
\end{eqnarray}
where the differential operators $\partial_{\bf E}$ and $\partial_{\bf B}$ are defined as
\begin{equation}
\partial_{\bf E} \;\equiv\; {\sf T}^{-1}\left[\frac{\delta\left({\sf T}\ov{Z}^{\alpha}\right)}{\delta{\bf E}({\bf r})}\right]\pd{}{\ov{Z}^{\alpha}}
\;\;\;{\rm and}\;\;\; \partial_{\bf B} \;\equiv\; {\sf T}^{-1}\left[\frac{\delta\left({\sf T}\ov{Z}^{\alpha}\right)}{\delta{\bf B}({\bf r})}\right]
\pd{}{\ov{Z}^{\alpha}}.
\label{eq:partial_EB_def}
\end{equation}
If we now use the identity $\delta\ov{\cal F} \equiv \mathbb{T}[\delta(\mathbb{T}^{-1}\ov{\cal F})]$ in Eq.~\eqref{eq:meta_ovF}, we obtain the functional-variation relations
\begin{eqnarray}
{\sf T}^{-1}\left[ \frac{1}{\cal J}\;\fd{(\mathbb{T}^{-1}\ov{\cal F})}{f}\right] & \equiv & \frac{1}{\ov{\cal J}}\;\fd{\ov{\cal F}}{\ov{f}},
\label{eq:ovF_del_f} \\
\fd{(\mathbb{T}^{-1}\ov{\cal F})}{{\bf E}({\bf r})} & \equiv & \fd{\ov{\cal F}}{{\bf E}({\bf r})} \;-\; \int_{\ov{\bf Z}} \partial_{\bf E}\ov{f}\;
\fd{\ov{\cal F}}{\ov{f}} \;\equiv\; \fd{\ov{\cal F}}{{\bf E}({\bf r})} \;-\;  \Delta_{\bf E}\ov{\cal F}({\bf r}), \label{eq:ovF_del_E}  \\
\fd{(\mathbb{T}^{-1}\ov{\cal F})}{{\bf B}({\bf r})} & \equiv & \fd{\ov{\cal F}}{{\bf B}({\bf r})} \;-\; \int_{\ov{\bf Z}} \partial_{\bf B}\ov{f}\;
\fd{\ov{\cal F}}{\ov{f}} \;\equiv\; \fd{\ov{\cal F}}{{\bf B}({\bf r})} \;-\;  \Delta_{\bf B}\ov{\cal F}({\bf r}).
\label{eq:ovF_del_B} 
\end{eqnarray}
We note, here, that the meta-pull-back operator $\mathbb{T}^{-1}$ introduces electromagnetic shifts $(\Delta_{\bf E}, \Delta_{\bf B})$ in functional derivatives due to the dependence of the phase-space transformation on the electromagnetic fields $({\bf E}, {\bf B})$, as shown in Eq.~\eqref{eq:partial_EB_def}.

\subsection{Transformed Hamiltonian functional}

As an application of the transformed functional variations \eqref{eq:ovF_del_f}-\eqref{eq:ovF_del_B}, we consider the transformed Hamiltonian functional $\ov{\cal H}$ obtained as the meta-push-forward of the Hamiltonian functional \eqref{eq:Ham_def}:
\begin{equation}
\ov{\cal H} \;\equiv\; \mathbb{T}{\cal H} \;=\; \frac{1}{8\pi}\int_{\bf r} \left( |{\bf E}|^{2} + |{\bf B}|^{2}\right) \;+\; \int_{\ov{\bf Z}}\ov{\cal J}\,\ov{f}\;\ov{K},
\label{eq:ov_Ham}
\end{equation}
where $\ov{K} \equiv {\sf T}^{-1}K$ denotes the transformed kinetic energy. First, the functional variation of $\ov{\cal H}$ yields
\begin{equation} 
\delta\ov{\cal H} \;=\; \int_{\bf r} \left( \delta{\bf E}\bdot\frac{\bf E}{4\pi} + \delta{\bf B}\bdot\frac{\bf B}{4\pi} \right) \;+\;
\int_{\ov{\bf Z}} \left[ \delta\ov{f}\; \ov{\cal J}\,\ov{K} \;+\frac{}{} \ov{f}\;\delta(\ov{\cal J}\,\ov{K}) \right],
\label{eq:delta_ovHam_primitive}
\end{equation}
where $\delta(\ov{\cal J}\,\ov{K}) = \delta\ov{\cal J}\;\ov{K} + \ov{\cal J}\;\delta\ov{K}$. Here, we have [see Eq.~\eqref{eq:Jacobian_T_delta}]
\begin{equation}
\delta\ov{\cal J} \;=\; -\;\pd{}{\ov{Z}^{\alpha}} \left[ \ov{\cal J}\;{\sf T}^{-1}\left(\delta{\sf T}\ov{Z}^{\alpha}\right) \right],
\label{eq:delta_ovJac}
\end{equation}
which also follows from Eq.~\eqref{eq:time_ovJac}, while
\begin{equation}
\delta\ov{K} \;=\; \delta\left({\sf T}^{-1}K\right) \;=\; {\sf T}^{-1}(\delta K) \;-\; \left(\left[{\sf T}^{-1},\;\delta\right]{\sf T}\right){\sf T}^{-1}K
\;\equiv\; -\;{\sf T}^{-1}\left(\delta{\sf T}\ov{Z}^{\alpha}\right)\;\pd{\ov{K}}{\ov{Z}^{\alpha}},
\label{eq:ov_delta_K}
\end{equation}
where we used $\delta K \equiv 0$ (i.e., the particle kinetic energy $K$ does not depend on ${\bf E}$ and ${\bf B}$). 

Next, by combining the two expressions \eqref{eq:delta_ovJac}-\eqref{eq:ov_delta_K}, we find
\begin{equation} 
\delta(\ov{\cal J}\,\ov{K}) \;=\; -\;\pd{}{\ov{Z}^{\alpha}}\left[ \ov{\cal J}\,\ov{K}\;{\sf T}^{-1}\left(\delta{\sf T}\ov{Z}^{\alpha}\right)\right],
\label{eq:delta_JK}
\end{equation}
so that integration by parts of this term yields
\begin{eqnarray} 
\delta\ov{\cal H} & = & \int_{\bf r} \left( \delta{\bf E}\bdot\frac{\bf E}{4\pi} + \delta{\bf B}\bdot\frac{\bf B}{4\pi} \right) \;+\;
\int_{\ov{\bf Z}} \left[ \delta\ov{f}\; \ov{\cal J}\,\ov{K} + \ov{\cal J}\,\ov{K}\;\left[{\sf T}^{-1}\left(\delta{\sf T}\ov{Z}^{\alpha}\right)\right]
\pd{\ov{f}}{\ov{Z}^{\alpha}} \right] \nonumber \\
 & \equiv & \int_{\bf r} \left( \delta{\bf E}({\bf r})\bdot\fd{\ov{\cal H}}{{\bf E}({\bf r})} + \delta{\bf B}({\bf r})\bdot
\fd{\ov{\cal H}}{{\bf B}({\bf r})} \right) \;+\; \int_{\ov{\bf Z}}\delta\ov{f}(\ov{\bf Z})\;\fd{\ov{\cal H}}{\ov{f}(\ov{\bf Z})},
\label{eq:delta_ovHam}
\end{eqnarray}
where we find the functional variations
\begin{eqnarray}
\fd{\ov{\cal H}}{\ov{f}} & = & \ov{\cal J}\;\ov{K}, \label{eq:ovH_del_f} \\
\fd{\ov{\cal H}}{{\bf E}} & = & \frac{{\bf E}}{4\pi} \;+\; \int_{\ov{\bf Z}}\partial_{\bf E}\ov{f}\;\fd{\ov{\cal H}}{\ov{f}} \;\equiv\;
\frac{{\bf E}}{4\pi} \;+\; \Delta_{\bf E}\ov{\cal H}, \label{eq:ovH_del_E} \\
\fd{\ov{\cal H}}{{\bf B}} & = & \frac{{\bf B}}{4\pi} \;+\; \int_{\ov{\bf Z}}\partial_{\bf B}\ov{f}\;\fd{\ov{\cal H}}{\ov{f}} \;\equiv\;
\frac{{\bf B}}{4\pi} \;+\; \Delta_{\bf B}\ov{\cal H}. \label{eq:ovH_del_B}
\end{eqnarray}
Hence, substituting Eqs.~\eqref{eq:ovH_del_f}-\eqref{eq:ovH_del_B} into Eqs.~\eqref{eq:ovF_del_f}-\eqref{eq:ovF_del_B} yields
\begin{eqnarray}
{\sf T}^{-1}\left[ \frac{1}{\cal J}\;\fd{(\mathbb{T}^{-1}\ov{\cal H})}{f}\right] & \equiv & \frac{1}{\ov{\cal J}}\;\fd{\ov{\cal H}}{\ov{f}} \;=\;
\ov{K} \;\equiv\; {\sf T}^{-1}\left( \frac{1}{\cal J}\;\fd{{\cal H}}{f}\right), \label{eq:TovH_del_f} \\
\fd{(\mathbb{T}^{-1}\ov{\cal H})}{{\bf E}({\bf r})} & \equiv & \fd{\ov{\cal H}}{{\bf E}({\bf r})} \;-\; \Delta_{\bf E}\ov{\cal H}({\bf r}) \;=\;
\frac{{\bf E}({\bf r})}{4\pi} \;\equiv\; \fd{\cal H}{{\bf E}({\bf r})}, \label{eq:TovH_del_E}  \\
\fd{(\mathbb{T}^{-1}\ov{\cal H})}{{\bf B}({\bf r})} & \equiv & \fd{\ov{\cal H}}{{\bf B}({\bf r})} \;-\; \Delta_{\bf B}\ov{\cal H}({\bf r}) \;=\;
\frac{{\bf B}({\bf r})}{4\pi} \;\equiv\; \fd{\cal H}{{\bf B}({\bf r})}, \label{eq:TovH_del_B} 
\end{eqnarray}
where we see that the terms $(\Delta_{\bf E}\ov{\cal H}, \Delta_{\bf B}\ov{\cal H})$ are exactly cancelled in the functional derivatives
$(\delta\ov{\cal H}/\delta{\bf E}, \delta\ov{\cal H}/\delta{\bf B})$, which is not the case for a general transformed functional $\ov{\cal F}$.

\subsection{Transformed Vlasov-Maxwell bracket}

The transformed Vlasov-Maxwell bracket is now constructed in a three-step process with the help of the meta-operators $\mathbb{T}^{-1}$ and $\mathbb{T}$. First, we note that the Vlasov-Maxwell functional bracket \eqref{eq:MV_bracket} is itself a functional and, thus, transforms as a functional under the action of the meta-push-forward $\mathbb{T}$. Hence, we express the meta-push-forward of the Vlasov-Maxwell functional bracket \eqref{eq:MV_bracket} as
\begin{eqnarray}
\mathbb{T}\left(\left[{\cal F},\frac{}{}{\cal G}\right]\right) & = & \int_{\ov{\bf Z}}\,\ov{\cal J}\;\ov{f}\; \left[{\sf T}^{-1}\left(
\left\{ {\cal J}^{-1}\fd{\cal F}{f},\; {\cal J}^{-1}\fd{\cal G}{f} \right\}\right) \right] \;+\; \left[ {\sf Maxwell}\frac{}{}{\sf sub-bracket}\right] \nonumber \\
 &  &-\; 4\pi\,e\int_{\ov{\bf Z}}\,\ov{\cal J}\,\ov{f} \left[ {\sf T}^{-1}\left(\fd{\cal F}{{\bf E}({\bf x})}\right)\bdot
{\sf T}^{-1}\left(\left\{ {\bf x},\; {\cal J}^{-1}\fd{\cal G}{f} \right\}\right) \right. \nonumber \\
 &  &\left.\hspace*{0.7in}-\; {\sf T}^{-1}\left(\fd{\cal G}{{\bf E}({\bf x})}\right)\bdot{\sf T}^{-1}\left(\left\{ {\bf x},\; {\cal J}^{-1}\fd{\cal F}{f} \right\}\right) \right],
\label{eq:meta_MV_1}
\end{eqnarray}
where the Maxwell sub-bracket is unaffected by the meta-push-forward (since it is independent of the Vlasov distribution function) and we used the distributivity property of the push-forward operation 
${\sf T}^{-1}$ in the interaction sub-bracket. 

Next, we use the definition of the transformed single-particle Poisson bracket \eqref{eq:new_PB} in the Vlasov and interaction sub-brackets to obtain
\begin{eqnarray}
\mathbb{T}\left(\left[{\cal F},\frac{}{}{\cal G}\right]\right) & = & \int_{\ov{\bf Z}}\,\ov{\cal J}\;\ov{f}\;\ov{\left\{ {\sf T}^{-1}\left(
{\cal J}^{-1}\fd{\cal F}{f}\right),\; {\sf T}^{-1}\left({\cal J}^{-1}\fd{\cal G}{f}\right) \right\}} \;+\; \left[ {\sf Maxwell}\frac{}{}{\sf sub-bracket}\right] \nonumber \\
 &  &-\;4\pi\,e\;\int_{\ov{\bf Z}}\,\ov{\cal J}\;\ov{f}\;  \left[ {\sf T}^{-1}\left(\fd{\cal F}{{\bf E}({\bf x})}\right)\bdot
\ov{\left\{ {\sf T}^{-1}{\bf x},\; {\sf T}^{-1}\left({\cal J}^{-1}\fd{\cal G}{f}\right) \right\}} \right. \nonumber \\
 &  &\left.\hspace*{1in}-\; {\sf T}^{-1}\left(\fd{\cal G}{{\bf E}({\bf x})}\right)\bdot\ov{\left\{ {\sf T}^{-1}{\bf x},\; {\sf T}^{-1}\left(
{\cal J}^{-1}\fd{\cal F}{f}\right) \right\}} \right].
\label{eq:meta_MV_2}
\end{eqnarray}
Lastly, we use the meta-pull-back operation $\mathbb{T}^{-1}$ to replace the particle functionals ${\cal F}$ and ${\cal G}$ with the transformed functionals ${\cal F} = \mathbb{T}^{-1}\ov{\cal F}$ and ${\cal G} = \mathbb{T}^{-1}\ov{\cal G}$, which yields the transformed Vlasov-Maxwell bracket
\begin{eqnarray}
\ov{\left[\ov{\cal F},\frac{}{} \ov{\cal G}\right]} & \equiv & \mathbb{T}\left(\left[\mathbb{T}^{-1}\ov{\cal F},\frac{}{} \mathbb{T}^{-1}\ov{\cal G}
\right]\right) \nonumber \\
 & = & \int_{\ov{\bf Z}}\,\ov{\cal J}\;\ov{f}\;\ov{\left\{ \left[{\sf T}^{-1}\;
\left({\cal J}^{-1}\fd{}{f}\;(\mathbb{T}^{-1}\ov{\cal F})\right)\right],\; \left[{\sf T}^{-1}\;\left({\cal J}^{-1}\fd{}{f}\;
(\mathbb{T}^{-1}\ov{\cal G})\right)\right] \right\}} \label{eq:meta_MV_red} \\
 &  &+\; 4\pi\,c\; \int_{\bf r}\;\left[ \fd{(\mathbb{T}^{-1}\ov{\cal F})}{{\bf E}({\bf r})}\bdot\nabla\btimes
\fd{(\mathbb{T}^{-1}\ov{\cal G})}{{\bf B}({\bf r})} \;-\; \fd{(\mathbb{T}^{-1}\ov{\cal G})}{{\bf E}({\bf r})}\bdot\nabla\btimes
\fd{(\mathbb{T}^{-1}\ov{\cal F})}{{\bf B}({\bf r})} \right] \nonumber \\
 &  &-\;4\pi\,e\;\int_{\ov{\bf Z}}\,\ov{\cal J}\;\ov{f}\;  \left[ {\sf T}^{-1}\left(\fd{(\mathbb{T}^{-1}\ov{\cal F})}{{\bf E}({\bf x})}\right)\bdot
\ov{\left\{ {\sf T}^{-1}{\bf x},\; \left[{\sf T}^{-1}\;\left({\cal J}^{-1}\fd{}{f}\;(\mathbb{T}^{-1}\ov{\cal G})\right)\right] \right\}} \right. 
\nonumber \\ 
 &  &\left.\hspace*{1in}-\; {\sf T}^{-1}\left(\fd{(\mathbb{T}^{-1}\ov{\cal G})}{{\bf E}({\bf x})}\right)\bdot\ov{\left\{ 
{\sf T}^{-1}{\bf x},\; \left[{\sf T}^{-1}\;\left({\cal J}^{-1}\fd{}{f}\;(\mathbb{T}^{-1}\ov{\cal F})\right)\right] \right\}} \right].
\nonumber
\end{eqnarray}
If we now substitute the transformed functional variations \eqref{eq:ovF_del_f}-\eqref{eq:ovF_del_B}, we finally obtain the transformed Vlasov-Maxwell bracket
\begin{eqnarray}
\ov{\left[\ov{\cal F},\frac{}{}\ov{\cal G}\right]}  & = & \int_{\ov{\bf Z}}\,\ov{\cal J}\;\ov{f}\; \ov{\left\{ \frac{1}{\ov{\cal J}}\,
\fd{\ov{\cal F}}{\ov{f}},\; \frac{1}{\ov{\cal J}}\,\fd{\ov{\cal G}}{\ov{f}} \right\}} \nonumber \\
 &  &+\; 4\pi\,c\;\int_{\bf r}\;\left[\left( \fd{\ov{\cal F}}{{\bf E}({\bf r})} - \Delta_{\bf E}\ov{\cal F}({\bf r})\right)\bdot\nabla\btimes\left( \fd{\ov{\cal G}}{{\bf B}({\bf r})} - 
\Delta_{\bf B}\ov{\cal G}({\bf r})\right) \right. \nonumber \\
 &  &\hspace*{0.5in}\left.-\; \left( \fd{\ov{\cal G}}{{\bf E}({\bf r})} - \Delta_{\bf E}\ov{\cal G}({\bf r})\right)\bdot\nabla\btimes\left( 
\fd{\ov{\cal F}}{{\bf B}({\bf r})} - \Delta_{\bf B}\ov{\cal F}({\bf r})\right) \right] \nonumber \\
 &  &-\;4\pi\,e\;\int_{\ov{\bf Z}}\,\ov{\cal J}\;\ov{f} \left[ {\sf T}^{-1}\left( \fd{\ov{\cal F}}{{\bf E}({\bf x})} - \Delta_{\bf E}\ov{\cal F}({\bf x})\right)\bdot\ov{\left\{ {\sf T}^{-1}{\bf x},\; \frac{1}{\ov{\cal J}}\,\fd{\ov{\cal G}}{\ov{f}} \right\}} \right. \nonumber \\
 &  &\hspace*{0.5in}\left.-\; {\sf T}^{-1}\left( \fd{\ov{\cal G}}{{\bf E}({\bf x})} - \Delta_{\bf E}\ov{\cal G}({\bf x})\right)\bdot\ov{\left\{ 
{\sf T}^{-1}{\bf x},\; \frac{1}{\ov{\cal J}}\,\fd{\ov{\cal F}}{\ov{f}} \right\}} \;\right].
\label{eq:MV_bracket_transform}
\end{eqnarray}
The bracket \eqref{eq:MV_bracket_transform} is of the form of that of Eq.~(47) of \cite{pjm13}, but is restricted since it is essentially a restatement of Vlasov-Maxwell theory, while the bracket of \cite{pjm13} applies to a larger class of theories.  We immediately note that the transformed Vlasov-Maxwell bracket \eqref{eq:MV_bracket_transform} automatically satisfies the Jacobi identity
\begin{eqnarray}
\ov{\left[\ov{[\ov{\cal F},\ov{\cal G}]},\frac{}{} \ov{\cal H}\right]} + \ov{\left[\ov{[\ov{\cal G},\ov{\cal H}]},\frac{}{} \ov{\cal F}\right]} +
\ov{\left[\ov{[\ov{\cal H},\ov{\cal F}]},\frac{}{} \ov{\cal G}\right]} & \equiv & \mathbb{T}\left( \left[[{\cal F}, {\cal G}],\frac{}{} {\cal H}\right]
+ \left[[{\cal G}, {\cal H}],\frac{}{} {\cal F}\right] + \left[[{\cal H}, {\cal F}],\frac{}{} {\cal G}\right] \right) \nonumber \\
 & \equiv & 0,
\label{eq:Jacobi_transform}
\end{eqnarray}
since the original Vlasov-Maxwell bracket \eqref{eq:MV_bracket} satisfies the Jacobi identity.

We now show that the transformed Vlasov-Maxwell equations \eqref{eq:transformed_Vlasov}, \eqref{eq:Edot_new}, and \eqref{eq:B_dot} can be expressed in Hamiltonian form as $\ov{[\ov{\cal F},\; \ov{\cal H}]}$, where
\begin{eqnarray}
\ov{\left[\ov{\cal F},\frac{}{}\ov{\cal H}\right]} & = & \int_{\ov{\bf Z}}\,\ov{\cal J}\;\ov{f}\; \ov{\left\{ \frac{1}{\ov{\cal J}}\,
\fd{\ov{\cal F}}{\ov{f}},\; \ov{K} \right\}} \nonumber \\
 &  &+\; c\int_{\bf r}\;\left[\left( \fd{\ov{\cal F}}{{\bf E}} - \Delta_{\bf E}\ov{\cal F}\right)\bdot\nabla\btimes{\bf B} \;-\; \left( 
\fd{\ov{\cal F}}{{\bf B}} - \Delta_{\bf B}\ov{\cal F}\right)\bdot\nabla\btimes {\bf E}\right] \nonumber \\
 &  &-\;4\pi\,e\;\int_{\ov{\bf Z}}\,\ov{\cal J}\;\ov{f} \left[ {\sf T}^{-1}\left( \fd{\ov{\cal F}}{{\bf E}({\bf x})} - \Delta_{\bf E}\ov{\cal F}({\bf x})\right)\bdot\ov{\left\{ {\sf T}^{-1}{\bf x},\; \ov{K} \right\}} \right. \nonumber \\
 &  &\left.\hspace*{0.7in}-\; {\sf T}^{-1}\left( \frac{{\bf E}({\bf x})}{4\pi}\right)\bdot\ov{\left\{ 
{\sf T}^{-1}{\bf x},\; \frac{1}{\ov{\cal J}}\,\fd{\ov{\cal F}}{\ov{f}} \right\}} \;\right],
\label{eq:MV_bracket_transform_H}
\end{eqnarray}
where we made use of Eqs.~\eqref{eq:TovH_del_f}-\eqref{eq:TovH_del_B}. Upon integration by parts, using Eq.~\eqref{eq:fgh_id_new}, we obtain
\begin{eqnarray}
\ov{\left[\ov{\cal F},\frac{}{}\ov{\cal H}\right]} & = & -\int_{\ov{\bf Z}}\fd{\ov{\cal F}}{\ov{f}}\left( \ov{\left\{\ov{f},\frac{}{} \ov{K} \right\}} + 
e\,{\sf T}^{-1}{\bf E}\bdot\ov{\left\{ {\sf T}^{-1}{\bf x}, \ov{f} \right\}} \right) - \int_{\bf r}\left( \fd{\ov{\cal F}}{{\bf B}} - 
\Delta_{\bf B}\ov{\cal F}\right)\bdot\left(c\frac{}{}\nabla\btimes {\bf E}\right)  \nonumber \\
 &  &+ \int_{\bf r}\left( \fd{\ov{\cal F}}{{\bf E}} - \Delta_{\bf E}\ov{\cal F}\right)\bdot\left[ c\;\nabla\btimes{\bf B} \;-\frac{}{} 4\pi\,
\mathbb{T}{\bf J}({\bf r}) \right] \nonumber \\
 & \equiv & \int_{\ov{\bf Z}}\fd{\ov{\cal F}}{\ov{f}}\;\pd{\ov{f}}{t} \;+\; \int_{\bf r} \left( \fd{\ov{\cal F}}{\bf E}\bdot\pd{\bf E}{t} \;+\;
\fd{\ov{\cal F}}{\bf B}\bdot\pd{\bf B}{t} \right),
\label{eq:MV_bracket_transform_H_2}
\end{eqnarray}
where we used
\begin{eqnarray*}
 &  &\int_{\bf r}\left[\Delta_{\bf E}\ov{\cal F}\bdot\left(c\,\nabla\btimes{\bf B} \;-\frac{}{} 4\pi\,\mathbb{T}{\bf J}\right) \;+\; \Delta_{\bf B}\ov{\cal F}\bdot\left(-\,c\frac{}{}\nabla\btimes{\bf E}\right) \right] \\
&  &\hspace*{0.3in}=\; \int_{\bf r}\int_{\ov{\bf Z}}\left( \pd{\bf E}{t}\bdot\partial_{\bf E}\ov{f} \;+\;
\pd{\bf B}{t}\bdot\partial_{\bf B}\ov{f}\right) \fd{\ov{\cal F}}{\ov{f}} \\
&  &\hspace*{0.3in}=\; \int_{\ov{\bf Z}}{\sf T}^{-1}\left(\pd{({\sf T}\ov{Z}^{\alpha})}{t}\right)\;\pd{\ov{f}}{\ov{Z}^{\alpha}}\,\fd{\ov{\cal F}}{\ov{f}} \;\equiv\; \int_{\ov{\bf Z}}\frac{\ov{\partial}\ov{Z}^{\alpha}}{\partial t}\;\pd{\ov{f}}{\ov{Z}^{\alpha}}\,\fd{\ov{\cal F}}{\ov{f}}.
\end{eqnarray*}

We have thus shown that, as a result of a general phase-space transformation ${\cal T}: {\bf z} \rightarrow \ov{\bf Z}$, the transformed Vlasov-Maxwell equations possess a Hamiltonian structure that is constructed by operators $({\sf T}, {\sf T}^{-1})$ and meta-operators $(\mathbb{T}, \mathbb{T}^{-1})$ derived from the phase-space transformation. Here, the transformed Hamiltonian functional \eqref{eq:ov_Ham} is defined as the meta-push-forward of the Hamiltonian functional $\ov{\cal H} \equiv \mathbb{T}{\cal H}$, while the transformed Vlasov-Maxwell bracket \eqref{eq:MV_bracket_transform} is defined as $\ov{[\ov{\cal F},\;\ov{\cal G}]} \equiv \mathbb{T}([\mathbb{T}^{-1}\ov{\cal F},\; \mathbb{T}^{-1}\ov{\cal G}])$. 

\section{\label{sec:local}Local Phase-space Coordinates}

Reduced Vlasov-Maxwell equations are often derived through a preliminary phase-space transformation ${\cal T}_{0}$ from the particle phase-space coordinates $({\bf x},{\bf p})$ to the local phase-space coordinates 
$({\bf x},p_{\|},p_{\bot},\zeta)$ derived from the local magnetic field ${\bf B}({\bf x}) \equiv B\;\bhat$, where $p_{\|} \equiv {\bf p}\bdot\bhat$ denotes the parallel component of the particle's momentum along the (local) magnetic-field unit vector at the particle's position ${\bf x}$, and $p_{\bot} \equiv |{\bf p}\btimes\bhat|$ denotes the magnitude of the perpendicular component of the particle's momentum. It is also convenient to express $p_{\bot} \equiv (2m\mu\,B)^{\frac{1}{2}}$ in terms of the lowest-order magnetic moment $\mu({\bf x},{\bf p}) \equiv |{\bf p}\btimes\bhat|^{2}/(2m\,B)$ for guiding-center \citep{Littlejohn_1983} and gyrocenter applications \citep{Brizard_Hahm_2007}. The local gyroangle $\zeta$ denotes the orientation of the particle's perpendicular momentum in the (local) plane perpendicular to $\bhat$, which is defined in terms of the differential equation $\wh{\bot} \equiv \partial\wh{\rho}/\partial\zeta$, where the plane locally perpendicular to $\bhat \equiv \wh{\bot}\btimes\wh{\rho}$ is defined in terms of two arbitrary rotating unit vectors 
$(\wh{\bot},\wh{\rho})$. The local momentum is, thus, decomposed under the action of ${\cal T}_{0}$ as
\begin{equation}
{\cal T}_{0}({\bf x},{\bf p}) \;\equiv\; \left({\bf x},\frac{}{} p_{\|}({\bf x},{\bf p})\,\bhat({\bf x}) \;+\; p_{\bot}({\bf x},{\bf p})\,
\wh{\bot}({\bf x},\zeta)\right).
\label{eq:local_pst}
\end{equation}
With the local phase-space coordinates $({\bf x},p_{\|},\mu,\zeta)$, the Jacobian is ${\cal J}_{0} = m\,B$. Lastly, we note that, at constant $({\bf x},
{\bf p})$, the local momentum coordinates $(p_{\|},\mu,\zeta)$ are time-dependent functions through their dependence on ${\bf B}$.

\subsection{Local Hamiltonian dynamics}

Local particle Hamiltonian dynamics is expressed in terms of the local kinetic energy $K_{0} = p_{\|}^{2}/2m + \mu\,B$ and a local Poisson bracket constructed as follows. We begin with the local Lagrange one-form 
\begin{equation} 
\gamma_{0} \;=\; \left[\frac{e}{c}\,{\bf A}({\bf x}) \;+\; \left( p_{\|}\,\bhat({\bf x}) \;+\frac{}{} p_{\bot}(\mu,{\bf x})\,\wh{\bot}(\zeta,{\bf x})\right)\right]\bdot\exd{\bf x},
\label{eq:gamma_0}
\end{equation}
from which we obtain the local Lagrange two-form 
\begin{eqnarray}
\vb{\omega}_{0} & \equiv & \exd\gamma_{0} \;=\; \frac{e}{2c}\,B_{0}^{* k}\,\varepsilon_{ijk}\,\exd x^{i}\wedge\exd x^{j} \;+\; \exd p_{\|}\wedge
\bhat\bdot\exd{\bf x} \;+\; \pd{p_{\bot}}{\mu}\;\exd\mu\wedge\wh{\bot}\bdot\exd{\bf x} \;-\; p_{\bot}\,\exd\zeta \wedge \wh{\rho}\bdot\exd{\bf x} \nonumber \\
 & \equiv & \frac{1}{2}\;\omega_{0\,\alpha\beta}\;\exd z_{0}^{\alpha}\wedge\exd z_{0}^{\beta},
\label{eq:omega_0}
\end{eqnarray}
where the local {\it canonical} magnetic field is defined as
\begin{eqnarray} 
{\bf B}_{0}^{*} & \equiv & \nabla\btimes\left[{\bf A} \;+\; \frac{c}{e}\,\left(p_{\|}\,\bhat \;+\frac{}{} p_{\bot}(\mu,B)\,\wh{\bot}\right)\right] \nonumber \\
 & = & {\bf B} \;+\; \frac{cp_{\|}}{e}\;\nabla\btimes\bhat \;+\; \frac{cp_{\bot}}{e} \left(\nabla\btimes\wh{\bot} \;-\; \frac{1}{2}\,\wh{\bot}\btimes
\nabla\ln B \right).
\label{eq:B0_star}
\end{eqnarray}
Next, we derive the local Poisson matrix (with components $J_{0}^{\alpha\beta}$) as the inverse of the Lagrange matrix (with components 
$\omega_{0\,\alpha\beta}$), which yields the local Poisson bracket
\begin{equation}
\{f,\; g\}_{0} \;\equiv\; \pd{f}{z_{0}^{\alpha}}\;J_{0}^{\alpha\beta}\;\pd{g}{z_{0}^{\beta}} \;=\; \nabla f\bdot\vb{\partial}g \;-\; \vb{\partial}f\bdot
\nabla g \;+\; \frac{e}{c}\,{\bf B}_{0}^{*}\bdot\left(\vb{\partial}f\btimes\vb{\partial}g\right),
\label{eq:PB_0}
\end{equation}
where $\vb{\partial}$ denotes the local momentum-space gradient:
\begin{equation} 
\vb{\partial}f \;\equiv\; \bhat\;\pd{f}{p_{\|}} \;+\; \wh{\bot}\;\frac{p_{\bot}}{mB}\;\pd{f}{\mu} \;-\; \frac{\wh{\rho}}{p_{\bot}}\;\pd{f}{\zeta}
\;\equiv\; \bhat\,\partial_{\|}f \;+\; \vb{\partial}_{\bot}f.
\label{eq:local_pgrad}
\end{equation}
We note that the local Poisson bracket \eqref{eq:PB_0}, which automatically satisfies the Jacobi identity [see Eq.~\eqref{eq:d_omega_0}], can be expressed in phase-space divergence form (using the local Jacobian ${\cal J}_{0} = m\,B$)
\begin{equation}
\{ f,\; g\}_{0} \;\equiv\; \frac{1}{{\cal J}_{0}}\;\pd{}{z_{0}^{\alpha}} \left({\cal J}_{0}\,f\frac{}{} \left\{ z_{0}^{\alpha},\frac{}{} g \right\}_{0}
\right),
\label{eq:PB_0_div}
\end{equation}
so that we obtain the integral identity $\int_{{\bf z}_{0}}{\cal J}_{0}\,f\;\{ g,\; h\}_{0} \equiv -\;\int_{{\bf z}_{0}}{\cal J}_{0}\,g\;\{ f,\; 
h\}_{0}$, where $\int_{{\bf z}_{0}} \equiv \int d^{3}x\,dp_{\|}d\mu\,d\zeta$.

The Hamilton equations \eqref{eq:dotz_alpha} are given in local particle phase space in terms of the local kinetic energy $K_{0}$ and the local Poisson bracket \eqref{eq:PB_0} as
\begin{eqnarray}
\dot{\bf x}_{0} & \equiv & \{ {\bf x},\; K_{0}\}_{0} \;=\; \frac{p_{\|}}{m}\;\bhat \;+\; \frac{p_{\bot}}{m}\;\wh{\bot}, \label{eq:x_dot_0} \\
\dot{p}_{\|0} & \equiv & \{ p_{\|},\; K_{0}\}_{0} \;+\; e\,{\bf E}\bdot\bhat \nonumber \\
 & = & \bhat\bdot\left( e\,{\bf E} \;-\; \mu\;\nabla B\right) \;+\; p_{\bot}\,\wh{\rho}\bdot\left[\frac{1}{m}\,\nabla\btimes\left(p_{\|}\,\bhat \;+\frac{}{} p_{\bot}\,\wh{\bot}\right)\right], \label{eq:p_dot_0} \\
\dot{\mu}_{0} & \equiv & \{ \mu,\; K_{0}\}_{0} \;+\; e\,{\bf E}\bdot\wh{\bot}\;\frac{p_{\bot}}{m\,B} \nonumber \\
 & = & \frac{p_{\bot}}{m\,B}\,\wh{\bot}\bdot\left( e\,{\bf E} \;-\; \mu\;\nabla B\right) \;-\; \frac{p_{\bot}p_{\|}}{m\,B}\,\wh{\rho}\bdot\left[\frac{1}{m}\,\nabla\btimes\left(p_{\|}\,\bhat \;+\frac{}{} p_{\bot}\,\wh{\bot}\right)\right], \label{eq:mu_dot_0} \\
\dot{\zeta}_{0} & \equiv & \{ \zeta,\; K_{0}\}_{0} \;-\; e\,{\bf E}\bdot\frac{\wh{\rho}}{p_{\bot}} \nonumber \\
 & = & -\; \frac{\wh{\rho}}{p_{\bot}}\bdot\left( e\,{\bf E} \;-\; \mu\;\nabla B\right) \;+\; \frac{eB}{mc} \;-\; \frac{p_{\|}}{p_{\bot}}\,\wh{\bot}\bdot\left[\frac{1}{m}\,\nabla\btimes\left(p_{\|}\,\bhat \;+\frac{}{} p_{\bot}\,\wh{\bot}\right)\right], \label{eq:zeta_dot_0}
\end{eqnarray}
where we used ${\sf T}_{0}^{-1}{\bf x} \equiv {\bf x}$ and ${\sf T}_{0}^{-1}{\bf E} \equiv {\bf E}$. We note that Eqs.~\eqref{eq:x_dot_0}-\eqref{eq:zeta_dot_0} satisfy the local Liouville Theorem
\begin{equation}
\nabla\bdot\left({\cal J}_{0}\dot{\bf x}_{0}\right) \;+\; \pd{}{p_{\|}}\left({\cal J}_{0}\,\dot{p}_{\|0}\right) \;+\; \pd{}{\mu}\left({\cal J}_{0}\,
\dot{\mu}_{0}\right) \;+\; \pd{}{\zeta}\left({\cal J}_{0}\,\dot{\zeta}_{0}\right) \;\equiv\; \pd{}{z_{0}^{\alpha}}\left(\dot{z}_{0}^{\alpha}\frac{}{}
{\cal J}_{0} \right) \;=\; 0,
\label{eq:local_Liouville}
\end{equation}
which is a necessary requirement from any set of Hamilton equations. We also note that when Eq.~\eqref{eq:mu_dot_0} is averaged with respect to the local gyroangle $\zeta$: $\langle\dot{\mu}_{0}\rangle \equiv \oint \dot{\mu}_{0}\, d\zeta/(2\pi)$, we find 
\[ \left\langle \dot{\mu}_{0}\right\rangle \;=\; -\;\mu_{0}\;\left[\frac{p_{\|}}{mB}\;(\nabla\bdot{\bf B})\right] \;=\; 0, \]
which is a necessary requirement for the construction of the guiding-center magnetic moment as an adiabatic invariant \citep{Cary_Brizard_2009}.

Using the local Hamilton equations \eqref{eq:x_dot_0}-\eqref{eq:zeta_dot_0}, the local Vlasov equation is now expressed as
\begin{equation}
\frac{\partial_{0}f_{0}}{\partial t} \;\equiv\; \pd{f_{0}}{t} \;+\; \frac{\partial_{0}z_{0}^{\alpha}}{\partial t}\;\pd{f_{0}}{z_{0}^{\alpha}} \;=\; -\,\dot{\bf x}_{0}\bdot\nabla f_{0} \;-\; \dot{p}_{\|0}\;\pd{f_{0}}{p_{\|}} \;-\; \dot{\mu}_{0}\;\pd{f_{0}}{\mu} \;-\; \dot{\zeta}_{0}\;\pd{f_{0}}{\zeta},
\label{eq:Vlasov_0}
\end{equation}
where the local time partial derivative $\partial_{0}/\partial t \equiv {\sf T}_{0}^{-1}(\partial{\sf T}_{0}/\partial)$ is derived below [see Eq.~\eqref{eq:time_T0}], with $\partial_{0} z_{0}^{\alpha}/\partial t$ evaluated at constant $({\bf x},{\bf p})$. We note that, since the local Hamilton equations \eqref{eq:x_dot_0}-\eqref{eq:zeta_dot_0} depend explicitly on the local gyroangle $\zeta$ (through the rotating unit vectors $\wh{\bot}$ and $\wh{\rho}$), the local Vlasov function $f_{0}$ evolves rapidly on the gyromotion time scale.

Lastly, the Maxwell equations for the electromagnetic fields $({\bf E},{\bf B})$ are
\begin{eqnarray}
\pd{\bf E}{t} & = & c\,\nabla\btimes{\bf B} \;-\; 4\pi\,e\;\int_{{\bf z}_{0}}\;\delta^{3}({\bf x} - {\bf r})\;{\cal J}_{0}\,f_{0}\;\dot{\bf x}_{0}
\;\equiv\; c\,\nabla\btimes{\bf B} \;-\; 4\pi\,{\bf J}_{0}, 
\label{eq:Edot_0} \\
\pd{\bf B}{t} & = & -\;c\,\nabla\btimes{\bf E},
\label{eq:Bdot_0}
\end{eqnarray}
with 
\begin{equation}
\nabla\bdot{\bf E}({\bf r}) \;=\; 4\pi\,e\int_{{\bf z}_{0}}\delta^{3}({\bf x} - {\bf r})\,{\cal J}_{0}\,f_{0} \;\equiv\; 4\pi\,\varrho_{0}
\label{eq:divE_0}
\end{equation} 
and $\nabla\bdot{\bf B}({\bf r}) = 0$. We will now show that the local Vlasov-Maxwell equations \eqref{eq:Vlasov_0}-\eqref{eq:Bdot_0} possess a Hamiltonian formulation.

\subsection{Local coordinate and field variations}

The particle velocity is represented in terms of components of the local magnetic field and, thus, the local particle phase-space coordinates acquire an explicit dependence on the magnetic field \citep{pjmVG13}.

\subsubsection{Local coordinate variations}

Since the definition of the local phase-space coordinates depends on the local magnetic field ${\bf B} \equiv B\;\bhat$, they are susceptible to variations $\delta{\bf B} \equiv \delta B\,\bhat + B\,\delta\bhat$ in the magnetic field ${\bf B} = B\,\bhat$. Hence, at fixed $({\bf x},{\bf p})$, we can calculate ${\sf T}_{0}^{-1}\left[\delta\left({\sf T}_{0}z_{0}^{\alpha}\right)\right] \equiv \delta_{0}z_{0}^{\alpha}$ for each local coordinate 
$z_{0}^{\alpha}$, with $\delta_{0}{\bf x} \equiv 0$. 

We begin with $\delta_{0}p_{\|}$: first, ${\sf T}_{0}p_{\|} \equiv {\bf p}\bdot\bhat$, so that $\delta({\sf T}_{0}p_{\|}) \equiv {\bf p}
\bdot\delta\bhat$; next, 
\[ {\sf T}_{0}^{-1}[\delta({\sf T}_{0}p_{\|})] \;\equiv\; ({\sf T}_{0}^{-1}{\bf p})\bdot\delta\bhat \;=\; \left( p_{\|}\,\bhat \;+\; p_{\bot}\,\wh{\bot}
\right)\bdot\delta\bhat. \]
By using the identity $\bhat\bdot\delta\bhat \equiv 0$, where $\delta\bhat \equiv \delta({\bf B}/B) = ({\bf I} - \bhat\bhat)\bdot\delta{\bf B}/B$, we thus find
\begin{equation}
\delta_{0} p_{\|} \;=\; p_{\bot}\;\wh{\bot}\bdot\delta\bhat.
\label{eq:delta_p||}
\end{equation}
Similarly, we find
\begin{equation}
\delta_{0} p_{\bot} \;=\; {\sf T}_{0}^{-1}[\delta({\sf T}_{0}p_{\bot})] \;=\; {\sf T}_{0}^{-1}\left[{\bf p}\btimes\delta\bhat\bdot\left({\bf p}\btimes\bhat/|{\bf p}\btimes\bhat|\right)\right] \;=\; -\;p_{\|}\;\wh{\bot}\bdot\delta\bhat, \label{eq:delta_pbot}
\end{equation}
and
\begin{equation}
\delta_{0}\mu \;=\; \delta_{0}\left(\frac{p_{\bot}^{2}}{2m\,B}\right) \;=\; \frac{p_{\bot}}{m\,B}\;\delta_{0} p_{\bot} \;-\; \mu\,\bhat\bdot\frac{\delta{\bf B}}{B} \;\equiv\; -\;\left( \frac{p_{\bot}p_{\|}}{m\,B}\;\wh{\bot} \;+\; \mu\,\bhat\right)\bdot\frac{\delta{\bf B}}{B}. \label{eq:delta_mu}
\end{equation}
We note that the variations \eqref{eq:delta_p||}-\eqref{eq:delta_mu} satisfy the energy conservation law 
\begin{equation}
\delta_{0}K_{0} \;=\; \frac{p_{\|}}{m} \;\delta_{0}p_{\|} \;+\; \delta_{0}\mu\;B \;+\; \mu\;\delta B \;\equiv\; 0.
\label{eq:delta_0_K0}
\end{equation}

The expression for the gyroangle variation $\delta_{0}\zeta$ is derived as follows. From the identity $\delta_{0}({\sf T}_{0}^{-1}{\bf p}) \equiv 0$, we find the vector identity
\begin{equation}
0 \;\equiv\; \delta_{0} p_{\|}\;\bhat \;+\; p_{\|}\;\delta\bhat \;+\; \delta_{0}p_{\bot}\;\wh{\bot} \;+\; p_{\bot}\;\delta_{0}\wh{\bot}.
\label{eq:delta_p_id}
\end{equation}
Using Eq.~\eqref{eq:delta_p||}, the parallel component of Eq.~\eqref{eq:delta_p_id} yields the identity $p_{\bot} (\wh{\bot}\bdot\delta\bhat + \bhat\bdot\delta_{0}\wh{\bot}) \equiv
p_{\bot} \delta_{0}(\bhat\bdot\wh{\bot}) \equiv 0$, while its component along $\wh{\bot}$ is identically zero
from Eq.~\eqref{eq:delta_pbot}. The remaining component is along $\wh{\rho}$: $p_{\|}\,\delta\bhat\bdot\wh{\rho} + p_{\bot}\,\delta_{0}\wh{\bot}\bdot
\wh{\rho} \equiv 0$. Using the definition $\wh{\rho} \equiv -\;\partial\wh{\bot}/\partial\zeta$, we thus find
\begin{equation}
\delta_{0}\zeta \;\equiv\; \frac{p_{\|}}{p_{\bot}}\;\wh{\rho}\bdot\delta\bhat. \label{eq:delta_zeta}
\end{equation}
In addition, we note the local-coordinate variations \eqref{eq:delta_p||}, \eqref{eq:delta_mu}, and \eqref{eq:delta_zeta} satisfy the divergence property
\begin{equation}
\pd{\delta_{0}p_{\|}}{p_{\|}} \;+\; \pd{\delta_{0}\mu}{\mu} \;+\; \pd{\delta_{0}\zeta}{\zeta} \;\equiv\; -\,\frac{\bhat}{B}\bdot\delta{\bf B} \;=\; -\;
\delta\ln B.
\label{eq:local_variation_div}
\end{equation}
Lastly, we also have the local partial-time derivatives
\begin{eqnarray}
\frac{\partial_{0}p_{\|}}{\partial t} & = & p_{\bot}\,\wh{\bot}\bdot\pd{\bhat}{t}, \\
\frac{\partial_{0}\mu}{\partial t} & = & -\;\left( \frac{p_{\bot}p_{\|}}{mB}\,\wh{\bot} \;+\; \mu\,\bhat\right)\bdot\frac{1}{B}\,\pd{\bf B}{t}, \\
\frac{\partial_{0}\zeta}{\partial t} & = & \frac{p_{\|}}{p_{\bot}}\,\wh{\rho}\bdot\pd{\bhat}{t},
\end{eqnarray}
which satisfy the divergence property
\begin{equation}
\pd{}{p_{\|}}\left(\frac{\partial_{0}p_{\|}}{\partial t}\right) \;+\; \pd{}{\mu}\left(\frac{\partial_{0}\mu}{\partial t}\right) \;+\; 
\pd{}{\zeta}\left(\frac{\partial_{0}\zeta}{\partial t}\right) \;\equiv\; -\,\frac{\bhat}{B}\bdot\pd{\bf B}{t} \;=\; -\;\pd{\ln B}{t}.
\label{eq:local_time_div}
\end{equation}
If we multiply this equation with ${\cal J}_{0} = m\,B$, we obtain 
\begin{equation} 
\pd{{\cal J}_{0}}{t} \;\equiv\; -\;\pd{}{z_{0}^{\alpha}}\left({\cal J}_{0}\;\frac{\partial_{0}z_{0}^{\alpha}}{\partial t}\right),
\label{eq:time_Jac_0}
\end{equation}
which is a special case of the general equation \eqref{eq:time_ovJac} for the transformed Jacobian $\ov{\cal J}$.

\subsubsection{Local field variations}

We now introduce the pull-back and push-forward operators associated with the local phase-space transformation: $f \equiv {\sf T}_{0}f_{0}$ and $f_{0} \equiv {\sf T}_{0}^{-1}f$, which both satisfy the scalar-covariance property $f({\bf x},{\bf p}) = f_{0}({\bf x},p_{\|},\mu,\zeta)$. Using these operators, we construct the relation between the variation $\delta f$ in particle phase space and the variation $\delta f_{0}$ in local particle phase space. 

For this purpose, we construct the push-forward ${\sf T}_{0}^{-1}(\delta f)$ of the particle phase-space variation $\delta f$:
\begin{equation}
{\sf T}_{0}^{-1}(\delta f) \;=\; \delta\left({\sf T}_{0}^{-1}f\right) \;+\; \left(\left[{\sf T}_{0}^{-1},\frac{}{}\delta\right]\,
{\sf T}_{0}\right)\;{\sf T}_{0}^{-1}f \;\equiv\; \delta f_{0} \;+\; \left(\left[{\sf T}_{0}^{-1},\frac{}{}\delta\right]\,
{\sf T}_{0}\right)\;f_{0},
\label{eq:delta_T0}
\end{equation}
where
\begin{eqnarray} 
\left(\left[{\sf T}_{0}^{-1},\frac{}{}\delta\right]\,{\sf T}_{0}\right)\;f_{0} & \equiv & \delta_{0}f_{0} \;=\;  \delta_{0} p_{\|}\,\pd{f_{0}}{p_{\|}} 
\;+\; \delta_{0}\mu\;\pd{f_{0}}{\mu} \;+\; \delta_{0}\zeta\;\pd{f_{0}}{\zeta} \label{eq:partialB_def} \\
 & = & \delta{\bf B}\bdot\left( \frac{\delta_{0}p_{\|}}{\delta{\bf B}}\,
\pd{f_{0}}{p_{\|}} \;+\; \frac{\delta_{0}\mu}{\delta{\bf B}}\,\pd{f_{0}}{\mu} \;+\; \frac{\delta_{0}\zeta}{\delta{\bf B}}\,\pd{f_{0}}{\zeta} \right) \nonumber \\
 & = & \frac{\delta{\bf B}}{B}\bdot\left[ \left( p_{\bot}\wh{\bot}\;\partial_{\|}f_{0} \;-\; p_{\|}\,\vb{\partial}_{\bot}f_{0} \right) \;-\;
\bhat\;\mu\,\pd{f_{0}}{\mu} \right] \;\equiv\; \delta{\bf B}\bdot\partial_{\bf B}^{(0)}f_{0},
\nonumber
\end{eqnarray}
where we used Eqs.~\eqref{eq:delta_p||}-\eqref{eq:delta_zeta}. By inserting Eq.~\eqref{eq:partialB_def} into Eq.~\eqref{eq:delta_T0}, the push-forward 
${\sf T}_{0}^{-1}(\delta f)$ is now expressed in terms of $\delta f_{0}$ and $\delta{\bf B}$ as
\begin{equation}
{\sf T}_{0}^{-1}(\delta f) \;=\; \delta f_{0} \;+\; \delta{\bf B}\bdot\partial_{\bf B}^{(0)}f_{0},
\label{eq:delta_T0_final}
\end{equation}
which agrees with Eqs.~(12) and (32) of \cite{pjmVG13}.

Another application of the push-forward relation \eqref{eq:delta_T0} involves replacing the operator $\delta$ with $\partial/\partial t$ in order to derive an expression for the local partial time derivative $\partial_{0}/\partial t$ used in Eq.~\eqref{eq:Vlasov_0}:
\begin{equation}
\frac{\partial_{0}f_{0}}{\partial t} \;\equiv\; {\sf T}_{0}^{-1}\left(\pd{f}{t}\right) \;\equiv\; \pd{f_{0}}{t} \;+\; \left( \left[{\sf T}_{0}^{-1},\; 
\pd{}{t}\right]\,{\sf T}_{0}\right) f_{0} \;\equiv\; \pd{f_{0}}{t} \;+\; \pd{\bf B}{t}\bdot\partial_{\bf B}^{(0)}f_{0}.
\label{eq:time_T0}
\end{equation}
Returning to the local Vlasov equation \eqref{eq:Vlasov_0}, we therefore find
\[ \pd{f_{0}}{t} \;+\; \pd{\bf B}{t}\bdot\partial_{\bf B}^{(0)}f_{0} \;=\; -\,\dot{\bf x}_{0}\bdot\nabla f_{0} \;-\; \dot{p}_{\|0}\;\pd{f_{0}}{p_{\|}} \;-\; \dot{\mu}_{0}\;\pd{f_{0}}{\mu} \;-\; \dot{\zeta}_{0}\;\pd{f_{0}}{\zeta}. \]
Next, we use the identity
\begin{equation}
\pd{\bf B}{t}\bdot\partial_{\bf B}^{(0)}f_{0} \;=\; \frac{\partial_{0}p_{\|}}{\partial t}\;\pd{f_{0}}{p_{\|}} \;+\; \frac{\partial_{0}\mu}{\partial t}\;
\pd{f_{0}}{\mu} \;+\; \frac{\partial_{0}\zeta}{\partial t}\;\pd{f_{0}}{\zeta} \;\equiv\; \frac{\partial_{0}z_{0}^{\alpha}}{\partial t}\;
\pd{f_{0}}{z_{0}^{\alpha}},
\label{eq:B_dot_z0}
\end{equation}
and define the total local time derivatives
\begin{equation}
\frac{d_{0}z_{0}^{\alpha}}{dt} \;\equiv\; \frac{\partial_{0}z_{0}^{\alpha}}{\partial t} \;+\; \dot{z}_{0}^{\alpha} \;\equiv\; 
\frac{\partial_{0}z_{0}^{\alpha}}{\partial t} \;+\; \left\{ z_{0}^{\alpha},\frac{}{} K_{0}\right\}_{0} \;+\; e\,{\bf E}\bdot\left\{ {\bf x},\frac{}{} 
z_{0}^{\alpha}\right\}_{0},
\end{equation}
with $\partial{\bf x}/\partial t \equiv 0$, so that the local Vlasov equation becomes
\begin{equation}
\pd{f_{0}}{t} \;=\; -\;\frac{d_{0}z_{0}^{\alpha}}{dt}\;\pd{f_{0}}{z_{0}^{\alpha}},
\label{eq:local_Vlasov}
\end{equation}
which can also be expressed in divergence form as follows.

If we now multiply the local Vlasov equation \eqref{eq:Vlasov_0} by the local Jacobian ${\cal J}_{0}$ and, use the local Liouville theorem 
\eqref{eq:local_Liouville}, we obtain
\begin{equation}
{\cal J}_{0}\;\pd{f_{0}}{t} \;+\; {\cal J}_{0}\;\pd{\bf B}{t}\bdot\partial_{\bf B}^{(0)}f_{0} \;=\; -\;\pd{}{z_{0}^{\alpha}}\left( {\cal J}_{0}\,
\dot{z}_{0}^{\alpha}\frac{}{}f_{0} \right).
\label{eq:local_Vlasov_J}
\end{equation}
Next, using Eq.~\eqref{eq:time_Jac_0}, we rewrite Eq.~\eqref{eq:B_dot_z0} as
\[ {\cal J}_{0}\;\pd{\bf B}{t}\bdot\partial_{\bf B}^{(0)}f_{0} \;\equiv\; \pd{{\cal J}_{0}}{t}\;f_{0} + \pd{}{p_{\|}}\left( {\cal J}_{0}\,
\frac{\partial_{0}p_{\|}}{\partial t}\;f_{0}\right) + \pd{}{\mu}\left( {\cal J}_{0}\,\frac{\partial_{0}\mu}{\partial t}\;f_{0}\right) + 
\pd{}{\zeta}\left( {\cal J}_{0}\,\frac{\partial_{0}\zeta}{\partial t}\;f_{0}\right), \]
and the local Vlasov equation \eqref{eq:local_Vlasov} can now be expressed in local phase-space divergence form
\begin{equation}
\pd{({\cal J}_{0}f_{0})}{t} \;=\; -\;\pd{}{z_{0}^{\alpha}}\left( \frac{d_{0}z_{0}^{\alpha}}{dt}\;{\cal J}_{0}\,f_{0} \right).
\label{eq:local_Vlasov-div}
\end{equation}

\subsection{Local Functionals}

Our next step is now to define a transformation from functionals ${\cal F}[f,{\bf E},{\bf B}]$ to functionals ${\cal F}_{0}[f_{0},{\bf E},{\bf B}]$ based on the scalar-covariance property: ${\cal F}[f,{\bf E},{\bf B}] = {\cal F}_{0}[f_{0},{\bf E},{\bf B}]$. Using 
Eqs.~\eqref{eq:ovF_del_f}-\eqref{eq:ovF_del_B}, we find
\begin{eqnarray}
{\sf T}_{0}^{-1}\left[\frac{1}{\cal J}\;\fd{{\cal F}_{0}[{\sf T}_{0}^{-1}f,{\bf E},{\bf B}]}{f({\bf z})} \right] & \equiv & \frac{1}{{\cal J}_{0}}\;
\fd{{\cal F}_{0}}{f_{0}({\bf z}_{0})}, 
\label{eq:deltaF0_f0} \\
\fd{{\cal F}_{0}[{\sf T}_{0}^{-1}f,{\bf E},{\bf B}]}{{\bf E}({\bf r})} & \equiv & \fd{{\cal F}_{0}}{{\bf E}({\bf r})}, \label{eq:deltaF0_E} \\
\fd{{\cal F}_{0}[{\sf T}_{0}^{-1}f,{\bf E},{\bf B}]}{{\bf B}({\bf r})} & \equiv & \fd{{\cal F}_{0}}{{\bf B}({\bf r})} \;-\; \int_{{\bf z}_{0}}\;
\delta^{3}({\bf x} - {\bf r}) \;\fd{{\cal F}_{0}}{f_{0}({\bf z}_{0})}\;\partial_{\bf B}^{(0)}f_{0} \nonumber \\
 & \equiv & \fd{{\cal F}_{0}}{{\bf B}({\bf r})} \;-\; \Delta_{\bf B}^{(0)}{\cal F}_{0}({\bf r}).
\label{eq:deltaF0_B}
\end{eqnarray}
The more general case of a local phase-space transformation that depends on ${\bf E}$ and ${\bf B}$ would introduce an additional term 
$\Delta_{\bf E}^{(0)}{\cal F}_{0}({\bf r})$ on the right side of Eq.~\eqref{eq:deltaF0_E} but would leave Eq.~\eqref{eq:deltaF0_f0} invariant.

As a specific application of the local functional variations \eqref{eq:deltaF0_f0}-\eqref{eq:deltaF0_B}, we consider the local Hamiltonian functional
\begin{equation} 
{\cal H}_{0} \;=\; \int_{\bf r} \left( \frac{|{\bf E}|^{2}}{8\pi} +  \frac{|{\bf B}|^{2}}{8\pi} \right) \;+\; \int_{{\bf z}_{0}}\,{\cal J}_{0}\,f_{0}
\;K_{0}.
\label{eq:local_Ham_0}
\end{equation}
The local functional variations of Eq.~\eqref{eq:local_Ham_0} are
\begin{eqnarray} 
\fd{{\cal H}_{0}}{{\bf B}({\bf r})} & = & \frac{{\bf B}({\bf r})}{4\pi} \;+\; \int_{{\bf z}_{0}}\,\delta^{3}({\bf x} - {\bf r})\;f_{0}\;
\fd{({\cal J}_{0}\,K_{0})}{\bf B} \nonumber \\
 & = & \frac{{\bf B}({\bf r})}{4\pi} \;+\; \int_{{\bf z}_{0}}\,\delta^{3}({\bf x} - {\bf r})\;{\cal J}_{0}f_{0} 
\left[\frac{\bhat}{B}\frac{}{}\left(K_{0} \;+\; \mu\,B \right) \right],
\label{eq:Ham_var_0}
\end{eqnarray}
with $\delta{\cal H}_{0}/\delta f_{0} = {\cal J}_{0}\,K_{0}$ and $\delta{\cal H}_{0}/\delta{\bf E} = {\bf E}/4\pi = \delta{\cal H}_{0}[{\sf T}_{0}^{-1}f,
{\bf E},{\bf B}]/\delta{\bf E}$. On the other hand, Eq.~\eqref{eq:deltaF0_B} yields
\begin{equation} 
\fd{{\cal H}_{0}[{\sf T}_{0}^{-1}f,{\bf E},{\bf B}]}{{\bf B}({\bf r})} \;=\; \fd{{\cal H}_{0}}{{\bf B}({\bf r})} \;-\; 
\Delta_{\bf B}^{(0)}{\cal H}_{0}({\bf r}),
\label{eq:THam_var_0_1}
\end{equation}
and thus Eqs.~\eqref{eq:Ham_var_0}-\eqref{eq:THam_var_0_1} yield 
\begin{equation} 
\fd{{\cal H}_{0}[{\sf T}_{0}^{-1}f,{\bf E},{\bf B}]}{{\bf B}({\bf r})} \;=\; \fd{{\cal H}_{0}}{{\bf B}({\bf r})} \;-\; 
\Delta_{\bf B}^{(0)}{\cal H}_{0}({\bf r}) \;\equiv\; \frac{{\bf B}}{4\pi}.
\label{eq:THam_var_0}
\end{equation}

\subsection{Local Vlasov-Maxwell bracket}

Now that we have calculated the local functional variations \eqref{eq:deltaF0_f0}-\eqref{eq:deltaF0_B}, we can transform the Vlasov-Maxwell bracket
\eqref{eq:MV_bracket} and obtain the local Vlasov-Maxwell bracket
\begin{eqnarray}
\left[{\cal F}_{0},\frac{}{}{\cal G}_{0}\right]_{0} & \equiv & \mathbb{T}_{0}\left(\left[\mathbb{T}_{0}^{-1}{\cal F}_{0},\frac{}{}
\mathbb{T}_{0}^{-1}{\cal G}_{0}\right]\right) \label{eq:MV_bracket_0} \\
 & = & \int_{{\bf z}_{0}}\,{\cal J}_{0}\;f_{0} \left\{ \frac{1}{{\cal J}_{0}}\,
\fd{{\cal F}_{0}}{f_{0}}, \frac{1}{{\cal J}_{0}}\,\fd{{\cal G}_{0}}{f_{0}} \right\}_{0} \nonumber \\
 &  &+ 4\pi\,c\int_{\bf r}\left[ \fd{{\cal F}_{0}}{{\bf E}}\bdot\nabla\btimes\left( \fd{{\cal G}_{0}}{{\bf B}} - \Delta_{\bf B}^{(0)}{\cal G}_{0}\right) \;-\; \fd{{\cal G}_{0}}{{\bf E}}\bdot\nabla\btimes\left( \fd{{\cal F}_{0}}{{\bf B}} - \Delta_{\bf B}^{(0)}{\cal F}_{0}\right)\right] \nonumber \\
 &  &-\;4\pi\,e\;\int_{{\bf z}_{0}}\,{\cal J}_{0}\;f_{0} \left( \fd{{\cal F}_{0}}{{\bf E}}\bdot\left\{ {\bf x},\; \frac{1}{{\cal J}_{0}}\,
\fd{{\cal G}_{0}}{f_{0}} \right\}_{0} \;-\; \fd{{\cal G}_{0}}{{\bf E}}\bdot\left\{ {\bf x},\; \frac{1}{{\cal J}_{0}}\,\fd{{\cal F}_{0}}{f_{0}} 
\right\}_{0} \right).
\nonumber
\end{eqnarray}
We note that this form is generic to all phase-space transformations that depend on the magnetic field only. In addition, because the local Poisson bracket $\{\;\,\;\}_{0}$ also depends on magnetic-field gradients (e.g., $\nabla\btimes\bhat$ in ${\bf B}_{0}^{*}$), the Jacobi identity 
\eqref{eq:MV_Jacobi} for the local bracket $[\;,\;]_{0}$ might be difficult to prove explicitly. We will show below that the local Jacobi identity will be greatly simplified by defining operators on functionals.

We now show that the local Vlasov-Maxwell equations \eqref{eq:Vlasov_0}-\eqref{eq:Bdot_0} can be formulated in Hamiltonian form as
\begin{equation}
\pd{{\cal F}_{0}}{t} \;=\; \left[{\cal F}_{0},\frac{}{} {\cal H}_{0}\right]_{0} \;\equiv\; \int_{{\bf z}_{0}} \fd{{\cal F}_{0}}{f_{0}}\;\pd{f_{0}}{t}
\;+\; \int_{\bf r} \left( \fd{{\cal F}_{0}}{\bf E}\bdot\pd{\bf E}{t} \;+\; \fd{{\cal F}_{0}}{\bf B}\bdot\pd{\bf B}{t} \right).
\label{eq:F0_dot_def}
\end{equation}
First, we calculate the local bracket with the local Hamiltonian functional \eqref{eq:local_Ham_0}:
\begin{eqnarray*} 
\left[{\cal F}_{0},\frac{}{}{\cal H}_{0}\right]_{0} & = & \int_{{\bf z}_{0}}\,{\cal J}_{0}\;f_{0} \left\{ \frac{1}{{\cal J}_{0}}\,
\fd{{\cal F}_{0}}{f_{0}},\; K_{0} \right\}_{0} \\
 &  &+\; 4\pi\,c\;\int_{\bf r}\;\left[ \fd{{\cal F}_{0}}{{\bf E}({\bf r})}\bdot\nabla\btimes\frac{{\bf B}({\bf r})}{4\pi} \;-\; 
\frac{{\bf E}({\bf r})}{4\pi}\bdot\nabla\btimes\left( \fd{{\cal F}_{0}}{{\bf B}({\bf r})} \;-\; \Delta_{\bf B}^{(0)}{\cal F}_{0}({\bf r}) \right)\right] \\
 &  &-\;4\pi\,e\;\int_{{\bf z}_{0}}\,{\cal J}_{0}\;f_{0} \left( \fd{{\cal F}_{0}}{{\bf E}({\bf x})}\bdot\left\{ {\bf x},\; K_{0} \right\}_{0} \;-\; 
\frac{{\bf E}({\bf x})}{4\pi}\bdot\left\{ {\bf x},\; \frac{1}{{\cal J}_{0}}\,\fd{{\cal F}_{0}}{f_{0}} \right\}_{0} \right),
\end{eqnarray*}
where the term $\Delta_{\bf B}^{(0)}{\cal H}_{0}$ is cancelled by $\delta{\cal H}_{0}/\delta{\bf B}$ [see Eq.~\eqref{eq:Ham_var_0}], with 
$\delta{\cal H}_{0}/\delta{\bf B} - \Delta_{\bf B}^{(0)}{\cal H}_{0} \equiv {\bf B}/4\pi$.

Next, after integrating by parts, we obtain
\begin{eqnarray}
\left[{\cal F}_{0},\frac{}{}{\cal H}_{0}\right]_{0} & = & -\int_{{\bf z}_{0}}\fd{{\cal F}_{0}}{f_{0}({\bf z}_{0})} \left( \{ f_{0},\; K_{0}\}_{0} 
\;+\frac{}{} e{\bf E}({\bf x})\bdot\{ {\bf x},\; f_{0} \}_{0} \;-\; c\,\nabla\btimes{\bf E}({\bf x})\bdot\partial_{\bf B}^{(0)}f_{0} \right) \nonumber \\
 &  &+\; \int_{\bf r} \left[ \fd{{\cal F}_{0}}{{\bf E}}\bdot\left(c\,\nabla\btimes{\bf B} \;-\; 4\pi\,{\bf J}_{0}\right) \;+\; 
\fd{{\cal F}_{0}}{{\bf B}}\bdot\left(-\frac{}{}c\,\nabla\btimes{\bf E}\right) \right],
\label{eq:MV_bracket_Ham_0}
\end{eqnarray}
from which we recover Eq.~\eqref{eq:F0_dot_def} when the local Vlasov-Maxwell equations \eqref{eq:Vlasov_0}-\eqref{eq:Bdot_0} are substituted. Here, we used the relation
\[ -\,c\,\nabla\btimes{\bf E}({\bf x})\bdot\partial_{\bf B}^{(0)}f_{0} \;=\; \pd{\bf B}{t}\bdot\partial_{\bf B}^{(0)}f_{0} \;\equiv\; 
\frac{\partial_{0}z_{0}^{\alpha}}{\partial t}\;\pd{f_{0}}{z_{0}^{\alpha}} \]
to recover the local Vlasov equation \eqref{eq:local_Vlasov}.

\subsection{\label{subsec:alternate_local}Alternate local Hamiltonian formulation}

It is sometimes convenient to formulate the local Vlasov-Maxwell equations in terms of the local Vlasov distribution $F_{0} \equiv 
{\cal J}_{0}\,f_{0}$, defined as the component of the Vlasov (differential) six-form $F_{0}\,d^{6}z_{0}$ (where $d^{6}z_{0}$ excludes the local Jacobian ${\cal J}_{0} \equiv m\,B$). In this case, the local Vlasov equation \eqref{eq:local_Vlasov-div} becomes
\begin{equation}
\pd{F_{0}}{t} \;=\; -\;\pd{}{z_{0}^{\alpha}}\left( \frac{d_{0}z_{0}^{\alpha}}{dt}\;F_{0} \right),
\label{eq:local_Vlasov_div}
\end{equation}
while the Maxwell equation \eqref{eq:Edot_0} becomes
\begin{equation}
\pd{\bf E}{t} \;=\; c\,\nabla\btimes{\bf B} \;-\; 4\pi\,e\int_{{\bf z}_{0}} F_{0}\;\dot{\bf x}_{0}.
\label{eq:Edot_0_J}
\end{equation}
Next, under this field transformation, the functional derivatives of ${\cal F}_{0}^{\prime}[F_{0},{\bf E},{\bf B}] \equiv {\cal F}_{0}[f_{0},{\bf E},
{\bf B}]$ become
\begin{equation}
\fd{{\cal F}_{0}}{f_{0}} \;=\; {\cal J}_{0}\;\fd{{\cal F}_{0}^{\prime}}{F_{0}} \;\;\;{\rm and}\;\;\; \fd{{\cal F}_{0}}{\bf B} \;=\; 
\fd{{\cal F}_{0}^{\prime}}{\bf B} \;+\; \frac{\bhat}{B}\;F_{0}\; \fd{{\cal F}_{0}^{\prime}}{F_{0}},
\label{eq:func_var_local_J}
\end{equation}
so that the local Vlasov-Maxwell bracket \eqref{eq:MV_bracket_0} becomes
\begin{eqnarray}
\left[{\cal F}_{0}^{\prime},\frac{}{}{\cal G}_{0}^{\prime}\right]_{0}^{\prime} & \equiv & \int_{{\bf z}_{0}}\,F_{0} \left\{ 
\fd{{\cal F}_{0}^{\prime}}{F_{0}}, \fd{{\cal G}_{0}^{\prime}}{F_{0}} \right\}_{0} \nonumber \\
 &  &+ 4\pi\,c\int_{\bf r}\left[ \fd{{\cal F}_{0}^{\prime}}{{\bf E}}\bdot\nabla\btimes\left( \fd{{\cal G}_{0}^{\prime}}{{\bf B}} - 
\Delta_{\bf B}^{\prime(0)}{\cal G}_{0}^{\prime}\right) - \fd{{\cal G}_{0}^{\prime}}{{\bf E}}\bdot\nabla\btimes\left( \fd{{\cal F}_{0}^{\prime}}{{\bf B}} - \Delta_{\bf B}^{\prime(0)}{\cal F}_{0}^{\prime}\right)\right] \nonumber \\
 &  &-\;4\pi\,e\;\int_{{\bf z}_{0}}\,F_{0} \left( \fd{{\cal F}_{0}^{\prime}}{{\bf E}}\bdot\left\{ {\bf x},\; \fd{{\cal G}_{0}^{\prime}}{F_{0}} 
\right\}_{0} \;-\; \fd{{\cal G}_{0}^{\prime}}{{\bf E}}\bdot\left\{ {\bf x},\; \fd{{\cal F}_{0}^{\prime}}{F_{0}} 
\right\}_{0} \right).
\label{eq:MV_bracket_0_J}
\end{eqnarray}
where
\begin{eqnarray}
\Delta_{\bf B}^{\prime(0)}{\cal F}_{0}^{\prime} & \equiv & \int_{{\bf z}_{0}}\delta^{3}({\bf x} - {\bf r})\; \fd{{\cal F}_{0}^{\prime}}{F_{0}}\;
\partial_{\bf B}^{\prime(0)}F_{0} \;\equiv\; \int_{{\bf z}_{0}}\delta^{3}({\bf x} - {\bf r})\; \fd{{\cal F}_{0}^{\prime}}{F_{0}}\;\left(
\partial_{\bf B}^{(0)}F_{0} \;-\; F_{0}\,\frac{\bhat}{B} \right) \nonumber \\
 & \equiv & \int_{{\bf z}_{0}}\delta^{3}({\bf x} - {\bf r})\; \fd{{\cal F}_{0}^{\prime}}{F_{0}}\;\left[ \pd{}{z_{0}^{\alpha}}\left(
\frac{\delta_{0}z_{0}^{\alpha}}{\delta{\bf B}}\;F_{0}\right) \right]. 
\end{eqnarray}
By considering the local Hamiltonian functional
\begin{equation}
{\cal H}_{0}^{\prime} \;=\; \int_{\bf r} \left( \frac{|{\bf E}|^{2}}{8\pi} +  \frac{|{\bf B}|^{2}}{8\pi} \right) \;+\; \int_{{\bf z}_{0}}\,F_{0}\;K_{0},
\label{eq:local_Ham_0_J}
\end{equation}
we obtain the functional derivative
\[ \fd{{\cal H}_{0}^{\prime}}{{\bf B}({\bf r})} \;=\; \frac{{\bf B}({\bf r})}{4\pi} \;+\; \int_{{\bf z}_{0}}\,\delta^{3}({\bf x} - {\bf r})\;F_{0}\;
\mu\,\bhat \;\equiv\; \frac{{\bf B}({\bf r})}{4\pi} \;+\; \Delta_{\bf B}^{\prime(0)}{\cal H}_{0}^{\prime}, \]
where
\begin{eqnarray}
\Delta_{\bf B}^{\prime(0)}{\cal H}_{0}^{\prime} & = & -\; \int_{{\bf z}_{0}}\delta^{3}({\bf x} - {\bf r})\; F_{0}\left(
\frac{\delta_{0}z_{0}^{\alpha}}{\delta{\bf B}}\;\pd{K_{0}}{z_{0}^{\alpha}}\right) \nonumber \\
 & \equiv & \int_{{\bf z}_{0}}\delta^{3}({\bf x} - {\bf r})\; F_{0}\;\fd{K_{0}}{\bf B} \;=\; \int_{{\bf z}_{0}}\delta^{3}({\bf x} - {\bf r})\; F_{0}\;
\mu\,\bhat.
\label{eq:Delta0_H_B}
\end{eqnarray}
Hence, we also find
\[ \fd{{\cal H}_{0}^{\prime}}{{\bf B}({\bf r})} \;-\; \Delta_{\bf B}^{\prime(0)}{\cal H}_{0}^{\prime}({\bf r}) \;\equiv\; \frac{{\bf B}({\bf r})}{4\pi}. \]
It is interesting to note that Eq.~\eqref{eq:Delta0_H_B} exactly yields the lowest-order guiding-center magnetization 
\begin{equation}
{\bf M}_{\rm gc} \;\equiv\; -\;\Delta_{\bf B}^{\prime(0)}{\cal H}_{0}^{\prime} \;=\; -\;\int\; F_{0}\;\mu\,\bhat\;dp_{\|}\,d\mu\,d\zeta.
\label{eq:M_gc}
\end{equation}
This is a special case of the general form for magnetizations and  polarizations arising as  functional derivatives of a term of the Hamiltonian (here $\int_{{\bf z}_{0}}F_{0} K_{0}$) that  appeared in \cite{pjm13}, which is a generalization of the expressions of \cite{pfirsch84} and \cite{pjmP85} for guiding-center kinetic theories. For recent work, see \cite{Brizard_2013} and \cite{Brizard_Tronci_2016}.

Lastly, we show that the local Vlasov-Maxwell equations \eqref{eq:local_Vlasov_div}-\eqref{eq:Edot_0_J}, as well as Eq.~\eqref{eq:B_dot}, possess a Hamiltonian structure in terms of the local Vlasov-Maxwell bracket \eqref{eq:MV_bracket_0_J} and the local Hamiltonian functional \eqref{eq:local_Ham_0_J}:
\begin{eqnarray}
\left[{\cal F}_{0}^{\prime},\frac{}{}{\cal H}_{0}^{\prime}\right]_{0}^{\prime} & = & -\int_{{\bf z}_{0}}\fd{{\cal F}_{0}^{\prime}}{F_{0}({\bf z}_{0})} \left[ \pd{}{z_{0}^{\alpha}}\left(F_{0}\frac{}{} \dot{z}_{0}^{\alpha}\right) \;-\; c\,\nabla\btimes{\bf E}({\bf x})\bdot\partial_{\bf B}^{\prime(0)}F_{0} \right] \nonumber \\
 &  &+\; \int_{\bf r} \left[ \fd{{\cal F}_{0}^{\prime}}{{\bf E}}\bdot\left(c\,\nabla\btimes{\bf B} \;-\frac{}{} 4\pi\,{\bf J}_{0}\right) \;+\; 
\fd{{\cal F}_{0}^{\prime}}{{\bf B}}\bdot\left(-\frac{}{}c\,\nabla\btimes{\bf E}\right) \right],
\label{eq:MV_bracket_Ham_0_J}
\end{eqnarray}
where the local Vlasov equation \eqref{eq:local_Vlasov_div} is recovered through the identities
\begin{eqnarray*} 
 &  &\int_{{\bf z}_{0}} F_{0} \left( \left\{ \fd{{\cal F}_{0}^{\prime}}{F_{0}}, K_{0}\right\}_{0} \;+\; e\,{\bf E}\bdot\left\{ {\bf x},\; 
\fd{{\cal F}_{0}^{\prime}}{F_{0}}\right\}_{0} \right) \\
 & = & -\;\int_{{\bf z}_{0}}{\cal J}_{0}\,\fd{{\cal F}_{0}^{\prime}}{F_{0}} \left( 
\left\{ F_{0},\frac{}{} K_{0}\right\}_{0} \;+\; e\,{\bf E}\bdot\left\{ {\bf x},\frac{}{} F_{0}\right\}_{0} \right) \equiv -\;\int_{{\bf z}_{0}}
\fd{{\cal F}_{0}^{\prime}}{F_{0}}\;\pd{}{z_{0}^{\alpha}}\left( F_{0}\frac{}{} \dot{z}_{0}^{\alpha}\right),
\end{eqnarray*}
and
\[ -\,c\,\nabla\btimes{\bf E}({\bf x})\bdot\partial_{\bf B}^{\prime(0)}F_{0} \;=\; \pd{\bf B}{t}\bdot\partial_{\bf B}^{\prime(0)}F_{0} \;\equiv\; 
\pd{}{z_{0}^{\alpha}}\left(\frac{\partial_{0}z_{0}^{\alpha}}{\partial t}\;F_{0}\right). \]

\section{\label{sec:Dyn_Red}Dynamical Reduction by Phase-space Transformation}

We have seen in  Sec.~\ref{sec:local} that, while the local Vlasov-Maxwell equations \eqref{eq:Vlasov_0}-\eqref{eq:Bdot_0} possess a Hamiltonian structure defined in terms of the local Hamiltonian functional \eqref{eq:local_Ham_0} and the local Vlasov-Maxwell bracket \eqref{eq:MV_bracket_0}, the local Vlasov function $f_{0}$ evolves rapidly on a short gyromotion time scale that needs to be removed for practical (e.g., numerical) applications. The asymptotic elimination of the gyromotion time scale from the local Vlasov-Maxwell equations proceeds with the guiding-center transformation $({\bf x},p_{\|},\mu,\zeta) \rightarrow (\ov{\bf X},\ov{p}_{\|},\ov{\mu},\ov{\zeta})$, where the guiding-center Hamiltonian dynamics is now decoupled from the gyromotion time scale \citep{Littlejohn_1983, Cary_Brizard_2009}. 

The guiding-center phase-space transformation, however, is a non-local transformation involving displacements in local particle phase space that are expressed as asymptotic expansions in powers of an ordering parameter $\epsilon \ll 1$. It is precisely this nonlocal feature of the guiding-center transformation that allows the explicit introduction of guiding-center polarization and magnetization effects in guiding-center Vlasov-Maxwell theory \citep{Brizard_2013, Tronko_Brizard_2015}. Since the guiding-center transformation becomes the identity transformation in the limit $\epsilon \rightarrow 0$, we now consider the transformation of the local Vlasov-Maxwell Hamiltonian structure under the action of a near-identity phase-space transformation.

\subsection{Near-identity phase-space transformation}

The dynamical reduction of the Vlasov-Maxwell equations \eqref{eq:Vlasov_eq}-\eqref{eq:div_B} is carried out by a near-identity phase-space transformation ${\cal T}^{\epsilon}: {\bf z}_{0} \rightarrow \ov{\bf Z} = {\cal T}^{\epsilon}{\bf z}_{0}$, where $\epsilon \ll 1$ denotes an ordering parameter associated with the dynamical reduction. The ordering parameter $\epsilon \equiv \omega\,\tau$ is often chosen by comparing a short orbital time scale $\tau$ with a long dynamical time scale $\omega^{-1} \gg \tau$ of interest. 

Each reduced phase-space coordinate $\ov{Z}^{\alpha}$ is expressed as an asymptotic expansion in powers of $\epsilon$ involving components of the generating vector fields $({\sf G}_{1}, {\sf G}_{2}, \cdots)$ on particle phase space:
\begin{equation}
\ov{Z}^{\alpha}({\bf z}_{0}; \epsilon) \;=\; z_{0}^{\alpha} \;+\; \epsilon\;G_{1}^{\alpha}({\bf z}_{0}) \;+\; \epsilon^{2} \left( 
G_{2}^{\alpha}({\bf z}_{0}) \;+\; \frac{1}{2}\; G_{1}^{\beta}({\bf z}_{0})\;\pd{G_{1}^{\alpha}({\bf z}_{0})}{z^{\beta}} \right) \;+\; \cdots,
\label{eq:z_ovz}
\end{equation}
where $\ov{Z}^{\alpha}({\bf z}_{0}; \epsilon = 0) = z_{0}^{\alpha}$ (i.e., to lowest order, the reduced phase-space coordinates are local). We note that this transformation is invertible (since $\epsilon \ll 1$), i.e., ${\cal T}^{-\epsilon}: \ov{\bf Z} \rightarrow {\bf z}_{0} = {\cal T}^{-\epsilon}\ov{\bf Z}$, where each particle coordinate $z_{0}^{\alpha}$ is expressed as an asymptotic expansion in terms of the same generating vector fields $({\sf G}_{1}, {\sf G}_{2}, \cdots)$ on reduced phase space:
\begin{equation}
z_{0}^{\alpha}(\ov{\bf Z}; \epsilon) \;=\; \ov{Z}^{\alpha} \;-\; \epsilon\;G_{1}^{\alpha}(\ov{\bf Z}) \;-\; \epsilon^{2} \left( G_{2}^{\alpha}(\ov{\bf Z}) 
\;-\; \frac{1}{2}\; G_{1}^{\beta}(\ov{\bf Z})\;\pd{G_{1}^{\alpha}(\ov{\bf Z})}{\ov{Z}^{\beta}} \right) \;+\; \cdots.
\label{eq:ovz_z}
\end{equation}
The reduced Jacobian of the transformation \eqref{eq:z_ovz} is constructed from the local Jacobian ${\cal J}_{0} = m\,B$ and the generating vector fields $({\sf G}_{1}, {\sf G}_{2}, \cdots)$ as
\begin{equation}
\ov{\cal J} \equiv {\cal J}_{0} - \pd{}{z_{0}^{\alpha}}\left[ {\cal J}_{0}\frac{}{} \left(\epsilon\,G_{1}^{\alpha} \;+\frac{}{} \epsilon^{2}\,
G_{2}^{\alpha} + \cdots\right) \;-\; \frac{\epsilon^{2}}{2}\;G_{1}^{\alpha}\;\pd{}{z_{0}^{\beta}}\left({\cal J}_{0}\frac{}{} G_{1}^{\beta} + 
\cdots\right) + \cdots \right].
\label{eq:ov_Jac}
\end{equation}
We note that, while the local particle phase-space coordinates ${\bf z}_{0}$ are independent of the electric field ${\bf E}$, the reduced phase-space coordinates $\ov{\bf Z}({\bf z}_{0}; {\bf E}, {\bf B})$, and the reduced Jacobian \eqref{eq:ov_Jac}, generally depend on the electromagnetic fields (but not the local Vlasov distribution $f_{0}$).

\subsection{Push-forward and Pull-back Operators}

The reduction phase-space transformations \eqref{eq:z_ovz}-\eqref{eq:ovz_z} induce transformations in function space \citep{Littlejohn_1982}, where the push-forward operator
\begin{equation}
{\sf T}^{-\epsilon}:\; f_{0} \;\rightarrow\; \ov{f} \;=\; {\sf T}^{-\epsilon}\,f_{0} \;\equiv\; f_{0}\;\circ\;{\cal T}^{-\epsilon}
\label{eq:push}
\end{equation}
transforms a function $f_{0}$ on local particle phase space into a function $\ov{f}$ on reduced phase space, while the pull-back operator
\begin{equation}
{\sf T}^{\epsilon}:\; \ov{f} \;\rightarrow\; f_{0} \;=\; {\sf T}^{\epsilon}\ov{f} \;\equiv\; \ov{f}\;\circ\;{\cal T}^{\epsilon}
\label{eq:pull}
\end{equation}
transforms a function $\ov{f}$ on reduced phase space into a function $f_{0}$ on local particle phase space. These induced transformations satisfy the scalar-covariance property 
\begin{equation}
\left. \begin{array}{rcl}
f_{0}({\bf z}_{0}) & = & {\sf T}^{\epsilon}\ov{f}({\bf z}_{0}) \;=\; \ov{f}({\cal T}^{\epsilon}{\bf z}_{0}) \;=\; \ov{f}(\ov{\bf Z}) \\
 &  & \\
\ov{f}(\ov{\bf Z}) & = & {\sf T}^{-\epsilon}f_{0}(\ov{\bf Z}) \;=\; f_{0}({\cal T}^{-\epsilon}\ov{\bf Z}) \;=\; f_{0}({\bf z}_{0})
\end{array} \right\}.
\label{eq:scalar_inv}
\end{equation}
Moreover, the action of the push-forward operator on the phase-space infinitesimal volume yields the Jacobian transformation
\begin{equation}
{\sf T}^{-\epsilon}({\cal J}_{0}\,d^{6}z_{0}) \;\equiv\; \ov{\cal J}\,d^{6}\ov{Z}, 
\label{eq:Jac_push}
\end{equation}
from which we obtain the Jacobian expansion \eqref{eq:ov_Jac}. It is also useful to express the reduced Jacobian \eqref{eq:ov_Jac} as
\begin{equation}
\ov{\cal J} \;\equiv\; \left({\sf T}^{-\epsilon}{\cal J}_{0}\right)\;\left[ 1 - \epsilon\,\exd\cdot{\sf G}_{1} - \epsilon^{2} \exd\cdot\left( 
{\sf G}_{2} - \frac{1}{2}\,{\sf G}_{1}\cdot\exd{\sf G}_{1}\right) + \cdots \right] \;\equiv\; \left({\sf T}^{-\epsilon}{\cal J}_{0}\right)\;
\ov{\cal J}_{\epsilon},
\label{eq:ovJac_def}
\end{equation}
where $\ov{\cal J}_{\epsilon}$ is defined by the identity ${\sf T}^{-\epsilon}(d^{6}z_{0}) \equiv \ov{\cal J}_{\epsilon}\,d^{6}\ov{Z}$.

The reduced Poisson bracket is defined in terms of the push-forward and pull-back operators:
\begin{eqnarray}
\left\{ \ov{f},\frac{}{} \ov{g}\right\}_{\epsilon} & = & {\sf T}^{-\epsilon}\left(\left\{f,\frac{}{} g\right\}_{0}\right) \;=\; 
{\sf T}^{-\epsilon}\left[\left\{{\sf T}^{\epsilon}\;\left({\sf T}^{-\epsilon}f\right),\frac{}{} {\sf T}^{\epsilon}\;\left({\sf T}^{-\epsilon}g\right)\right\}_{0}\right] \;=\; {\sf T}^{-\epsilon}\left( \left\{ {\sf T}^{\epsilon}\ov{f},\frac{}{} {\sf T}^{\epsilon}\ov{g}\right\}_{0} \right) \nonumber \\
 & \equiv & \pd{\ov{f}}{\ov{Z}^{\alpha}}\;\ov{J}^{\alpha\beta}\;\pd{\ov{g}}{\ov{Z}^{\beta}} \;\equiv\; \frac{1}{\ov{\cal J}}\,
\pd{}{\ov{Z}^{\alpha}}\left(\ov{\cal J}\;\ov{f}\;\left\{ \ov{Z}^{\alpha},\frac{}{} \ov{g}\right\}_{\epsilon}\right),
\label{eq:PB_red}
\end{eqnarray}
where the phase-space divergence form makes use of the reduced Liouville property
\begin{equation}
\frac{1}{\ov{\cal J}}\,\pd{}{\ov{Z}^{\alpha}}\left(\ov{\cal J}\frac{}{} \ov{J}^{\alpha\beta} \right) \;\equiv\; 0.
\label{eq:red_Liouville}
\end{equation}
Because of the divergence form \eqref{eq:PB_red} of the reduced Poisson bracket, we have the reduced phase-space integral identity:
\[ \int_{\ov{\bf Z}}\,\ov{\cal J}\; \ov{h}\;\left\{ \ov{f},\frac{}{} \ov{g}\right\}_{\epsilon} \;\equiv\; -\;\int_{\ov{\bf Z}}\,\ov{\cal J}\; \ov{g}\;\left\{ \ov{f},\frac{}{} \ov{h}\right\}_{\epsilon}, \]
which is valid for any three reduced phase-space functions $(\ov{f},\ov{g},\ov{h})$. 

Lastly, since the reduced Poisson matrix (with components $\ov{J}^{\alpha\beta} \equiv \{ \ov{Z}^{\alpha},\;\ov{Z}^{\beta}\}_{\epsilon}$) is defined as the inverse of the reduced Lagrange matrix (with components $\ov{\omega}_{\alpha\beta}$ defined as the components of an exact two-form $\ov{\vb{\omega}} = {\sf T}^{-\epsilon}\vb{\omega}_{0}$ in reduced phase space), the Jacobi property of the reduced Poisson bracket \eqref{eq:PB_red} is guaranteed by the identity $\exd\ov{\vb{\omega}} \equiv 0$, i.e., the exterior derivative of an exact two-form $\ov{\vb{\omega}} \equiv \exd\ov{\Gamma}$ is automatically zero. A direct proof of the Jacobi identity for the reduced Poisson bracket \eqref{eq:PB_red} follows from the push-forward transformation
\begin{eqnarray}
{\sf T}^{-\epsilon}\left\{ f,\frac{}{} \{ g,\;h\}_{0} \right\}_{0} & = & {\sf T}^{-\epsilon}\left\{ {\sf T}^{\epsilon}({\sf T}^{-\epsilon}f),\frac{}{} 
{\sf T}^{\epsilon}\left[ {\sf T}^{-\epsilon}\{ {\sf T}^{\epsilon}({\sf T}^{-\epsilon}g),\frac{}{} {\sf T}^{\epsilon}({\sf T}^{-\epsilon}h)\}_{0} \right]\right\}_{0} \nonumber \\
 & = & \left\{ \left({\sf T}^{-\epsilon}f\right),\frac{}{} \left\{ \left({\sf T}^{-\epsilon}g\right),\; \left({\sf T}^{-\epsilon}h\right)\right\}_{\epsilon}\;\right\}_{\epsilon} \;\equiv\; \left\{ \ov{f},\frac{}{} \left\{ \ov{g},\; \ov{h}\right\}_{\epsilon}\;\right\}_{\epsilon},
\label{eq:push_PBPB}
\end{eqnarray}
where the definition \eqref{eq:PB_red} for the reduced Poisson bracket has been used twice. Hence, since the local Poisson bracket $\{\;,\;\}_{0}$ satisfies the Jacobi identity, then so does the reduced Poisson bracket \eqref{eq:PB_red}.

\subsection{Reduced phase-space displacements}

We now define two complementary phase-space displacements in terms of push-forward and pull-back operators as follows. The local particle phase-space displacement $\Delta^{\alpha}$ is defined in terms of the pull-back operator as
\begin{equation}
\Delta^{\alpha}({\bf z}_{0}) \;\equiv\; z_{0}^{\alpha} \;-\; {\sf T}^{\epsilon}\ov{Z}^{\alpha} \;=\; -\;\epsilon\;G_{1}^{\alpha}({\bf z}_{0}) \;-\; 
\epsilon^{2} \left( G_{2}^{\alpha}({\bf z}_{0}) \;+\; \frac{1}{2}\; G_{1}^{\beta}({\bf z}_{0})\;\pd{G_{1}^{\alpha}({\bf z}_{0})}{z_{0}^{\beta}} \right) 
\;+\; \cdots,
\label{eq:Delta}
\end{equation}
while the reduced phase-space displacement $\ov{\Delta}^{\alpha}$ is defined in terms of the push-forward operator as
\begin{equation}
\ov{\Delta}^{\alpha}(\ov{\bf Z}) \;\equiv\; {\sf T}^{-\epsilon}z_{0}^{\alpha} \;-\; \ov{Z}^{\alpha} \;=\; -\; \epsilon\;G_{1}^{\alpha}(\ov{\bf Z}) 
\;-\; \epsilon^{2} \left( G_{2}^{\alpha}(\ov{\bf Z}) \;-\; \frac{1}{2}\; G_{1}^{\beta}(\ov{\bf Z})\;\pd{G_{1}^{\alpha}(\ov{\bf Z})}{\ov{Z}^{\beta}} \right)
\;+\; \cdots.
\label{eq:Delta_bar}
\end{equation}
These two definitions are related by push-forward or pull-back transformations $\ov{\Delta}^{\alpha} \equiv {\sf T}^{-\epsilon}\Delta^{\alpha}$ 
and $\Delta^{\alpha} \equiv {\sf T}^{\epsilon}\ov{\Delta}^{\alpha}$.

Lastly, we note that the Jacobian $\ov{\cal J}_{\epsilon}$ defined in Eq.~\eqref{eq:ovJac_def} can be expressed as
\begin{equation}
\ov{\cal J}_{\epsilon} \;\equiv\; 1 \;+\; \exd\cdot\ov{\vb{\Delta}} \;=\; 1 \;+\; \pd{\ov{\Delta}^{\alpha}}{\ov{Z}^{\alpha}} \;\equiv\; 
\pd{({\sf T}^{-\epsilon}z_{0}^{\alpha})}{\ov{Z}^{\alpha}},
\label{eq:Jac_epsilon}
\end{equation}
and the reduced Jacobian \eqref{eq:ovJac_def} becomes $\ov{\cal J} = ({\sf T}^{-\epsilon}{\cal J}_{0})\;(1 + \exd\cdot\ov{\vb{\Delta}})$.

\subsection{Reduced partial time derivative}

Since the phase-space transformations considered here are time-dependent transformations (which nonetheless preserve time), the operation of time differentiation does not commute with the push-forward and pull-back operators \eqref{eq:push}-\eqref{eq:pull}. Hence, we define the reduced partial time derivative
\begin{equation}
\frac{\partial_{\epsilon}}{\partial t} \;\equiv\; {\sf T}^{-\epsilon}\left(\frac{\partial_{0}}{\partial t}\;{\sf T}^{\epsilon}\right) \;\equiv\; 
\pd{}{t} \;+\; \frac{\partial_{\epsilon}\ov{Z}^{\alpha}}{\partial t}\;\pd{}{\ov{Z}^{\alpha}},
\label{eq:time_epsilon}
\end{equation}
where
\begin{equation}
\frac{\partial_{\epsilon}\ov{Z}^{\alpha}}{\partial t} \;\equiv\; {\sf T}^{-\epsilon}\left[
\frac{\partial_{0}({\sf T}^{\epsilon}\ov{Z}^{\alpha})}{\partial t}\right] \;=\; {\sf T}^{-\epsilon}\left( \frac{\partial_{0}z_{0}^{\alpha}}{\partial t}
\;-\; \frac{\partial_{0}\Delta^{\alpha}}{\partial t}\right),
\end{equation}
and the partial-time derivative of the reduced Jacobian \eqref{eq:ovJac_def} is expressed as
\begin{equation}
\pd{\ov{\cal J}}{t} \;=\; -\;\pd{}{\ov{Z}^{\alpha}}\left( \ov{\cal J}\;\frac{\partial_{\epsilon}\ov{Z}^{\alpha}}{\partial t} \right),
\label{eq:ov_Jacobian_dot}
\end{equation}
which follows from Eq.~\eqref{eq:time_ovJac}.

\subsection{Reduced Hamiltonian dynamics}

Next, we transform the noncanonical Hamilton equations \eqref{eq:dotz_alpha} to obtain
\begin{eqnarray}
\frac{d_{\epsilon}\ov{Z}^{\alpha}}{dt} & \equiv & \frac{\partial_{\epsilon}\ov{Z}^{\alpha}}{\partial t} \;+\; \left\{ \ov{Z}^{\alpha},\frac{}{} \ov{K}\right\}_{\epsilon} \;+\; e\,{\sf T}^{-\epsilon}{\bf E}\bdot\left\{ {\sf T}^{-\epsilon}{\bf x},\; \ov{Z}^{\alpha} \right\}_{\epsilon}
\label{eq:red_Ham_eq}
\end{eqnarray}
where the push-forward of the kinetic energy is
\begin{equation}
\ov{K} \;\equiv\; {\sf T}^{-\epsilon}K \;=\; K \;-\; \epsilon\,G_{1}^{\alpha}\,\pd{K}{z^{\alpha}} \;-\; \epsilon^{2} \left[ G_{2}^{\alpha}\,
\pd{K}{z^{\alpha}} \;-\; \frac{1}{2}\,G_{1}^{\beta}\,\pd{}{z^{\beta}}\left(G_{1}^{\alpha}\,\pd{K}{z^{\alpha}}\right) \right] \;+\; \cdots.
\label{eq:ovH_Lie}
\end{equation}
The reduced Hamilton equations \eqref{eq:red_Ham_eq} satisfy the reduced Liouville theorem
\begin{equation}
\pd{\ov{\cal J}}{t} \;+\; \pd{}{\ov{Z}^{\alpha}}\left(\ov{\cal J}\;\frac{d_{\epsilon}\ov{Z}^{\alpha}}{dt} \right) \;=\; \pd{}{\ov{Z}^{\alpha}}\left[
\ov{\cal J}\;\left(\frac{d_{\epsilon}\ov{Z}^{\alpha}}{dt} \;-\; \frac{\partial_{\epsilon}\ov{Z}^{\alpha}}{\partial t}\right) \right] \;\equiv\; 0,
\label{eq:Liouville}
\end{equation}
which follows from Eq.~\eqref{eq:red_Ham_eq} and the reduced Liouville property \eqref{eq:red_Liouville}.

\section{\label{sec:Red_VM}Reduced Vlasov-Maxwell Equations}

The phase-space transformations \eqref{eq:z_ovz}-\eqref{eq:ovz_z}, and their associated induced operators and meta-operators, allow us to derive the set of reduced Vlasov-Maxwell equations \citep{Brizard_2008}. Examples of such reduced Vlasov-Maxwell equations include the nonlinear gyrokinetic Vlasov-Maxwell equations \citep{Brizard_Hahm_2007, Krommes_2012}, which yield important numerical advantages \citep{Garbet_2010} in the computer simulations of the turbulent evolution of magnetized fusion plasmas.

\subsection{Reduced Vlasov Equation}

The reduced Vlasov equation is defined as the push-forward of the Vlasov equation \eqref{eq:Vlasov_Ham}
\begin{eqnarray}
0 & = & {\sf T}^{-\epsilon}\left( \frac{\partial_{0}f_{0}}{\partial t} \;+\; \{ f_{0},\; K_{0}\}_{0} \;+\; e\,{\bf E}\bdot\{{\bf x},\; f_{0}\}_{0} 
\right) \nonumber \\
 & \equiv & \frac{\partial_{\epsilon}\ov{f}}{\partial t} \;+\; \left\{ \ov{f},\frac{}{} \ov{K}\right\}_{\epsilon} \;+\; e\,{\sf T}^{-\epsilon}{\bf E}\bdot\left\{ {\sf T}^{-\epsilon}{\bf x},\frac{}{} \ov{f}\right\}_{\epsilon},
\label{eq:Vlasov_push}
\end{eqnarray}
which, using Eq.~\eqref{eq:time_epsilon}, yields
\begin{equation}
\pd{\ov{f}}{t} \;=\; -\;\frac{d_{\epsilon}\ov{Z}^{\alpha}}{dt}\;\pd{\ov{f}}{\ov{Z}^{\alpha}} \;\equiv\; -\;\left(
\frac{\partial_{\epsilon}\ov{Z}^{\alpha}}{\partial t} \;+\; \left\{ \ov{Z}^{\alpha},\frac{}{} \ov{K}\right\}_{\epsilon} \;+\; 
e\,{\sf T}^{-\epsilon}{\bf E}\bdot\left\{ {\sf T}^{-\epsilon}{\bf x},\; \ov{Z}^{\alpha} \right\}_{\epsilon}\right)\pd{\ov{f}}{\ov{Z}^{\alpha}},
\label{eq:Vlasov_red}
\end{equation}
where the reduced Vlasov distribution $\ov{f} \equiv {\sf T}^{-\epsilon}f_{0}$ is defined as the push-forward of the local particle Vlasov distribution 
$f_{0}$. The reduced Vlasov equation \eqref{eq:Vlasov_red} can also be written in divergence form as
\begin{equation}
0 \;=\; \pd{(\ov{\cal J}\,\ov{f})}{t} \;+\; \pd{}{\ov{Z}^{\alpha}}\left(\ov{\cal J}\,\ov{f}\;\frac{d_{\epsilon}\ov{Z}^{\alpha}}{dt} \right),
\label{eq:red_Vlasov_div}
\end{equation}
where we used the reduced Liouville theorem \eqref{eq:Liouville}.

\subsection{Reduced Maxwell Equations}

The reduced Maxwell equations are obtained from Eqs.~\eqref{eq:Edot_0} and \eqref{eq:divE_0} by transforming the local current and charge densities by meta-push-forward operation:
\begin{eqnarray}
\pd{{\bf E}({\bf r})}{t} \;-\; c\;\nabla\btimes{\bf B}({\bf r}) & = & -\; 4\pi\,e\;\int_{\ov{\bf Z}}\,\ov{\cal J}(\ov{\bf Z})\;\ov{f}(\ov{\bf Z})\;
\delta^{3}(\ov{\bf X} + \ov{\vb{\rho}}_{\epsilon} - {\bf r})\;\left( \frac{d_{\epsilon}\ov{\bf X}}{dt} \;+\; \frac{d_{\epsilon}
\ov{\vb{\rho}}_{\epsilon}}{dt} \right) \nonumber \\
 & \equiv & -\,4\pi\;\mathbb{T}^{\epsilon}{\bf J}_{0}({\bf r}), 
\label{eq:E_dot_red} \\
\nabla\bdot{\bf E}({\bf r}) & = & 4\pi\,e\;\int_{\ov{\bf Z}}\,\ov{\cal J}(\ov{\bf Z})\;\ov{f}(\ov{\bf Z})\;\delta^{3}(\ov{\bf X} + \ov{\vb{\rho}}_{\epsilon} - {\bf r}) \;\equiv\; 4\pi\;\mathbb{T}^{\epsilon}\varrho_{0}({\bf r}),
\label{eq:div_E_red}
\end{eqnarray}
while the remaining Maxwell equations \eqref{eq:B_dot} and \eqref{eq:div_B} are unaffected by the dynamical reduction. In Eq.~\eqref{eq:E_dot_red}, we have introduced the push-forward of the particle velocity ${\bf v} = d{\bf x}/dt$:
\begin{equation}
{\sf T}^{-\epsilon}\left(\frac{d_{0}{\bf x}}{dt}\right) \;=\; {\sf T}^{-\epsilon}\left[\frac{d_{0}}{dt}\;{\sf T}^{\epsilon}\left(\ov{\bf X} \;+\frac{}{}
\ov{\vb{\rho}}_{\epsilon}\right)\right] \;\equiv\; \frac{d_{\epsilon}\ov{\bf X}}{dt} \;+\; \frac{d_{\epsilon}\ov{\vb{\rho}}_{\epsilon}}{dt} \;\equiv\;
\left\{ \ov{\bf X} + \ov{\vb{\rho}}_{\epsilon},\frac{}{} \ov{K}\right\}_{\epsilon}.
\label{eq:velocity_push}
\end{equation}
In Eqs.~\eqref{eq:E_dot_red}-\eqref{eq:velocity_push}, the reduced spatial displacement 
\begin{equation}
\ov{\vb{\rho}}_{\epsilon} \;\equiv\; {\sf T}^{-\epsilon}{\bf x} \;-\; \ov{\bf X}
\label{eq:rho_def} 
\end{equation}
plays an important part in the definition of the reduced polarization and magnetization \citep{Brizard_2008} in the reduced Maxwell equations \eqref{eq:E_dot_red}-\eqref{eq:div_E_red}. The meta-push-forward of the (local) charge density in Eq.~\eqref{eq:div_E_red} yields the transformation
\begin{equation}
\mathbb{T}^{\epsilon}\varrho_{0} \;\equiv\; \varrho_{\epsilon} \;-\; \nabla\bdot{\bf P}_{\epsilon},
\label{eq:meta_push_rho}
\end{equation}
where $\varrho_{\epsilon}$ denotes the reduced charge density, defined as
\begin{equation}
\varrho_{\epsilon}({\bf r}) \;\equiv\; e\;\int_{\ov{\bf Z}}\delta^{3}(\ov{\bf X} - {\bf r})\;\ov{\cal J}(\ov{\bf Z})\;\ov{f}(\ov{\bf Z}),
\label{eq:red_charge}
\end{equation} 
and the reduced polarization charge density $\varrho_{\rm pol} \equiv -\nabla\bdot{\bf P}_{\epsilon}$ is expressed in terms of the reduced polarization 
${\bf P}_{\epsilon}$, defined as \citep{Brizard_2008, Brizard_2009}
\begin{equation}
{\bf P}_{\epsilon}({\bf r}) \;\equiv\; e\;\int_{\ov{\bf Z}}\delta^{3}(\ov{\bf X} - {\bf r})\;\ov{\vb{\rho}}_{\epsilon}\;\ov{\cal J}(\ov{\bf Z})\;
\ov{f}(\ov{\bf Z}) \;+\; \cdots. 
\label{eq:red_pol}
\end{equation}
The meta-push-forward of the (local) current density in Eq.~\eqref{eq:E_dot_red} yields the transformation
\begin{equation}
\mathbb{T}^{\epsilon}{\bf J}_{0} \;\equiv\; {\bf J}_{\epsilon} \;+\; \pd{{\bf P}_{\epsilon}}{t} \;+\; c\;\nabla\btimes{\bf M}_{\epsilon},
\label{eq:meta_push_J}
\end{equation}
where ${\bf J}_{\epsilon}$ denotes the reduced current density, defined as
\begin{equation}
{\bf J}_{\epsilon}({\bf r}) \;\equiv\; e\;\int_{\ov{\bf Z}}\delta^{3}(\ov{\bf X} - {\bf r})\;\frac{d_{\epsilon}\ov{\bf X}}{dt}\;\ov{\cal J}(\ov{\bf Z})\;\ov{f}(\ov{\bf Z}),
\label{eq:red_current}
\end{equation}  
${\bf J}_{\rm pol} \equiv \partial{\bf P}_{\epsilon}/\partial t$ denotes the reduced polarization current density, and the reduced magnetization current density ${\cal J}_{\rm mag} \equiv c\;\nabla\btimes{\bf M}_{\epsilon}$ is expressed in terms of the reduced magnetization ${\bf M}_{\epsilon}$, defined as \citep{Brizard_2008, Brizard_2009}
\begin{equation}
{\bf M}_{\epsilon}({\bf r}) \;\equiv\; \frac{e}{c}\;\int_{\ov{\bf Z}}\delta^{3}(\ov{\bf X} - {\bf r})\;\ov{\vb{\rho}}_{\epsilon}\btimes\left(
\frac{1}{2}\,\frac{d_{\epsilon}\ov{\vb{\rho}}_{\epsilon}}{dt} + \frac{d_{\epsilon}\ov{\bf X}}{dt}\right)\;\ov{\cal J}(\ov{\bf Z})\;\ov{f}(\ov{\bf Z})
\;+\; \cdots.
\label{eq:red_mag}
\end{equation} 
Hence, the reduced Maxwell equations \eqref{eq:E_dot_red}-\eqref{eq:div_E_red} may also be written as
\begin{eqnarray}
\pd{{\bf D}_{\epsilon}}{t} \;-\; c\,\nabla\btimes{\bf H}_{\epsilon} & = & -\;4\pi\;{\bf J}_{\epsilon}, \label{eq:D_dot} \\
\nabla\bdot{\bf D}_{\epsilon} & = & 4\pi\;\varrho_{\epsilon}, \label{eq:div_D}
\end{eqnarray}
where the reduced electromagnetic fields are ${\bf D}_{\epsilon} \equiv {\bf E} + 4\pi\,{\bf P}_{\epsilon}$ and ${\bf H}_{\epsilon} \equiv {\bf B} -
4\pi\,{\bf M}_{\epsilon}$. The remaining two Maxwell equations \eqref{eq:B_dot} and \eqref{eq:div_B} are unaffected by the phase-space transformation.

\subsection{Reduced Hamiltonian Formulation}

Before we discuss the lifting transformation of the local Vlasov-Maxwell bracket \eqref{eq:MV_bracket_0} by Lie-transform method, we introduce the reduced Hamiltonian functional obtained by the meta-push-forward transformation of the local Hamiltonian functional \eqref{eq:local_Ham_0}:
\begin{equation}
\ov{\cal H}[\ov{f},{\bf E},{\bf B}] \;\equiv\; \mathbb{T}^{\epsilon}{\cal H}_{0} \;=\; \frac{1}{8\pi}\;\int_{\bf r}\;\left( |{\bf E}({\bf r})|^{2} 
\;+\frac{}{} |{\bf B}({\bf r})|^{2} \right) \;+\; \int_{\ov{\bf Z}}\,\ov{\cal J}(\ov{\bf Z})\;\ov{f}(\ov{\bf Z})\;\ov{K}(\ov{\bf Z};\;{\bf E},{\bf B}).
\label{eq:Ham_red}
\end{equation}
The reduced Hamiltonian functional \eqref{eq:Ham_red} can be derived by Noether method from the Lagrangian formulation of gyrokinetic Vlasov-Maxwell theory
\citep{Brizard_2000_prl, Brizard_2000_pop}, where $\ov{K} \equiv \ov{H} - e\,{\sf T}^{-\epsilon}\Phi$ is expressed in terms of the gyrocenter Hamiltonian $\ov{H}$ and the push-forward of the electric potential $\Phi$.

The time derivative of the reduced Hamiltonian \eqref{eq:Ham_red} yields
\begin{eqnarray}
\pd{\ov{\cal H}}{t} & = & \int_{\bf r}\;\left( \frac{\bf E}{4\pi}\bdot\pd{\bf E}{t} \;+\; \frac{\bf B}{4\pi}\bdot\pd{\bf B}{t}\right) \;+\; 
\int_{\ov{{\bf Z}}}\left( \pd{(\ov{\cal J}\ov{f})}{t}\;\ov{K} \;+\; \ov{\cal J}\,\ov{f}\;\pd{\ov{K}}{t}\right) \\
 & = & \int_{\ov{\bf Z}}\,\ov{\cal J}\; \ov{f}\;\left[\frac{d_{\epsilon}\ov{K}}{dt} \;-\; e\,{\sf T}^{-\epsilon}\left({\bf E}\bdot\frac{d_{0}{\bf x}}{dt}\right)\right] \nonumber \\
 & = & \int_{\ov{\bf Z}}\,\ov{\cal J}\; \ov{f}\;{\sf T}^{-\epsilon}\left[{\sf T}_{0}^{-1}\left(\frac{dK}{dt} \;-\; e\,{\bf E}\bdot{\bf v} \right)\right] \;=\; 0, \nonumber
\end{eqnarray}
where Eqs.~\eqref{eq:H_dot} and \eqref{eq:red_Vlasov_div} were used and exact divergences (in space and phase space) vanish upon integration. In Sec.~\ref{sec:Lift_VM}, we will construct a reduced Vlasov-Maxwell bracket $[\;,\;]_{\epsilon}$ in terms of which the reduced Vlasov-Maxwell equations
\eqref{eq:Vlasov_red} and \eqref{eq:E_dot_red}-\eqref{eq:div_E_red} (with Eqs.~\eqref{eq:B_dot} and \eqref{eq:div_B}) are expressed in Hamiltonian form as
\begin{equation}
\pd{\ov{\cal F}}{t} \;=\; \left[\ov{\cal F},\frac{}{} \ov{\cal H}\right]_{\epsilon},
\label{eq:ovF_dot}
\end{equation}
with the reduced Hamiltonian functional $\ov{\cal H}$ given by Eq.~\eqref{eq:Ham_red}.

\section{\label{sec:Lift_VM}Lie-transform Lift of the Vlasov-Maxwell Bracket}

In complete analogy with the definition of the reduced Poisson bracket \eqref{eq:PB_red}, we now define the reduced Vlasov-Maxwell bracket 
$[\;,\;]_{\epsilon}$: 
\begin{equation}
\left[\ov{\cal F},\frac{}{} \ov{\cal G}\right]_{\epsilon} \;\equiv\; \mathbb{T}^{\epsilon}\left(\left[\left(\mathbb{T}^{-\epsilon}\ov{\cal F}\right),
\frac{}{}\left(\mathbb{T}^{-\epsilon}\ov{\cal G}\right)\right]_{0}\right), 
\label{eq:meta_MV}
\end{equation}
where $\ov{\cal F}$ and $\ov{\cal G}$ are two arbitrary functionals depending on the reduced Vlasov distribution $\ov{f}$ and the electromagnetic fields
$({\bf E},{\bf B})$. The reduced bracket \eqref{eq:meta_MV} satisfies the standard antisymmetry and Leibniz properties. By following the same procedure leading to the bracket transformation \eqref{eq:push_PBPB}, we obtain the functional-bracket transformation
\begin{eqnarray}
\mathbb{T}^{\epsilon}\left(\left[ {\cal F}_{0},\frac{}{} [{\cal G}_{0},\; {\cal H}_{0}]_{0} \right]_{0}\right) & = & \mathbb{T}^{\epsilon}\left(
\left[ \mathbb{T}^{-\epsilon}(\mathbb{T}^{\epsilon}{\cal F}_{0}),\frac{}{} \mathbb{T}^{-\epsilon}\left( \mathbb{T}^{\epsilon}[ 
\mathbb{T}^{-\epsilon}(\mathbb{T}^{\epsilon}{\cal G}_{0}),\frac{}{} \mathbb{T}^{-\epsilon}(\mathbb{T}^{\epsilon}{\cal H}_{0})]_{0} \right)
\right]_{0}\right) \nonumber \\
 & = & \left[ \mathbb{T}^{\epsilon}{\cal F}_{0},\frac{}{} \left[ \mathbb{T}^{\epsilon}{\cal G}_{0},\; \mathbb{T}^{\epsilon}{\cal H}_{0}
\right]_{\epsilon}\;\right]_{\epsilon} \;\equiv\; \left[ \ov{\cal F},\frac{}{} \left[\, \ov{\cal G},\; \ov{\cal H}\right]_{\epsilon}\;\right]_{\epsilon},
\label{eq:metapush_MVMV}
\end{eqnarray} 
which shows that since the Vlasov-Maxwell bracket $[\;,\;]$ and the local Vlasov-Maxwell bracket $[\;,\;]_{0}$ satisfy the Jacobi identity (see  App.~\ref{sec:Jacobi_MV}), then so does the reduced Vlasov-Maxwell bracket $[\;,\;]_{\epsilon}$:
\begin{equation}
\left[ \ov{\cal F},\frac{}{} [\ov{\cal G},\; \ov{\cal K}]_{\epsilon}\right]_{\epsilon} \;+\; \left[ \ov{\cal G},\frac{}{} [\ov{\cal K},\; 
\ov{\cal F}]_{\epsilon}\right]_{\epsilon} \;+\; \left[ \ov{\cal K},\frac{}{} [\ov{\cal F},\; \ov{\cal G}]_{\epsilon}\right]_{\epsilon} \;=\; 0.
\label{eq:MV_Jacobi_red}
\end{equation}

\subsection{Reduced Vlasov-Maxwell Bracket}

In what follows, we will now combine the local phase-space transformation ${\cal T}_{0}$ and the near-identity phase-space transformation 
${\cal T}^{\epsilon}$ into a single phase-space transformation ${\cal T}_{\epsilon} \equiv {\cal T}^{\epsilon}\circ{\cal T}_{0}$. Hence, we introduce the operators ${\sf T}_{\epsilon} \equiv {\sf T}_{0}\,{\sf T}^{\epsilon}$ and ${\sf T}_{\epsilon}^{-1} \equiv {\sf T}^{-\epsilon}\,{\sf T}_{0}^{-1}$ as well as the meta-operators $\mathbb{T}_{\epsilon} \equiv \mathbb{T}^{\epsilon}\,\mathbb{T}_{0}$ and $\mathbb{T}_{\epsilon}^{-1} \equiv \mathbb{T}_{0}^{-1}\,
\mathbb{T}^{-\epsilon}$.

The reduced Vlasov-Maxwell bracket \eqref{eq:MV_bracket_transform} now becomes
\begin{eqnarray}
\left[\ov{\cal F},\frac{}{} \ov{\cal G}\right]_{\epsilon} & \equiv & \mathbb{T}_{\epsilon}\left(\left[\mathbb{T}_{\epsilon}^{-1}\ov{\cal F},\frac{}{}
\mathbb{T}_{\epsilon}^{-1}\ov{\cal G}\right]\right)  \label{eq:MV_red} \\
 & = & \int_{\ov{\bf Z}}\,\ov{\cal J}\;\ov{f}\;\left\{ \frac{1}{\ov{\cal J}}\;\fd{\ov{\cal F}}{\ov{f}},\; \frac{1}{\ov{\cal J}}\;\fd{\ov{\cal G}}{\ov{f}} \right\}_{\epsilon} \nonumber \\
 &  &+\; 4\pi\,c\; \int_{\bf r}\;\left[ \fd{(\mathbb{T}_{\epsilon}^{-1}\ov{\cal F})}{{\bf E}({\bf r})}\bdot\nabla\btimes\fd{(\mathbb{T}_{\epsilon}^{-1}
\ov{\cal G})}{{\bf B}({\bf r})} \;-\; \fd{(\mathbb{T}_{\epsilon}^{-1}\ov{\cal G})}{{\bf E}({\bf r})}\bdot\nabla\btimes\fd{(\mathbb{T}_{\epsilon}^{-1}
\ov{\cal F})}{{\bf B}({\bf r})} \right] \nonumber \\
 &  &-\;4\pi\,e\;\int_{\bf r}\;\fd{(\mathbb{T}_{\epsilon}^{-1}\ov{\cal F})}{{\bf E}({\bf r})}\bdot \left[\int_{\ov{\bf Z}}\,\ov{\cal J}\;\ov{f}\;  
\delta^{3}(\ov{\bf X} + \ov{\vb{\rho}}_{\epsilon} - {\bf r})\;\left\{ \ov{\bf X} + \ov{\vb{\rho}}_{\epsilon},\; 
\frac{1}{\ov{\cal J}}\;\fd{\ov{\cal G}}{\ov{f}} \right\}_{\epsilon} \right] \nonumber \\ 
 &  &+\;4\pi\,e\;\int_{\bf r}\;\fd{(\mathbb{T}_{\epsilon}^{-1}\ov{\cal G})}{{\bf E}({\bf r})}\bdot \left[\int_{\ov{\bf Z}}\,\ov{\cal J}\;\ov{f}\;  
\delta^{3}(\ov{\bf X} + \ov{\vb{\rho}}_{\epsilon} - {\bf r})\;\left\{ \ov{\bf X} + \ov{\vb{\rho}}_{\epsilon},\; 
\frac{1}{\ov{\cal J}}\;\fd{\ov{\cal F}}{\ov{f}} \right\}_{\epsilon} \right],
\nonumber
\end{eqnarray}
where the reduced functional derivatives are defined as
\begin{eqnarray}
\fd{\left(\mathbb{T}_{\epsilon}^{-1}\ov{\cal F}\right)}{{\bf E}({\bf r})} & = & \fd{\ov{\cal F}}{{\bf E}({\bf r})} \;-\; \int_{\ov{\bf Z}}\;
{\sf T}^{-1}_{\epsilon}\left[\fd{({\sf T}_{\epsilon}\ov{Z}^{\alpha})}{{\bf E}({\bf r})}\right]\;\pd{\ov{f}}{\ov{Z}^{\alpha}}\;
\fd{\ov{\cal F}}{\ov{f}(\ov{\bf Z})} \nonumber \\
 & \equiv & \fd{\ov{\cal F}}{{\bf E}({\bf r})} \;-\; \Delta_{\bf E}^{(\epsilon)}\ov{\cal F}({\bf r}),
\label{eq:fd_ovE} \\
\fd{\left(\mathbb{T}_{\epsilon}^{-1}\ov{\cal F}\right)}{{\bf B}({\bf r})} & = & \fd{\ov{\cal F}}{{\bf B}({\bf r})} \;-\; \int_{\ov{\bf Z}}\;
{\sf T}^{-1}_{\epsilon}\left[\fd{({\sf T}_{\epsilon}\ov{Z}^{\alpha})}{{\bf B}({\bf r})}\right]\;\pd{\ov{f}}{\ov{Z}^{\alpha}}\;
\fd{\ov{\cal F}}{\ov{f}(\ov{\bf Z})} \nonumber \\
 & \equiv & \fd{\ov{\cal F}}{{\bf B}({\bf r})} \;-\; \Delta_{\bf B}^{(\epsilon)}\ov{\cal F}({\bf r}),
\label{eq:fd_ovB}
\end{eqnarray}
with
\begin{equation}
{\sf T}_{\epsilon}^{-1}\left(\fd{(\mathbb{T}_{\epsilon}^{-1}\ov{\cal F})}{{\bf E}({\bf x})}\right) \;\equiv\; \int_{\bf r}\;\delta^{3}(\ov{\bf X} + 
\ov{\vb{\rho}}_{\epsilon} - {\bf r})\;\fd{\left(\mathbb{T}_{\epsilon}^{-1}\ov{\cal F}\right)}{{\bf E}({\bf r})}.
\label{eq:fd_ovE_x}
\end{equation}
We will now show that the reduced Vlasov-Maxwell bracket \eqref{eq:MV_red} can be used to derive the reduced Vlasov equation
\eqref{eq:Vlasov_red} and the reduced Maxwell equations \eqref{eq:B_dot} and \eqref{eq:E_dot_red}.

\subsection{Hamiltonian Formulation of the Reduced Vlasov-Maxwell Equations}

By inserting the reduced Hamiltonian functional \eqref{eq:Ham_red} in the reduced Vlasov-Maxwell bracket \eqref{eq:MV_red}, and integrating by parts, we obtain
\begin{eqnarray}
\left[\ov{\cal F},\frac{}{} \ov{\cal H}\right]_{\epsilon}  & = & -\int_{\ov{\bf Z}}\;\fd{\ov{\cal F}}{\ov{f}} \left[ \left\{ \ov{f},
\frac{}{} \ov{K} \right\}_{\epsilon} + 4\pi\,e\left(\int_{\bf r}\;\delta^{3}(\ov{\bf X} + \ov{\vb{\rho}}_{\epsilon} - {\bf r})\;
\left\{ \ov{\bf X} + \ov{\vb{\rho}}_{\epsilon},\frac{}{} \ov{f} \right\}_{\epsilon}\bdot\fd{\left(\mathbb{T}_{\epsilon}^{-1}
\ov{\cal H}\right)}{{\bf E}({\bf r})}\right) \right] \nonumber \\
 &  &+\; 4\pi\;\int_{\bf r} \fd{(\mathbb{T}_{\epsilon}^{-1}\ov{\cal F})}{{\bf E}({\bf r})}\bdot \left[ c\;\nabla\btimes\left(
\fd{\left(\mathbb{T}_{\epsilon}^{-1}\ov{\cal H}\right)}{{\bf B}({\bf r})}\right) \;-\; \mathbb{T}_{\epsilon}{\bf J}({\bf r}) \right] \nonumber \\
 &  &- 4\pi c\;\int_{\bf r} \fd{(\mathbb{T}_{\epsilon}^{-1}\ov{\cal F})}{{\bf B}({\bf r})}\bdot \nabla\btimes\left(
\fd{\left(\mathbb{T}_{\epsilon}^{-1}\ov{\cal H}\right)}{{\bf E}({\bf r})}\right),
\label{eq:MV_red_Ham}
\end{eqnarray}
where the meta push-forward $\mathbb{T}_{\epsilon}{\bf J} \equiv {\bf J}_{\epsilon} + \partial{\bf P}_{\epsilon}/\partial t + c\,\nabla\btimes
{\bf M}_{\epsilon}$ of the particle-current density is given by Eq.~\eqref{eq:E_dot_red}, with $\delta\ov{\cal H}/\delta\ov{f} = \ov{\cal J}\,\ov{K}$ and $\{\ov{\bf X} + \ov{\vb{\rho}}_{\epsilon},\; \ov{K}\}_{\epsilon} \equiv d_{\epsilon}\ov{\bf X}/dt + d_{\epsilon}\ov{\vb{\rho}}_{\epsilon}/dt$. In addition, the functional derivatives of $\mathbb{T}_{\epsilon}^{-1}\ov{\cal H}$ are 
\begin{eqnarray}
\fd{\left(\mathbb{T}_{\epsilon}^{-1}\ov{\cal H}\right)}{{\bf E}({\bf r})} & = & \left[ \frac{{\bf E}({\bf r})}{4\pi} \;+\; \int_{\ov{\bf Z}}\;
{\sf T}_{\epsilon}^{-1}\left(\fd{({\sf T}_{\epsilon}\ov{Z}^{\alpha})}{{\bf E}({\bf r})}\right)\;\pd{\ov{f}}{\ov{Z}^{\alpha}}\;\ov{\cal J}\,\ov{K} 
\right] - \Delta_{\bf E}^{(\epsilon)}\ov{\cal H} \;\equiv\; \frac{{\bf E}({\bf r})}{4\pi},
\label{eq:fd_ovHam_ovE} \\
\fd{\left(\mathbb{T}_{\epsilon}^{-1}\ov{\cal H}\right)}{{\bf B}({\bf r})} & = & \left[ \frac{{\bf B}({\bf r})}{4\pi} \;+\; \int_{\ov{\bf Z}}\;
{\sf T}_{\epsilon}^{-1}\left(\fd{({\sf T}_{\epsilon}\ov{Z}^{\alpha})}{{\bf B}({\bf r})}\right)\;\pd{\ov{f}}{\ov{Z}^{\alpha}}\;\ov{\cal J}\,\ov{K} \right] - \Delta_{\bf B}^{(\epsilon)}\ov{\cal H} \;\equiv\; \frac{{\bf B}({\bf r})}{4\pi},
\label{eq:fd_ovHam_ovB}
\end{eqnarray}
where we used the functional derivatives \eqref{eq:ovH_varE_C}-\eqref{eq:ovH_varB_C} of the reduced Hamiltonian functional $\ov{\cal H}$, with
\begin{equation}
{\sf T}_{\epsilon}^{-1}\left(\fd{(\mathbb{T}_{\epsilon}^{-1}\ov{\cal H})}{{\bf E}({\bf x})}\right) \;=\; \frac{1}{4\pi}\;{\sf T}_{\epsilon}^{-1}{\bf E}.
\end{equation}

By combining these expressions, we find
\begin{eqnarray}
\left[\ov{\cal F},\frac{}{} \ov{\cal H}\right]_{\epsilon}  & = & -\;\int_{\ov{\bf Z}}\;\fd{\ov{\cal F}}{\ov{f}} \left[ \left\{ \ov{f},
\frac{}{} \ov{K} \right\}_{\epsilon} \;+\frac{}{} e\;{\sf T}_{\epsilon}^{-1}{\bf E}\bdot\left\{ \ov{\bf X} + \ov{\vb{\rho}}_{\epsilon},\frac{}{} \ov{f} 
\right\}_{\epsilon} \right] \nonumber \\
 & + & \int_{\bf r} \left[ \left(\fd{\ov{\cal F}}{{\bf E}} - \Delta_{\bf E}^{(\epsilon)}\ov{\cal F}\right)\bdot \left(c\;\nabla\btimes{\bf B} 
\;-\frac{}{} 4\pi\;\mathbb{T}_{\epsilon}{\bf J} \right) - \left(\fd{\ov{\cal F}}{{\bf B}} - \Delta_{\bf B}^{(\epsilon)}\ov{\cal F}\right)\bdot \left(c
\frac{}{}\nabla\btimes{\bf E}\right) \right] \nonumber \\
 & \equiv & \int_{\ov{\bf Z}}\;\fd{\ov{\cal F}}{\ov{f}}\;\pd{\ov{f}}{t} \;+\; \int_{\bf r}\;
\left( \fd{\ov{\cal F}}{{\bf E}}\bdot\pd{\bf E}{t} \;+\; \fd{\ov{\cal F}}{{\bf B}}\bdot\pd{\bf B}{t} \right),
\label{eq:MV_red_Ham_final}
\end{eqnarray}
where we used the reduced Maxwell equations \eqref{eq:B_dot} and \eqref{eq:E_dot_red}. We also substituted the reduced Vlasov equation 
\eqref{eq:Vlasov_red}:
\begin{equation}
\pd{\ov{f}}{t} \;=\; -\;\left\{ \ov{f},\frac{}{} \ov{K} \right\}_{\epsilon} \;-\frac{}{} e\;{\sf T}_{\epsilon}^{-1}{\bf E}\bdot\left\{ \ov{\bf X} + 
\ov{\vb{\rho}}_{\epsilon},\frac{}{} \ov{f} \right\}_{\epsilon} \;-\; \frac{\partial_{\epsilon}\ov{Z}^{\alpha}}{\partial t}\;
\pd{\ov{f}}{\ov{Z}^{\alpha}},
\end{equation}
where
\begin{eqnarray}
\frac{\partial_{\epsilon}\ov{Z}^{\alpha}}{\partial t}\;\pd{\ov{f}}{\ov{Z}^{\alpha}} & \equiv & {\sf T}_{\epsilon}^{-1}\left[\pd{({\sf T}_{\epsilon}
\ov{Z}^{\alpha})}{t}\right]\;\pd{\ov{f}}{\ov{Z}^{\alpha}} \\
 & = & \int_{\bf r} \left[ \pd{\bf E}{t}\bdot{\sf T}^{-\epsilon}\left(\fd{({\sf T}_{\epsilon}\ov{Z}^{\alpha})}{{\bf E}({\bf r})}\right)\;
\pd{\ov{f}}{\ov{Z}^{\alpha}} \;+\; \pd{\bf B}{t}\bdot{\sf T}^{-\epsilon}\left(\fd{({\sf T}_{\epsilon}\ov{Z}^{\alpha})}{{\bf B}({\bf r})}\right)
\;\pd{\ov{f}}{\ov{Z}^{\alpha}} \right], \nonumber 
\end{eqnarray}
so that we made use of the identity
\begin{equation}
\int_{\bf r} \left( \Delta_{\bf E}^{(\epsilon)}\ov{\cal F}\bdot\pd{\bf E}{t} \;+\; \Delta_{\bf B}^{(\epsilon)}\ov{\cal F}\bdot\pd{\bf B}{t} \right) 
\;\equiv\; \int_{\ov{\bf Z}}\;\frac{\partial_{\epsilon}\ov{Z}^{\alpha}}{\partial t}\;\pd{\ov{f}}{\ov{Z}^{\alpha}}\;\fd{\ov{\cal F}}{\ov{f}}
\end{equation}
in Eq.~\eqref{eq:MV_red_Ham}. Hence, the reduced Vlasov-Maxwell equations can be expressed as Eq.~\eqref{eq:MV_red_Ham_final} in terms of the reduced Hamiltonian functional \eqref{eq:Ham_red} and the reduced Vlasov-Maxwell bracket \eqref{eq:MV_red}.

\section{Summary}

The reduced Vlasov-Maxwell bracket \eqref{eq:MV_red} has been derived from the local Vlasov-Maxwell bracket \eqref{eq:MV_bracket_0} by Lie-transform methods based on the dynamical reduction associated with a near-identity phase-space transformation ${\cal T}^{\epsilon}$ and its inverse ${\cal T}^{-\epsilon}$. These phase-space transformations induce transformations on functions denoted by the push-forward operator 
${\sf T}^{-\epsilon}f_{0} \equiv f_{0} \circ {\cal T}^{-\epsilon}$ and the pull-back operator ${\sf T}^{\epsilon}\ov{f} \equiv \ov{f} \circ {\cal T}^{\epsilon}$. These pull-back and push-forward operators, in turn, induce transformations on functionals denoted by the meta-push-forward operator $\mathbb{T}^{\epsilon}$ and meta-pull-back operator $\mathbb{T}^{-\epsilon}$, which guarantee the Jacobi property 
\eqref{eq:MV_Jacobi_red} for the reduced Vlasov-Maxwell bracket.

In future work, we will explore the Hamiltonian formulation of the guiding-center Vlasov-Maxwell equations based on recent works by \cite{Bur_BMQ_2015} and \cite{Bur_BQ_2015}, as well as the variational formulations of guiding-center Vlasov-Maxwell theory derived by \cite{Brizard_Tronci_2016}. In particular, we will focus on investigating how the Hamiltonian properties of the reduced Vlasov-Maxwell bracket \eqref{eq:MV_red} survive (1) the {\it closure} problem: the process of truncation of the guiding-center Vlasov-Maxwell bracket at a finite order in $\epsilon$ (so far expressions have been derived at all orders in $\epsilon$) and (2) the {\it averaging} problem: the process by which the gyroangle is eliminated from the guiding-center Vlasov-Maxwell bracket (since guiding-center Vlasov-Maxwell equations do not involve the fast gyromotion time scale). In Eqs.~\eqref{eq:MV_red}-\eqref{eq:fd_ovB}, since the terms $\ov{\vb{\rho}}_{\epsilon}$ and $\Delta_{{\bf E},{\bf B}}^{(\epsilon)}$ are expected to contain gyroangle-independent and gyroangle-dependent contributions resulting from the guiding-center transformation, the gyroangle-averaging and closure problems of the guiding-center Vlasov-Maxwell bracket will be addressed explicitly.

\acknowledgments

Work by AJB, PJM, and J.W. Burby were supported by U.~S.~Dept.~of Energy contracts No.~DE-SC0006721 (AJB) and DE-SC0014032 (AJB), No.~DE-FG02-04ER-54742 (PJM), and No.~DE-FG02-
86ER53223 (JWB). Also, AJB and PJM have greatly benefited from the kind hospitality of the Centre de Physique Th\'{e}orique where this work was started. Work by MV was supported as part of the A*MIDEX
project (ANR-11-IDEX-0001-02), funded by the "Investissements d'Avenir", French Government program, managed by the FrenchNational Research Agency (ANR).

\appendix

\section{\label{sec:Jacobi_PB}Jacobi Property for the single-particle Poisson Bracket}

In this Appendix, we prove that the noncanonical particle Poisson bracket \eqref{eq:PB_def_div} satisfies the Jacobi identity
\begin{equation} 
{\rm JAC}(f,g,h) \;\equiv\; \left\{ f,\frac{}{} \{ g,\; h\}\right\} \;+\; \left\{ g,\frac{}{} \{ h,\; f\}\right\} \;+\; \left\{ h,\frac{}{} 
\{ f,\; g\}\right\} \;=\; 0,
\label{eq:PB_Jacobi}
\end{equation}
if the Poisson tensor $J^{\alpha\beta} \equiv \{ z^{\alpha},\; z^{\beta}\}$ satisfies the condition
\begin{equation}
J^{\alpha\sigma}\,\partial_{\sigma}J^{\beta\delta} \;+\; J^{\beta\sigma}\,\partial_{\sigma}J^{\delta\alpha} \;+\; J^{\delta\sigma}\,
\partial_{\sigma}J^{\alpha\beta} \;=\; 0.
\label{eq:Jacobi_condition}
\end{equation}
First, we write
\begin{equation}
\left\{ f,\frac{}{} \{ g,\; h\}\right\} \;=\; \partial_{\alpha}f\,J^{\alpha\sigma} \left[ \partial_{\sigma}J^{\beta\delta}\,\left(\partial_{\beta}g
\frac{}{}\partial_{\delta}h\right) + J^{\beta\delta} \left(\partial^{2}_{\beta\sigma}g\frac{}{}\partial_{\delta}h + \partial_{\beta}g\frac{}{}
\partial^{2}_{\delta\sigma}h\right)\right], 
\end{equation}
so that, with cyclic permutations of $(f,g,h)$, Eq.~\eqref{eq:PB_Jacobi} becomes
\begin{eqnarray}
{\rm JAC}(f,g,h) & = & \partial_{\alpha}f\,\partial_{\beta}g\,\partial_{\delta}h \left( J^{\alpha\sigma}\,\partial_{\sigma}J^{\beta\delta} \;+\frac{}{} 
J^{\beta\sigma}\,\partial_{\sigma}J^{\delta\alpha} + J^{\delta\sigma}\,\partial_{\sigma}J^{\alpha\beta} \right) \label{eq:JAC_def} \\
 &  &+\; \partial^{2}_{\alpha\sigma}f \left(\partial_{\beta}g\;K^{\beta\sigma|\delta\alpha}\; \partial_{\delta}h \right) \;+\; 
\partial^{2}_{\beta\sigma}g \left(\partial_{\delta}h\;K^{\delta\sigma|\alpha\beta}\; \partial_{\alpha}f \right) \nonumber \\
 &  &+\; \partial^{2}_{\delta\sigma}h \left(\partial_{\alpha}f\;K^{\alpha\sigma|\beta\delta}\; \partial_{\beta}g \right),
\nonumber
\end{eqnarray}
where we have introduced the fourth-rank tensor $K^{\alpha\mu|\beta\nu} \equiv J^{\alpha\mu}\,J^{\beta\nu} + J^{\beta\mu}\,J^{\nu\alpha}$. Next, we note that only the part of $K^{\alpha\mu|\beta\nu}$ that is symmetric with respect to the interchange $\mu\leftrightarrow\nu$ survives for each of the last terms in Eq.~\eqref{eq:JAC_def}, since it is associated with a second-order partial derivative $\partial^{2}_{\mu\nu}$. We thus replace $K^{\alpha\mu|\beta\nu}$ with its symmetric part
\begin{eqnarray}
\frac{1}{2}\;\left( K^{\alpha\mu|\beta\nu} \;+\frac{}{} K^{\alpha\nu|\beta\mu} \right) & = & \frac{1}{2} \left[ \left(J^{\alpha\mu}\,J^{\beta\nu} \;+\; 
J^{\beta\mu}\,J^{\nu\alpha}\right) \;+\frac{}{} \left(J^{\alpha\nu}\,J^{\beta\mu} \;+\; J^{\beta\nu}\,J^{\mu\alpha}\right) \right] \nonumber \\
 & = & \frac{1}{2} \left[ \left( J^{\alpha\mu} +\frac{}{} J^{\mu\alpha}\right)\,J^{\beta\nu} + 
J^{\beta\mu}\,\left( J^{\nu\alpha} +\frac{}{} J^{\alpha\nu}\right) \right] \equiv 0,
\label{eq:K_sym}
\end{eqnarray}
which vanishes identically as a result of the antisymmetry of the Poisson matrix. Hence, Eq.~\eqref{eq:JAC_def} now becomes
\[ {\rm JAC}(f,g,h) \;=\; \partial_{\alpha}f\,\partial_{\beta}g\,\partial_{\delta}h \left( J^{\alpha\sigma}\,\partial_{\sigma}J^{\beta\delta} \;+\frac{}{} 
J^{\beta\sigma}\,\partial_{\sigma}J^{\delta\alpha} + J^{\delta\sigma}\,\partial_{\sigma}J^{\alpha\beta} \right) \;=\; 0, \] 
which satisfies the Jacobi identity \eqref{eq:PB_Jacobi} if the Jacobi condition \eqref{eq:Jacobi_condition} is satisfied.

Another way to look at the Jacobi condition \eqref{eq:Jacobi_condition} is to consider instead the Lagrange tensor $\omega_{\alpha\beta}$, whose components are formally defined as the components of the inverse of the Poisson matrix (with components $J^{\alpha\beta}$), i.e., $J^{\alpha\sigma}\,\omega_{\sigma\beta} \equiv \delta^{\alpha}_{\beta}$. When expressed in terms of the Lagrange tensor, the Jacobi condition \eqref{eq:Jacobi_condition} is now expressed as
\begin{equation}
\partial_{\alpha}\omega_{\beta\sigma} \;+\; \partial_{\beta}\omega_{\sigma\alpha} \;+\; \partial_{\sigma}\omega_{\alpha\beta} \;=\; 0, 
\label{eq:three_form}
\end{equation}
which implies that the three-form 
\[ \exd\omega \;\equiv\; \frac{1}{3!} \left(\partial_{\alpha}\omega_{\beta\sigma} +\frac{}{} \partial_{\beta}\omega_{\sigma\alpha} + \partial_{\sigma}\omega_{\alpha\beta}\right) \exd z^{\alpha}\wedge\exd z^{\beta}\wedge\exd z^{\sigma} \;=\; 0 \]
is closed. This condition, in turn, is guaranteed by the definition
\begin{equation} 
\omega \;\equiv\; \exd\gamma \;=\; \frac{e}{2c}\,B^{i}\,\varepsilon_{ijk}\;\exd x^{j}\wedge\exd x^{k} \;+\; \exd p_{i}\wedge \exd x^{i},
\label{eq:PB_twoform}
\end{equation}
in terms of the one-form $\gamma \equiv [(e/c)\,{\bf A} + {\bf p}]\bdot\exd{\bf x}$, where the magnetic field ${\bf B} \equiv \nabla\btimes{\bf A}$ is defined in terms of the vector potential ${\bf A}$. It is now immediately clear that the closure of the two-form \eqref{eq:PB_twoform} is expressed in terms of the three-form
\begin{equation} 
\exd\omega \;=\; \frac{e}{c}\;(\nabla\bdot{\bf B})\; \exd x\wedge \exd y \wedge \exd z \;\equiv\; 0,
\label{eq:d_omega}
\end{equation}
which is guaranteed by Eq.~\eqref{eq:div_B}. The Jacobi condition \eqref{eq:three_form} based on the local two-form \eqref{eq:omega_0}:
\begin{eqnarray}
\exd\omega_{0} & = & \frac{e}{c}\,(\nabla\bdot{\bf B}_{0}^{*}) \exd x\wedge \exd y \wedge \exd z + \left( \left[ \frac{e}{c}\pd{{\bf B}_{0}^{*}}{p_{\|}} - \nabla\btimes\bhat\right]\,\exd p_{\|} + \left[ \frac{e}{c}\pd{{\bf B}_{0}^{*}}{\mu} - \nabla\btimes\left(\pd{p_{\bot}}{\mu}\wh{\bot}\right) \right]\exd\mu \right. \nonumber \\ 
 &  &\left.+ \left[ \frac{e}{c}\pd{{\bf B}_{0}^{*}}{\zeta} + \nabla\btimes\left(p_{\bot}\,\wh{\rho}\right) \right]\exd\zeta\right)^{k}\wedge \frac{1}{2}\varepsilon_{ijk}\,\exd x^{i}\wedge\exd x^{j} \;\equiv\; 0,
\label{eq:d_omega_0}
\end{eqnarray}
is automatically satisfied because $\nabla\bdot{\bf B}_{0}^{*} = 0$, $(e/c)\,\partial{\bf B}_{0}^{*}/\partial p_{\|} = \nabla\btimes\bhat$, $(e/c)\,
\partial{\bf B}_{0}^{*}/\partial\mu = \nabla\btimes(\wh{\bot}\,\partial p_{\bot}/\partial\mu)$, and $(e/c)\,\partial{\bf B}_{0}^{*}/\partial\zeta = -\,\nabla\btimes(p_{\bot}\,\wh{\rho})$.

\section{\label{sec:Jacobi_MV}Jacobi Property for the Vlasov-Maxwell Bracket}

In this Appendix, we prove the Jacobi identity \eqref{eq:MV_Jacobi} for the Vlasov-Maxwell bracket \eqref{eq:MV_bracket}, emphasizing the role that  is inherited from the Jacobi identity \eqref{eq:PB_Jacobi} for the noncanonical particle Poisson bracket \eqref{eq:PB_def}.  This is essentially the same proof as that given by \cite{pjm13}, which was first done in 1981, but here it is  considerably simplified. 

First, we write the Vlasov-Maxwell bracket in terms of functional derivatives with respect to $\psi^{a} = (f, {\bf E}, {\bf B})$ and the pairing  $\langle\ |\ \rangle$, casting it into the general form of  \cite{pjm82}:
\begin{equation}
\left[{\cal F},\frac{}{} {\cal G}\right] \;\equiv\; \left\langle \left. \fd{\cal F}{\psi^{a}}\;\right|\;\mathbb{J}^{ab}\;\fd{\cal G}{\psi^{b}}\right\rangle,
\label{eq:MV_O}
\end{equation}
where the antisymmetric Poisson matrix $J^{\alpha\beta}$ in the single-particle Poisson bracket is replaced with the anti-self-adjoint operator 
$\mathbb{J}^{ab}$ [see Eq.~\eqref{eq:PB_def}]. We note that, like the Jacobi condition \eqref{eq:Jacobi_condition} of the Poisson bracket, the Jacobi identity \eqref{eq:MV_Jacobi} for the Vlasov-Maxwell bracket only involves functional derivatives of the operator  $\mathbb{J}^{ab}$  (see the Bracket Theorem originally proven by \cite{pjm82}).  Next, we write
\begin{eqnarray}
\left[ {\cal K},\frac{}{} [{\cal F},\;{\cal G}]\right] & = & \int_{\bf z}\;f \left\{ {\cal K}_{f},\frac{}{} [{\cal F},\;{\cal G}]_{f} \right\} 
+ 4\pi\,c\;\int_{\bf r}\;\left( {\cal K}_{{\bf E}}\bdot\nabla\btimes [{\cal F},\;{\cal G}]_{\bf B} -\frac{}{} [{\cal F},\;{\cal G}]_{\bf E}\bdot\nabla\btimes{\cal K}_{{\bf B}} \right) \nonumber \\
 &  &-\;4\pi\,e\;\int_{{\bf z}_{0}}\;f \left( {\cal K}_{{\bf E}}\bdot\left\{ {\bf x},\frac{}{} [{\cal F},\;{\cal G}]_{f} \right\} \;-\frac{}{}
[{\cal F},\;{\cal G}]_{\bf E}\bdot\left\{ {\bf x},\frac{}{} {\cal K}_{f} \right\} \right),
\label{eq:MV_Jac_1}
\end{eqnarray}
where the functional derivative of the bracket $[{\cal F},\;{\cal G}]$, now denoted by subscripts,  involves the functional derivative of the anti-self-adjoint operator 
$\delta\mathbb{J}^{ab}/\delta\psi^{c}$. Hence, since the operator $\mathbb{J}^{ab}$ is independent of the electric field ${\bf E}$, we find
\begin{equation}
\left[{\cal F},\;{\cal G}\right]_{\bf E} \;=\; 0\,, 
\label{eq:FG_E}
\end{equation}
up to terms involving second functional  derivatives, which vanish by themselves in the Jacobi identity  by virtue of the Bracket Theorem.  Next, since the operator $\mathbb{J}^{ab}$ depends on the Vlasov distribution $f$ and the magnetic field ${\bf B}$, we find
\begin{eqnarray}
\left[{\cal F},\;{\cal G}\right]_{f} & = & \left\{ {\cal F}_{f},\frac{}{} {\cal G}_{f} \right\} \;-\; 4\pi\,e\; \left( {\cal F}_{{\bf E}}\bdot
\left\{ {\bf x},\frac{}{} {\cal G}_{f} \right\} \;-\; {\cal G}_{{\bf E}}\bdot\left\{ {\bf x},\frac{}{} {\cal F}_{f} \right\} \right), \label{eq:FG_f} \\
\left[{\cal F},\;{\cal G}\right]_{\bf B} & = & \frac{e}{c}\;\int_{\bf z}\,f\;\delta^{3}({\bf x} - {\bf r})\;\left\{{\bf x},\frac{}{} {\cal F}_{f}\right\}\btimes \left\{{\bf x},\frac{}{} {\cal G}_{f}\right\}, \label{eq:FG_B}
\end{eqnarray}
where Eq.~\eqref{eq:FG_B} follows from the noncanonical Poisson bracket \eqref{eq:PB_def} and the identity
\[ \pd{}{\bf B}\;\{ f,\; g\} \;=\; \frac{e}{m^{2}c}\;\pd{f}{\bf v}\btimes\pd{g}{\bf v} \;\equiv\; \frac{e}{c}\;\{{\bf x},\; f\}\btimes\{ {\bf x},\; g\}. \]

We now begin our proof by combining Eqs.~\eqref{eq:FG_E}-\eqref{eq:FG_B} into Eq.~\eqref{eq:MV_Jac_1} to obtain
\begin{eqnarray}
\left[ {\cal K},\frac{}{} [{\cal F},\;{\cal G}]\right] & = & \int_{\bf z}f \left\{ {\cal K}_{f},\frac{}{} [{\cal F},\;{\cal G}]_{f} \right\} 
+ 4\pi\,c\;\int_{\bf r}\left( {\cal K}_{{\bf E}}\bdot\nabla\btimes [{\cal F},\;{\cal G}]_{\bf B} \right) \nonumber \\
  & &- 4\pi\,e\;\int_{\bf z}f \left( {\cal K}_{{\bf E}}\bdot\left\{ {\bf x},\frac{}{} [{\cal F},\;{\cal G}]_{f} \right\} \right).
\label{eq:MV_Jac_2}
\end{eqnarray}
By using Eq.~\eqref{eq:FG_f}, we write the expression for the Vlasov sub-bracket
\begin{eqnarray}
\left\{ {\cal K}_{f},\frac{}{} [{\cal F},\;{\cal G}]_{f} \right\} & = & \left\{ {\cal K}_{f},\frac{}{} \left\{ {\cal F}_{f},\; {\cal G}_{f} \right\}
\right\} \;-\; 4\pi\,e\; \left\{ {\cal K}_{f},\frac{}{}  \left( {\cal F}_{{\bf E}}\bdot\frac{}{}\left\{ {\bf x},\; {\cal G}_{f} \right\}\right) \right\} 
\nonumber \\
 &  &+\; 4\pi\,e\; \left\{ {\cal K}_{f},\frac{}{}  \left( {\cal G}_{{\bf E}}\bdot\frac{}{}\left\{ {\bf x},\; {\cal F}_{f} \right\}\right) \right\},
\label{eq:JAC_1}
\end{eqnarray}
and the expression for the interaction sub-bracket
\begin{eqnarray}
&&{\cal K}_{{\bf E}}\bdot\left\{ {\bf x},\frac{}{} [{\cal F},\;{\cal G}]_{f} \right\} = {\cal K}_{{\bf E}}\bdot\left\{ {\bf x},\frac{}{}
\left\{ {\cal F}_{f},\; {\cal G}_{f} \right\} \right\} 
\nonumber\\
&& \hspace{ 4cm}
- 4\pi\,e\left[ {\cal K}_{{\bf E}}\bdot{\sf M}({\cal G}_{f})\bdot{\cal F}_{{\bf E}} -\frac{}{} 
{\cal K}_{{\bf E}}\bdot{\sf M}({\cal F}_{f})\bdot{\cal G}_{{\bf E}}\right],
\label{eq:JAC_2}
\end{eqnarray}
where the matrix ${\sf M}(\cdots) \equiv \left\{ {\bf x},\frac{}{} \left\{ {\bf x},\; (\cdots) \right\}\right\}$ is symmetric (which is proved by using the Jacobi identity for $\{\;,\;\}$ and $\{x^{i},\;x^{j}\} \equiv 0$). By using Eq.~\eqref{eq:FG_B}, on the other hand, we write the expression for the Maxwell sub-bracket
\begin{eqnarray}
\int_{\bf r}\;{\cal K}_{{\bf E}}\bdot\nabla\btimes [{\cal F},\;{\cal G}]_{\bf B} & = & \int_{\bf r}\;\nabla\btimes
{\cal K}_{{\bf E}}\bdot[{\cal F},\;{\cal G}]_{\bf B} \label{eq:JAC_3} \\
 & = & \frac{e}{c}\int_{\bf z}\;f \left( \left\{{\cal K}_{\bf E},\;{\cal F}_{f}\right\}\bdot\left\{{\bf x},\; {\cal G}_{f}\right\} \;-\frac{}{}
\left\{{\cal K}_{\bf E},\;{\cal G}_{f}\right\}\bdot\left\{{\bf x},\; {\cal F}_{f}\right\} \right) \nonumber \\
 & = & \frac{e}{c}\int_{\bf z} f \left[ {\cal K}_{\bf E}\bdot\left\{{\bf x},\frac{}{} \{ {\cal F}_{f}, {\cal G}_{f}\} \right\} + \left\{
\left( {\cal K}_{\bf E}\bdot\frac{}{}\{{\bf x}, {\cal G}_{f}\}\right),\frac{}{} {\cal F}_{f} \right\} \right. \nonumber \\
 &  &\left.-\; \left\{ \left( {\cal K}_{\bf E}\bdot\frac{}{}\{{\bf x}, {\cal F}_{f}\}\right),\frac{}{} {\cal G}_{f} \right\} \right],
\nonumber
\end{eqnarray}
where we first integrated by parts (with $\{(\cdots),\;{\bf x}\}\bdot\nabla{\cal K}_{\bf E} = \{ (\cdots),\; {\cal K}_{\bf E}\}$) and then used the Jacobi identity for $\{\;,\;\}$.

Next, we combine Eqs.~\eqref{eq:JAC_1}-\eqref{eq:JAC_3} into Eq.~\eqref{eq:MV_Jac_2} to obtain an expansion in powers of $(4\pi\,e)$:
\begin{equation}
\left[ {\cal K},\frac{}{} [{\cal F},\;{\cal G}]\right] \;\equiv\; \int_{\bf z} f \left[ {\cal J}_{0}({\cal K},{\cal F},{\cal G}) \;+\frac{}{} 
4\pi\,e\;{\cal J}_{1}({\cal K},{\cal F},{\cal G}) \;+\; (4\pi\,e)^{2}\;{\cal J}_{2}({\cal K},{\cal F},{\cal G}) \right], 
\label{eq:JAC_4}
\end{equation}
where the Jacobi terms ${\cal J}_{n}$ ($n = 0,1,2$) are
\begin{eqnarray}
{\cal J}_{0}({\cal K},{\cal F},{\cal G}) & = & \left\{ {\cal K}_{f},\frac{}{} \left\{ {\cal F}_{f},\; {\cal G}_{f} \right\}\right\}, \label{eq:J_0} \\
{\cal J}_{1}({\cal K},{\cal F},{\cal G}) & = & \left\{ {\cal K}_{f},\frac{}{} \left( {\cal G}_{\bf E}\bdot\frac{}{}\{{\bf x},\; {\cal F}_{f}\} \;-\;
{\cal F}_{\bf E}\bdot\frac{}{}\{{\bf x},\; {\cal G}_{f}\}\right)\right\} \;+\; \left\{ {\cal G}_{f},\frac{}{} \left( {\cal K}_{\bf E}\bdot\frac{}{}
\{{\bf x},\; {\cal F}_{f}\}\right)\right\} \nonumber \\
 &  &-\; \left\{ {\cal F}_{f},\frac{}{} \left( {\cal K}_{\bf E}\bdot\frac{}{}\{{\bf x},\; {\cal G}_{f}\}\right)\right\}, \label{eq:J_1} \\
{\cal J}_{2}({\cal K},{\cal F},{\cal G}) & = & {\cal K}_{{\bf E}}\bdot{\sf M}({\cal G}_{f})\bdot{\cal F}_{{\bf E}} \;-\; {\cal K}_{{\bf E}}\bdot
{\sf M}({\cal F}_{f})\bdot{\cal G}_{{\bf E}}. \label{eq:J_2}
\end{eqnarray}
The Jacobi identity \eqref{eq:MV_Jacobi} for the Vlasov-Maxwell bracket is now replaced by the Jacobi identities
\begin{equation}
{\cal J}_{n}({\cal K},{\cal F},{\cal G}) \;+\; {\cal J}_{n}({\cal F},{\cal G},{\cal K}) \;+\; {\cal J}_{n}({\cal G},{\cal K},{\cal F}) \;=\; 0,
\label{eq:JAC_n}
\end{equation}
which hold separately for $n = 0,1,2$. The Jacobi identity \eqref{eq:JAC_n} for $n = 0$ is immediately satisfied because of the particle phase-space Jacobi identity \eqref{eq:PB_Jacobi}. Lastly, it is a simple task to show that Jacobi identities \eqref{eq:JAC_n} are satisfied for $n = 1$ and 2 by pairwise cancellations.

\section{\label{sec:comm_op}Operator Commutation Relations}

In this Appendix, we derive the expression \eqref{eq:delta_T_def} for a general phase-space transformation. We first consider the case of a near-identity transformation and consider the push-forward ${\sf T}^{-\epsilon}\delta f$:
\begin{equation}
{\sf T}^{-\epsilon}\,\delta f \;=\; \delta\left({\sf T}^{-\epsilon}f\right) \;+\; \left(\left[{\sf T}^{-\epsilon},\frac{}{}\delta\right]\,
{\sf T}^{\epsilon}\right)\;{\sf T}^{-\epsilon}f \;\equiv\; \delta\ov{f} \;+\; \left(\left[{\sf T}^{-\epsilon},\frac{}{}\delta\right]\,
{\sf T}^{\epsilon}\right)\;\ov{f},
\label{eq:delta_T}
\end{equation}
where $\left[{\sf T}^{-\epsilon},\;\delta\right]\,{\sf T}^{\epsilon} \equiv {\sf T}^{-\epsilon}(\delta{\sf T}^{\epsilon}) - \delta$ is explicitly calculated up to third order as
\begin{eqnarray}
{\sf T}^{-\epsilon}(\delta{\sf T}^{\epsilon}) - \delta & \equiv & \left(\cdots {\sf T}_{3}^{-1}{\sf T}_{2}^{-1}{\sf T}_{1}^{-1}\right)\;\delta\;\left(
{\sf T}_{1}{\sf T}_{2}{\sf T}_{3}\frac{}{}\cdots\right) \;-\; \delta \label{eq:delta_comm} \\
 & = & \epsilon\,\delta{\sf G}_{1}\cdot\exd \;+\; \epsilon^{2}\,\left[ \delta{\sf G}_{2} \;+\; \frac{1}{2} \left( \delta{\sf G}_{1}\cdot\exd{\sf G}_{1} \;-\frac{}{} {\sf G}_{1}\cdot\exd\delta{\sf G}_{1}\right)\right]\cdot\exd \nonumber \\
 &  &+\; \epsilon^{3} \left\{ \delta{\sf G}_{3} \;+\frac{}{} 
\left( \delta{\sf G}_{1}\cdot\exd{\sf G}_{2} \;-\frac{}{} {\sf G}_{2}\cdot\exd\delta{\sf G}_{1}\right) \right. \nonumber \\
 &  &\left.+ \frac{1}{6} \left[ \delta{\sf G}_{1}\cdot\exd\left({\sf G}_{1}\cdot\exd{\sf G}_{1}\right) + {\sf G}_{1}\cdot\exd\left({\sf G}_{1}\cdot\exd\delta{\sf G}_{1}\right) -\frac{}{} 2\;{\sf G}_{1}\cdot\exd\left(\delta{\sf G}_{1}\cdot\exd{\sf G}_{1}\right) \right] \right\}\cdot\exd 
\nonumber
\end{eqnarray}
Here, we note that $\left[{\sf T}^{-\epsilon},\;\delta\right]\,{\sf T}^{\epsilon} \equiv (\cdots)^{\alpha}\;\partial/\partial\ov{Z}^{\alpha}$ is a phase-space vector field, whose components $(\cdots)^{\alpha}$ need to be calculated. 

For this purpose, we introduce the particle phase-space displacement
\begin{eqnarray}
\Delta^{\alpha} & \equiv & -\,\epsilon\,G_{1}^{\alpha} \;-\; \epsilon^{2} \left( G_{2}^{\alpha} \;+\; \frac{1}{2}\,{\sf G}_{1}\cdot\exd G_{1}^{\alpha}
\right) \nonumber \\
 &  &-\; \epsilon^{3} \left[ G_{3}^{\alpha} \;+\; {\sf G}_{1}\cdot\exd G_{2}^{\alpha} \;+\; \frac{1}{6}\; {\sf G}_{1}\cdot\exd\left({\sf G}_{1}\cdot\exd 
G_{1}^{\alpha}\right) \right] - \cdots,
\end{eqnarray}
and its variation (up to third order)
\begin{eqnarray}
\delta\Delta^{\alpha} & \equiv & -\,\epsilon\,\delta G_{1}^{\alpha} \;-\; \epsilon^{2} \left[ \delta G_{2}^{\alpha} \;+\; \frac{1}{2}\,\left(
\delta{\sf G}_{1}\cdot\exd G_{1}^{\alpha} \;+\frac{}{} {\sf G}_{1}\cdot\exd \delta G_{1}^{\alpha} \right) \right] \nonumber \\
 &  &-\; \epsilon^{3} \left\{ \delta G_{3}^{\alpha} \;+\; \left(\delta{\sf G}_{1}\cdot\exd G_{2}^{\alpha} \;+\frac{}{} {\sf G}_{1}\cdot\exd \delta G_{2}^{\alpha} \right) \right.   \\
 &  &\left.+\; \frac{1}{6}\; \left[ \delta{\sf G}_{1}\cdot\exd\left({\sf G}_{1}\cdot\exd G_{1}^{\alpha}\right) \;+\;
{\sf G}_{1}\cdot\exd\left(\delta{\sf G}_{1}\cdot\exd G_{1}^{\alpha}\right) \;+\frac{}{} {\sf G}_{1}\cdot\exd\left({\sf G}_{1}\cdot\exd \delta
G_{1}^{\alpha}\right)\right] \right\},
\nonumber
\end{eqnarray}
so that we obtain the push-forward relation (up to third order)
\begin{eqnarray}
{\sf T}^{-\epsilon}\left(\delta \Delta^{\alpha}\right) & = & -\;\epsilon\,\delta G_{1}^{\alpha} \;-\; \epsilon^{2}\,\left[ \delta G_{2}^{\alpha} \;+\; 
\frac{1}{2} \left( \delta{\sf G}_{1}\cdot\exd G_{1}^{\alpha} \;-\frac{}{} {\sf G}_{1}\cdot\exd \delta G_{1}^{\alpha}\right)\right] 
\label{eq:delta_comm_Delta} \\
 &  &-\; \epsilon^{3} \left\{ \delta G_{3}^{\alpha} \;+\frac{}{} \left( \delta{\sf G}_{1}\cdot\exd G_{2}^{\alpha} \;-\frac{}{} {\sf G}_{2}\cdot\exd
\delta G_{1}^{\alpha} \right) \right. \nonumber \\
 &  &\left.+\; \frac{1}{6} \left[ \delta{\sf G}_{1}\cdot\exd\left({\sf G}_{1}\cdot\exd G_{1}^{\alpha}\right) + {\sf G}_{1}\cdot\exd\left(
{\sf G}_{1}\cdot\exd\delta G_{1}^{\alpha}\right) -\frac{}{} 2\;{\sf G}_{1}\cdot\exd\left(\delta{\sf G}_{1}\cdot\exd G_{1}^{\alpha}\right) \right] \right\}.
\nonumber
\end{eqnarray}
By comparing Eq.~\eqref{eq:delta_comm} with Eq.~\eqref{eq:delta_comm_Delta}, we obtain the variation-commutation relation
\begin{equation}
\left[{\sf T}^{-\epsilon},\frac{}{} \delta\right]\,{\sf T}^{\epsilon} \;\equiv\; -\;{\sf T}^{-\epsilon}\left(\delta\Delta^{\alpha}\right)\;\pd{}{\ov{Z}^{\alpha}},
\label{eq:var_commute}
\end{equation}
and Eq.~\eqref{eq:delta_T} becomes
\begin{equation}
{\sf T}^{-\epsilon}(\delta f) \;\equiv\; \delta\ov{f} \;-\; \left({\sf T}^{-\epsilon}\delta\Delta^{\alpha}\right)\;\pd{\ov{f}}{\ov{Z}^{\alpha}},
\label{eq:push_deltaf}
\end{equation}
where
\begin{eqnarray}
{\sf T}^{-\epsilon}\delta\Delta^{\alpha} & \equiv & {\sf T}^{-\epsilon}\left( \delta{\bf E}({\bf r})\bdot\fd{\Delta^{\alpha}}{{\bf E}({\bf r})} 
\;+\; \delta{\bf B}({\bf r})\bdot\fd{\Delta^{\alpha}}{{\bf B}({\bf r})} \right) \nonumber \\
 & = & \delta{\bf E}({\bf r})\bdot\left[{\sf T}^{-\epsilon}\left( 
\fd{\Delta^{\alpha}}{{\bf E}({\bf r})} \right)\right] \;+\; \delta{\bf B}({\bf r})\bdot\left[{\sf T}^{-\epsilon}\left( 
\fd{\Delta^{\alpha}}{{\bf B}({\bf r})} \right)\right].
\label{eq:T_delta_Delta}
\end{eqnarray}
We note that ${\sf T}^{-\epsilon}\delta\Delta^{\alpha}$ may be expressed in terms of $\delta\ov{\Delta}^{\alpha}$ according to the identity
\[ {\sf T}^{-\epsilon}\delta\Delta^{\alpha} \;\equiv\; {\sf T}^{-\epsilon}\delta\left({\sf T}^{\epsilon}\frac{}{}\ov{\Delta}^{\alpha}\right) \;=\;
\delta\ov{\Delta}^{\alpha} \;+\; \left( \left[{\sf T}^{-\epsilon},\;\delta\right]\;{\sf T}^{\epsilon}\right)\;\ov{\Delta}^{\alpha}, \]
which yields the matrix relation
\begin{equation}
{\sf T}^{-\epsilon}\delta\vb{\Delta} \;=\; \left({\bf I} \;+\; \pd{\ov{\vb{\Delta}}}{\ov{\bf Z}}\right)^{-1}\bdot\delta\ov{\vb{\Delta}},
\label{eq:delta_Delta_ovDelta_C}
\end{equation}
where ${\bf I} \;+\; \partial\ov{\vb{\Delta}}/\partial\ov{\bf Z}$ is the matrix with components $\partial({\sf T}^{-\epsilon}z^{\alpha})/\partial
\ov{Z}^{\beta}$.

For a general phase-space transformation, we find Eq.~\eqref{eq:delta_T_def}
\begin{eqnarray}
{\sf T}^{-1}(\delta f) & = & \delta\left({\sf T}^{-1}f\right) \;+\; \left(\left[{\sf T}^{-1},\frac{}{} \delta\right]\,{\sf T}\right)\;{\sf T}^{-1}f \nonumber \\
 & \equiv & \delta\ov{f} \;+\; {\sf T}^{-1}\left[\delta\frac{}{}\left({\sf T}\ov{Z}^{\alpha}\right)\right]\;\pd{\ov{f}}{\ov{Z}^{\alpha}}
\label{eq:delta_T_f}
\end{eqnarray}
as an extension of Eq.~\eqref{eq:push_deltaf}.

\section{\label{sec:explicit_bracket}Explicit Reduced Functional Derivatives}

In this Appendix, we derive the definitions for the functional derivatives appearing in the reduced Vlasov-Maxwell bracket \eqref{eq:MV_red}. For this purpose, we consider an arbitrary reduced functional $\ov{\cal F}[\ov{f},{\bf E},{\bf B}]$, with the functional variation $\delta\ov{\cal F}$ defined as
\begin{equation}
\delta\ov{\cal F} \;\equiv\; \int_{\ov{\bf Z}}\;\delta\ov{f}(\ov{\bf Z})\;\fd{\ov{\cal F}}{\ov{f}(\ov{\bf Z})} \;+\; \int_{\bf r}
\left(\delta{\bf E}({\bf r})\bdot\fd{\ov{\cal F}}{{\bf E}({\bf r})} + \delta{\bf B}({\bf r})\bdot\fd{\ov{\cal F}}{{\bf B}({\bf r})}\right).
\label{eq:ovF_var_def}
\end{equation}
Using the meta-operations $\mathbb{T}_{\epsilon} \equiv \mathbb{T}^{\epsilon}\,\mathbb{T}_{0}$ and $\mathbb{T}_{\epsilon}^{-1} \equiv 
\mathbb{T}_{0}^{-1}\mathbb{T}^{-\epsilon}$, we also find
\begin{eqnarray}
\delta\ov{\cal F} \;=\; \mathbb{T}_{\epsilon}\left[\delta\left(\mathbb{T}_{\epsilon}^{-1}\ov{\cal F}\right)\right] & \equiv & 
\int_{\ov{\bf Z}}\;\ov{\cal J}\;{\sf T}^{-1}_{\epsilon}\left(\delta f\right)\;{\sf T}^{-1}_{\epsilon}\left[{\cal J}^{-1}
\fd{\left(\mathbb{T}_{\epsilon}^{-1}\ov{\cal F}\right)}{f} \right] \nonumber \\
 &  &+\; \int_{\bf r}\left(\delta{\bf E}({\bf r})\bdot
\fd{\left(\mathbb{T}_{\epsilon}^{-1}\ov{\cal F}\right)}{{\bf E}({\bf r})} + \delta{\bf B}({\bf r}) \bdot
\fd{\left(\mathbb{T}_{\epsilon}^{-1}\ov{\cal F}\right)}{{\bf B}({\bf r})}\right),
\label{eq:deltaF_primitive}
\end{eqnarray}
where ${\sf T}^{-1}_{\epsilon}f \equiv \ov{f}$ and the push-forward ${\sf T}^{-1}_{\epsilon}(\delta f)$ is given by Eq.~\eqref{eq:delta_T_f}.

\subsection{Reduced functional variations}

By substituting Eqs.~\eqref{eq:push_deltaf}-\eqref{eq:T_delta_Delta} into Eq.~\eqref{eq:deltaF_primitive}, we obtain
\begin{eqnarray}
\mathbb{T}_{\epsilon}\left[\delta\left(\mathbb{T}_{\epsilon}^{-1}\ov{\cal F}\right)\right] & = & \int_{\ov{\bf Z}}\;\ov{\cal J}\;
\left[ \delta\ov{f} \;+\; {\sf T}^{-1}_{\epsilon}\left[\delta\frac{}{}\left({\sf T}_{\epsilon}\ov{Z}^{\alpha}\right)\right]\;\pd{\ov{f}}{\ov{Z}^{\alpha}}\right]\;{\sf T}^{-1}_{\epsilon}\left[{\cal J}^{-1}\;\fd{\left(\mathbb{T}_{\epsilon}^{-1}\ov{\cal F}\right)}{f({\bf z}_{0})} \right] \nonumber \\
 &  &+\; \int_{\bf r}\left(\delta{\bf E}({\bf r})\bdot\fd{\left(\mathbb{T}_{\epsilon}^{-1}\ov{\cal F}\right)}{{\bf E}({\bf r})} + \delta{\bf B}({\bf r}) \bdot\fd{\left(\mathbb{T}_{\epsilon}^{-1}\ov{\cal F}\right)}{{\bf B}({\bf r})}\right).
\label{eq:delta_F_ovF}
\end{eqnarray}
By comparing Eq.~\eqref{eq:delta_F_ovF} with Eq.~\eqref{eq:ovF_var_def}, we find the reduced functional derivatives
\begin{eqnarray}
{\sf T}^{-1}_{\epsilon}\left[{\cal J}^{-1}\;\fd{\left(\mathbb{T}_{\epsilon}^{-1}\ov{\cal F}\right)}{f({\bf z}_{0})} \right] & \equiv & \ov{\cal J}^{-1}\;
\fd{\ov{\cal F}}{\ov{f}(\ov{\bf Z})}, 
\label{eq:fd_ovF_C} \\
\fd{\left(\mathbb{T}_{\epsilon}^{-1}\ov{\cal F}\right)}{{\bf E}({\bf r})} & \equiv & \fd{\ov{\cal F}}{{\bf E}({\bf r})} \;-\; \int_{\ov{\bf Z}}\;\left[
{\sf T}^{-1}_{\epsilon}\left(\fd{({\sf T}_{\epsilon}\ov{Z}^{\alpha})}{{\bf E}({\bf r})}\right)\right]\;\pd{\ov{f}}{\ov{Z}^{\alpha}}\;
\fd{\ov{\cal F}}{\ov{f}(\ov{\bf Z})},
\label{eq:fd_ovE_C} \\
\fd{\left(\mathbb{T}_{\epsilon}^{-1}\ov{\cal F}\right)}{{\bf B}({\bf r})} & \equiv & \fd{\ov{\cal F}}{{\bf B}({\bf r})} \;-\; \int_{\ov{\bf Z}}\;\left[
{\sf T}^{-1}_{\epsilon}\left(\fd{({\sf T}_{\epsilon}\ov{Z}^{\alpha})}{{\bf B}({\bf r})}\right)\right]\;\pd{\ov{f}}{\ov{Z}^{\alpha}}\;
\fd{\ov{\cal F}}{\ov{f}(\ov{\bf Z})},
\label{eq:fd_ovB_C}
\end{eqnarray}
which are used in the reduced Vlasov-Maxwell bracket \eqref{eq:MV_red}.

\subsection{Reduced Jacobian variation}

In the general case of a phase-space transformation that involves the fields $({\bf E},{\bf B})$, the variation of the reduced Jacobian 
$\delta\ov{\cal J}_{\epsilon}$ must be calculated. First, we consider the variation of the reduced Jacobian \eqref{eq:Jac_epsilon}:
\begin{eqnarray}
\delta\ov{\cal J}_{\epsilon} & \equiv & -\;\pd{}{z^{\alpha}}\left[ \frac{}{} \left(\epsilon\,\delta G_{1}^{\alpha} +\frac{}{} \epsilon^{2}\,
\delta G_{2}^{\alpha} + \cdots\right) - \frac{\epsilon^{2}}{2}\;\delta G_{1}^{\alpha}\;\pd{}{z^{\beta}}\left({\cal J}\frac{}{} G_{1}^{\beta} + \cdots\right) \right. \nonumber \\
 &  &\left.\hspace*{0.7in}- \frac{\epsilon^{2}}{2}\;G_{1}^{\alpha}\;\pd{}{z^{\beta}}\left(\frac{}{} \delta G_{1}^{\beta} + \cdots\right) + \cdots \right].
\label{eq:ov_Jac_delta}
\end{eqnarray}
Using Eq.~\eqref{eq:delta_comm_Delta}, we readily find for a near-identity transformation:
\begin{equation}
\delta\ov{\cal J}_{\epsilon} \;\equiv\; \pd{}{\ov{Z}^{\alpha}} \left[ \ov{\cal J}_{\epsilon}\frac{}{}\left({\sf T}^{-\epsilon}\delta{\Delta}^{\alpha}\right)\right].
\label{eq:Jacobian_delta}
\end{equation}
Next, the extension of Eq.~\eqref{eq:Jacobian_delta} to the case of a general phase-space transformation yields
\begin{equation}
\delta\ov{\cal J} \;=\; -\;\pd{}{\ov{Z}^{\alpha}}\left[\ov{\cal J}\;{\sf T}_{\epsilon}^{-1}\left(\delta\frac{}{}{\sf T}_{\epsilon}\ov{Z}^{\alpha}
\right)\right].
\label{eq:Jacobian_T_delta}
\end{equation}

\subsection{Reduced Hamiltonian functional variation}

Lastly, we note that, using the reduced Hamiltonian functional \eqref{eq:Ham_red}, we find the reduced Hamiltonian functional variation
\begin{eqnarray}
\delta\ov{\cal H} & = & \int_{\bf r} \left( \delta{\bf E}\bdot\frac{\bf E}{4\pi} \;+\; \delta{\bf B}\bdot\frac{\bf B}{4\pi}\right) \;+\; \int_{\ov{\bf Z}}
\left[ \delta\ov{f}\;\ov{\cal J}\,\ov{K} \;+\; \ov{f} \left( \delta\ov{\cal J}\;\ov{K} \;+\frac{}{} \ov{\cal J}\;\delta\ov{K} \right) \right] \nonumber \\
 & = & \int_{\bf r} \left( \delta{\bf E}\bdot\frac{\bf E}{4\pi} \;+\; \delta{\bf B}\bdot\frac{\bf B}{4\pi}\right) \;+\; \int_{\ov{\bf Z}}
\delta\ov{f}\;\ov{\cal J}\,\ov{K} \nonumber \\
 &  &-\;  \int_{\ov{\bf Z}}\ov{f}\;\pd{}{\ov{Z}^{\alpha}} \left[\ov{\cal J}\,\ov{K} \;
\left({\sf T}^{-1}_{\epsilon}\frac{}{}\delta({\sf T}_{\epsilon}\ov{Z}^{\alpha})\right) \right],
\label{eq:delta_ovH_def}
\end{eqnarray}
where we used Eq.~\eqref{eq:Jacobian_T_delta} and
\begin{eqnarray}
\delta\ov{K} & = & \delta\left({\sf T}^{-1}_{\epsilon}K\right) \;=\; {\sf T}^{-1}_{\epsilon}(\delta K) \;+\; \left(\left[\delta,\;{\sf T}^{-1}_{\epsilon}\right]\frac{}{}{\sf T}_{\epsilon}\right){\sf T}^{-1}_{\epsilon}K \nonumber \\
 & \equiv & -\;{\sf T}^{-1}_{\epsilon}\left(\delta({\sf T}_{\epsilon}\ov{Z}^{\alpha})\right)\;\pd{\ov K}{\ov{Z}^{\alpha}},
\end{eqnarray}
with $\delta K \equiv 0$ (i.e., the particle kinetic energy is field-independent). If we now integrate by parts the Vlasov part of 
Eq.~\eqref{eq:delta_ovH_def}, we find
\begin{eqnarray}
\delta\ov{\cal H} & = & \int_{\bf r} \left( \delta{\bf E}\bdot\frac{\bf E}{4\pi} \;+\; \delta{\bf B}\bdot\frac{\bf B}{4\pi}\right) \;+\; \int_{\ov{\bf Z}}
\left[ \delta\ov{f} \;+\; {\sf T}^{-1}_{\epsilon}\left(\delta({\sf T}_{\epsilon}\ov{Z}^{\alpha})\right)\;\pd{\ov{f}}{\ov{Z}^{\alpha}} \right]
\ov{\cal J}\,\ov{K} \nonumber \\
 & \equiv & \int_{\bf r} \left( \delta{\bf E}({\bf r})\bdot\fd{\ov{\cal H}}{{\bf E}({\bf r})} + \delta{\bf B}({\bf r})\bdot
\fd{\ov{\cal H}}{{\bf B}({\bf r})}\right) + \int_{\ov{\bf Z}} \delta\ov{f}(\ov{\bf Z})\;\fd{\ov{\cal H}}{\ov{f}(\ov{\bf Z})},
\end{eqnarray}
where $\delta\ov{\cal H}/\delta\ov{f} \equiv \ov{\cal J}\,\ov{K}$, and
\begin{eqnarray}
\fd{\ov{\cal H}}{{\bf E}({\bf r})} & = & \frac{{\bf E}({\bf r})}{4\pi} \;+\; \int_{\ov{\bf Z}}\;{\sf T}^{-1}_{\epsilon}\left(
\frac{\delta({\sf T}_{\epsilon}\ov{Z}^{\alpha})}{{\bf E}({\bf r})}\right)\;\pd{\ov{f}}{\ov{Z}^{\alpha}}\;\ov{\cal J}\,\ov{K} \;\equiv\;
\frac{{\bf E}({\bf r})}{4\pi} \;+\; \Delta_{\bf E}^{(\epsilon)}\ov{\cal H}, \label{eq:ovH_varE_C} \\
\fd{\ov{\cal H}}{{\bf B}({\bf r})} & = & \frac{{\bf B}({\bf r})}{4\pi} \;+\; \int_{\ov{\bf Z}}\;{\sf T}^{-1}_{\epsilon}\left(
\frac{\delta({\sf T}_{\epsilon}\ov{Z}^{\alpha})}{{\bf B}({\bf r})}\right)\;\pd{\ov{f}}{\ov{Z}^{\alpha}}\;\ov{\cal J}\,\ov{K} \;\equiv\;
\frac{{\bf B}({\bf r})}{4\pi} \;+\; \Delta_{\bf B}^{(\epsilon)}\ov{\cal H}. \label{eq:ovH_varB_C}
\end{eqnarray}
These expressions can now be used in Eqs.~\eqref{eq:fd_ovHam_ovE}-\eqref{eq:fd_ovHam_ovB}.

\bibliographystyle{jpp}

\bibliography{lift}

\end{document}